\documentclass[12pt]{article}
\usepackage{latexsym,epsf}

\makeatletter
\@addtoreset{equation}{section}

\makeatother

\begin{document}

\title{Corrections to Finite-Size Scaling in the Lattice $N$-Vector Model
for $N=\infty$}
\author{
  \\
  {\small Sergio Caracciolo}              \\[-0.2cm]
  {\small\it Scuola Normale Superiore and INFN -- Sezione di Pisa}  \\[-0.2cm]
  {\small\it I-56100 Pisa, ITALIA}          \\[-0.2cm]
  {\small Internet: {\tt sergio.caracciolo@sns.it}}     \\[-0.2cm]
  \\[-0.1cm]  \and
  {\small Andrea Pelissetto}              \\[-0.2cm]
  {\small\it Dipartimento di Fisica and INFN -- Sezione di Pisa}    \\[-0.2cm]
  {\small\it Universit\`a degli Studi di Pisa}        \\[-0.2cm]
  {\small\it I-56100 Pisa , ITALIA}          \\[-0.2cm]
  {\small Internet: {\tt pelissetto@sabsns.sns.it}}   \\[-0.2cm]
  {\protect\makebox[5in]{\quad}}  
  \\
}
\vspace{0.5cm}

\maketitle
\thispagestyle{empty}   

\vspace{0.2cm}

\begin{abstract}
We compute the corrections to finite-size scaling for the $N$-vector
model on the square lattice in the large-$N$ limit. 
We find that corrections behave as $\log L/L^2$. For tree-level
improved hamiltonians corrections behave as $1/L^2$. In
general $l$-loop improvement is expected to reduce this behaviour
to $1/(L^2 \log^l L)$. We show that the finite-size-scaling and 
the perturbative limit do not commute in the calculation of the
corrections to finite-size scaling. We present a detailed study
of the corrections for the $RP^\infty$-model.

\end{abstract}

\clearpage

\newcommand{\be}{\begin{equation}}
\newcommand{\ee}{\end{equation}}
\newcommand{\<}{\langle}
\renewcommand{\>}{\rangle}

\def\spose#1{\hbox to 0pt{#1\hss}}
\def\ltapprox{\mathrel{\spose{\lower 3pt\hbox{$\mathchar"218$}}
 \raise 2.0pt\hbox{$\mathchar"13C$}}}
\def\gtapprox{\mathrel{\spose{\lower 3pt\hbox{$\mathchar"218$}}
 \raise 2.0pt\hbox{$\mathchar"13E$}}}

\def\bsigma{\mbox{\protect\boldmath $\sigma$}}
\def\bpi{\mbox{\protect\boldmath $\pi$}}
\def\btau{\mbox{\protect\boldmath $\tau$}}
\def\hatp{\hat p}
\def\hatl{\hat l}

\def\msbar{ {\overline{\hbox{\scriptsize MS}}} }
\def\normalmsbar{ {\overline{\hbox{\normalsize MS}}} }

\newcommand{\R}{\hbox{{\rm I}\kern-.2em\hbox{\rm R}}}

\newcommand{\reff}[1]{(\ref{#1})}
\def\smfrac#1#2{{\textstyle\frac{#1}{#2}}}

\section{Introduction}

In the study of statistical models it is extremely
important to understand finite-size corrections. Indeed in 
experiments and in numerical work it is essential to take into
account the finite size of the system in order to extract correct 
infinite-volume predictions from the data. Finite-size scaling (FSS) 
\cite{Fisher_72,Fisher-Barber_72,Barber_83,Cardy_book,Privman_FSS_book}
concerns the critical 
behaviour of systems in which one or more directions are finite, even
though microscopically large, and it is therefore essential  
in the analysis of experimental data in many situations, for instance, 
for films of finite thickness. Numerically FSS can be used in a variety
of ways to extract informations on infinite-volume systems. A very interesting 
method to extract critical indices comparing data on lattices of 
different sizes was introduced by Nightingale 
\cite{Nightingale_76_82}, the so-called phenomenological 
renormalization group. Recently
FSS has been used to obtain precise predictions at very large 
values of the correlation length from simulations on small lattices.
This extrapolation technique was introduced by L\"uscher, Weisz and 
Wolff \cite{Luscher_91}
and subsequently applied to many different models
\cite{Kim,fss_greedy,o3_letter,MGMC_96,SU3_letter,SU3_paper}: a careful 
theoretical analysis (see Sect. 5.1.2 of Ref. \cite{SU3_paper})
shows that the method is extremely convenient for 
asymptotically free theories and indeed one was able to simulate 
the O(3) $\sigma$-model \cite{o3_letter} up to $\xi\approx 10^5$ and the 
$SU(3)$ chiral model \cite{SU3_letter,SU3_paper}
up to $\xi\approx 4 \cdot 10^5$ using relatively
small lattices ($L\le 512$). In order to use these techniques reliably
it is extremely important to have some theoretical prediction on the 
behaviour of the corrections to FSS. One can use this information in two 
different ways. A possibility is to take advantage of the theoretical 
prediction to extrapolate 
the Monte Carlo data to the FSS limit --- that still involves a limit
$L\to \infty$ --- in the spirit of Ref. \cite{Luscher_91}. One also needs
this information if one determines the FSS curve by comparing 
data from simulations on lattices of different sizes as proposed  in
Ref. \cite{fss_greedy}. For instance checking the absence (within error bars) 
of corrections to
FSS for lattices of sizes $64 \le L \le 256$ is enough if the corrections
vanish as $1/L^2$ while it can be totally misleading if corrections 
behaving as $1/\log L$ are present (this is the case 
for instance of the four-state Potts model, see Ref. 
\cite{Sokal-Salas}).

A second topic that will be extensively discussed in this paper
is the 
improvement of lattice hamiltonians 
\cite{Symanzik,LW,perfect,Luscher_LAT97,Hasenfratz_LAT97}. 
The idea behind all these attempts
is to modify the lattice hamiltonian with the addition of irrelevant operators
in order to reduce lattice artifacts: in this way one hopes to have scaling 
and FSS at shorter correlation lengths. For general statistical models
this is a non-trivial program. For asymptotically-free theories 
the idea is much simpler to implement since in this case improvement
can be discussed using perturbation theory.

In this paper we will study the problem of corrections to FSS and 
improvement in the context of the large-$N$ $N$-vector model. 
This theory provides the simplest example
for the realization of a nonabelian global symmetry. 
Its two-dimensional version has been 
extensively studied because it shares with four-dimensional gauge 
theories the property of being asymptotically free in the 
weak-coupling perturbative 
expansion~\cite{Polyakov_75,Brezin_76,Bardeen_76}.
This picture predicts a nonperturbative generation of a mass gap that
controls the exponential decay at large distance of the correlation 
functions. 

Besides perturbation theory, the two-dimensional 
$N$-vector model can be studied using different techniques. 
It can be solved in the $N=\infty$ limit~\cite{Stanley,DiVecchia}
and $1/N$ corrections can be systematically 
calculated~\cite{Muller,Flyvbjerg,Campostrini_90ab}. 
An exact $S$-matrix can be computed~\cite{Zamolodchikov_79,
Polyakov-Wiegmann_83}
and, using the thermodynamic Bethe ansatz, the exact mass-gap
of the theory in the limit $\beta\to\infty$ has been obtained 
\cite{Hasenfratz-Niedermayer_1,Hasenfratz-Niedermayer_2}.
The model has also been the object of extensive numerical work 
\cite{Wolff_O4_O8,MGMC_O4,o3_letter,CEMPS,Alles-Symanzik} 
mainly devoted to checking the 
correctness of the perturbative predictions 
\cite{Falcioni-Treves,CP-3loop,CP-4loop}.

In the large-$N$ limit FSS can be studied analytically
\cite{Barber-Fisher_73,Brezin_82,Luscher_82}. Here we will concentrate
on the corrections to FSS in two dimensions and we will compute the 
deviations from FSS for generic lattice interactions. We will
show that in general  corrections to FSS behave as $\log L/L^2$. This is in 
agreement with a general renormalization-group argument that shows that
corrections to FSS are controlled by the first subleading operator
\cite{Privman-Fisher_83}.
Tree-level improvement changes the behaviour by a logarithm of $L$:
these actions have corrections behaving as $1/L^2$. Subsequent improvement
should reduce the corrections to $1/(L^2 \log L)$ and so on. Full 
$O(a^2)$ improvement to all orders of perturbation theory 
provides an action with corrections behaving as $\log L/L^4$.

Besides the standard $N$-vector model we will also discuss a mixed
$N$-vector/$RP^{N-1}$-model \cite{DiVecchia,Magnoli-Ravanini}. 
There are two reasons why we decided to include this computation:
first of all, for large values of $N$, models with only two-spin interactions 
show many simplifying features: for instance in 
the $\beta$-function only the leading term does not vanish.
For this reason one may expect that the behaviour of the corrections
for this class of models is far simpler than in 
generic models. Instead the mixed
$N$-vector/$RP^{N-1}$-model shows a more complex behaviour and, for instance, 
the $\beta$-function is non trivial to all orders of perturbation theory. 
We find that in the mixed model 
the corrections behave as $(\log L/L^2) f(L)$ where $f(L)$ is 
a non-trivial function such that $f(\infty)$ is finite and that admits 
an asymptotic expansion in powers of $1/\log L$. The presence of powers
$1/(L^2 \log^n L)$ is somewhat unexpected from the point of view 
of perturbation theory. We will show that this is related to the 
non commutativity of the limits $L\to\infty$ and $\beta\to\infty$. 
In other words the perturbative limit $\beta\to\infty$ at $L$ fixed
followed by the limit $L\to\infty$ gives results that are 
different from the FSS limit.
The commutativity of these two limits  has been the object of an 
intense debate. The standard wisdom is that the two limits are identical,
but this point has been seriously questioned by Patrascioiu and Seiler
\cite{Pat-Seil_comment} (for an answer to these criticisms 
see Ref. \cite{CEPS_O3_reply_to_Pat-Seil}) together with many other 
assumptions derived from perturbation theory \cite{Pat-Seil}.
Our calculation shows that, if the standard assumption is true, it is 
a result far from obvious: indeed the limits are {\em different}
for the corrections to FSS. 

A second motivation 
lies in recent work \cite{Hasenbusch,Catteral-etal} 
on the $RP^{N-1}$ models that has shown the possibility 
that these models have very large FSS corrections. We wanted to understand 
if there is any sign of this phenomenon in the large-$N$ limit. 
Our explicit calculation shows that $RP^{\infty}$
has corrections that are larger than those of the $N$-vector
model. Depending on the observable, for reasonable lattice sizes,
we find an increase by a factor of 6-15. This is in qualitative 
agreement with the scenario of Ref. \cite{Hasenbusch}.

The paper is organized as follows: in Sect. 2 we define the models 
we consider, and compute various observables in the large-$N$ limit.
In Sect. 3 we discuss the corrections to FSS for the $N$-vector model
and in Sect. 4 we extend our results to the mixed 
$N$-vector/$RP^{N-1}$ model. In Sect. 5 we present our conclusions.

In the Appendices we report some general results on the FSS behaviour 
of lattice sums. These results are of general interest and may be applied
in many other contexts: in particular they may be used to study 
FSS properties of models that have a height (SOS) representation
(see Refs. \cite{Nijs-etal,Salas-Sokal-antiferromagnet} and 
references therein). 
In Appendix A we define a set of basic functions that appear in all
our results and we report some of their properties. We extend here 
the results of Ref. \cite{BF-Arch-Rat-Mech-Anal}.
In Appendix B 
we give an algorithmic procedure that allows to compute the expansion 
in powers of $1/L^2$ of any sum involving powers of the lattice 
propagator for a Gaussian model with arbitrary interaction in the FSS 
limit. As an example we report the explicit formulae that are needed 
in our main discussion. In Appendix C we report the asymptotic behaviour of 
some lattice integrals.

Preliminary results of this work were presented at the 
Lattice '96 Conference \cite{CP_LAT96}.

\section{The models} \label{sec2}

In this paper we will study the finite-size-scaling properties
of the classical $N$-vector model on a square lattice with
local, translation- and parity-invariant ferromagnetic interactions.
The hamiltonian is  given by
\be
{\cal H} =\, N \sum_{x,y} J(x-y) \bsigma_x \cdot \bsigma_y \; ,
\label{eq2.1}
\ee
where the fields $\bsigma_x$ satisfy $\bsigma_x^2=1$. The partition
function is simply
\be
Z =\, \int \prod_x d\bsigma_x \, e^{-\beta {\cal H}}.
\ee
We will consider general {\em local} interactions. If $\widehat{J}(p)$ is
the Fourier transform of $J(x)$, locality and parity invariance 
imply that
$\widehat{J}(p)$ is a continuous function of $p$, even under $p\to -p$.
We will
require invariance under rotations of $\pi/2$, that is we will
assume $\widehat{J}(p)$  symmetric under interchange of ${p}_1$ and
${p}_2$. Redefining $\beta$ we can normalize the
couplings so that
\be
\widehat{J}(q) = \widehat{J}(0) - {q^2 \over 2} + O(q^4)
\ee
for $q\to 0$. We also introduce the function
\be
    w(q) = - 2 (\widehat{J}(q) - \widehat{J}(0)),
\ee
that behaves as $q^2$ for $q\to 0$. 
Finally we will require the theory to have
the usual (formal) continuum limit: we will assume that
the equation $w(q) = 0$ has only one solution for
$-\pi \le q_i \le \pi$, namely $q=0$.
We will need the small-$q$ behaviour of $w(q)$: we will assume in this limit
the form
\be
w(q) =\, \hat{q}^2 + \alpha_1\, \sum_\mu \hat{q}^4_\mu + 
      \alpha_2 \ (\hat{q}^2)^2 + O(q^6),
\label{wqespansione}
\ee
where $\alpha_1$ and $\alpha_2$ are arbitrary constants. Here 
$\hat{q}^2 = \hat{q}^2_1 + \hat{q}^2_2$, $\hat{q} = 2 \sin (q/2)$.

Let us give some examples we will use in the following.
The standard $N$-vector model with hamiltonian 
\be
H^{std}\, =\,- N \sum_{x\mu} \bsigma_x \cdot \bsigma_{x+\mu}
\ee
corresponds to $w(q) = \hat{q}^2$ and thus we have $\alpha_1 = \alpha_2 = 0$.
Other possibilities are:
\begin{enumerate}
\item the Symanzik improved hamiltonian \cite{Symanzik}
\be
H^{Sym} = 
   - N \sum_{x\mu} \left( {4\over3} \bsigma_x \cdot \bsigma_{x+\mu} \, -\, 
         {1\over12} \bsigma_x \cdot \bsigma_{x+2\mu} \right),
\label{Symanzik}
\ee
for which we have 
\be
w(q) = \hat{q}^2 + {1\over12} \sum_\mu \hat{q}^4_\mu,
\ee
and $\alpha_1 = 1/12$ and $\alpha_2 = 0$;
\item the ``diagonal" hamiltonian \cite{Niedermayer_LAT96}
\be
H^{diag} = 
   - N \sum_{x} \left( {2\over3} \sum_\mu \bsigma_x \cdot \bsigma_{x+\mu} 
    \, +\, {1\over6} \sum_{\hat{d}}\bsigma_x \cdot \bsigma_{x+\hat{d}} \right)
\label{Symanzikdiag}
\ee
where $\hat{d}$ are the two diagonal vectors $(1,\pm1)$, for which we have
\be
w(q) = \, \hat{q}^2 - {1\over6} \hat{q}^2_1 \hat{q}_2^2,
\ee
and $\alpha_1 = 1/12$ and $\alpha_2 = -1/12$;
\item the perfect laplacian introduced in various works on the 
renormalization group \cite{Bell-Wilson}
and recently revived in connection with the 
perfect actions \cite{perfect}: 
\be
{1\over w(q)}\, =\, {1\over 3 \kappa} + 
   \sum_{l_1=-\infty}^\infty \sum_{l_2=-\infty}^\infty 
   {1\over (q_1 + 2 \pi l_1)^2 + (q_2 + 2 \pi l_2)^2}\, 
   {\hat{q}_1^2 \hat{q}_2^2 \over (q_1 + 2 \pi l_1)^2 (q_2 + 2 \pi l_2)^2}
\label{perfectlaplacian}
\ee
for which $\alpha_1 = 1/12$ and $\alpha_2 = (\kappa-4)/(12 \kappa)$. 
\end{enumerate}
In general we will speak of tree-level improved hamiltonians\footnote{Properly
speaking we should speak of $O(a^2)$ tree-level improved hamiltonians.
One can also consider $O(a^{2k})$ tree-level improved ones which are such that
\be
w(q) = q^2 + O(q^{4 + 2k})
\ee
for $q^2 \to 0$. We do not consider them here since tree-level improvement
beyond $O(a^2)$ does not have any effect on the corrections to FSS 
at order $1/L^2$. For a perturbative study of this class of hamiltonians
see Ref. \cite{RV}. Classically perfect hamiltonians are hamiltonians
improved to all orders in $a$ \cite{Hasenfratz-Niedermayer_97}.} 
whenever $\alpha_1 = 1/12$, $\alpha_2 = 0$: in this case,
for $q^2 \to 0$,
\be
w(q) = q^2 + O(q^{6}).
\ee
The hamiltonians \reff{Symanzik} and 
\reff{perfectlaplacian} for $\kappa = 4$ are examples of 
tree-level improved hamiltonians.
 
In order to study the finite-size-scaling properties we must specify
the geometry. We will consider here a square lattice of size $L\times T$ 
or a strip of width $L$
with periodic boundary conditions in the finite direction(s). 
The large-$N$ limit of this model is
well known \cite{Stanley}. 
The theory is parametrized by a mass parameter $m^2_{L,T}$
related to $\beta$ by the gap equation
\be
\beta = {1\over LT} \sum_{n_1,n_2} {1\over w(p) + m^2_{L,T}} \,
   \equiv {\cal I}_{L,T} (m^2_{L,T}),
\ee
where $p_1 = 2 \pi n_1/L$, $p_2 = 2 \pi n_2/T$ and the sum extends
over $0\le n_1 \le L-1$ and $0\le n_2 \le T-1$. The two-point
{\em isovector} Green's function is then given by
\be
G_V(x-y;L,T) \equiv \< \bsigma_x \cdot \bsigma_y\> =
      {1\over \beta} \, {1\over LT}
       \sum_{n_1,n_2} {e^{ip\cdot(x-y)} \over w(p) + m^2_{L,T}}.
\label{isovector}
\ee
All other Green's functions are obtained from
$G_V(x;L,T)$ using the factorization theorem
\be
\< (\bsigma_{x_1}\cdot \bsigma_{y_1})
   (\bsigma_{x_2}\cdot \bsigma_{y_2}) \ldots
   (\bsigma_{x_n}\cdot \bsigma_{y_n}) \> \,=\,
\< (\bsigma_{x_1}\cdot \bsigma_{y_1})   \>
\< (\bsigma_{x_2}\cdot \bsigma_{y_2})   \> \ldots
\< (\bsigma_{x_n}\cdot \bsigma_{y_n})   \>.
\label{factorization}
\ee
In particular we will consider the {\em isotensor} (spin-two)
two-point function
\be
G_T(x-y;L,T) \equiv \, \< (\bsigma_x\cdot\bsigma_y)^2\> - {1\over N}\, =\,
                   \< \bsigma_x\cdot\bsigma_y\>^2 + O(1/N).
\label{isotensor}
\ee
Beside the standard $N$-vector model we will also discuss a 
mixed $N$-vector/$RP^{N-1}$-model 
\cite{DiVecchia,Magnoli-Ravanini,CEPS_RP,CPS_RP_Dallas}.
We will restrict our attention to
nearest-neighbour interactions since only in this case the model is
easily solvable in the large-$N$ limit. The hamiltonian is given
by
\be
H^{mix} = -N\, \sum_{\< xy\>}\left[ (1-r) \bsigma_x\cdot\bsigma_y + 
{r\over 2} (\bsigma_x\cdot\bsigma_y)^2\right],
\label{mixedmodel}
\ee
where the sum is extended over all links $\<xy\>$ and $r$ is 
a free parameter varying between 0 and 1. For $r=0$ we have the 
nearest-neighbour $N$-vector model, while $r=1$ corresponds to 
the $RP^{N-1}$-model. Notice that for $r=1$ the theory is invariant under 
local transformations $\bsigma_x \to \epsilon_x \bsigma_x$, 
$\epsilon_x = \pm 1$. Therefore for $RP^{N-1}$ only isotensor 
observables are relevant. The limit we consider here corresponds to
$N\to\infty$ with $r$ fixed. We mention that this is not the only case
in which the model is solvable: a different large-$N$ limit is
considered in Ref. \cite{CPS_RP_Dallas}.

Also in this case the theory is parametrized by a mass parameter $m^2_{L,T}$
related\footnote{$\beta$ is related to $m_{L,T}$ by Eq. \reff{mixedgapeq} 
only for $\beta > \beta_c(r)$, where $\beta = \beta_c(r)$ is a first-order 
transition line \cite{Magnoli-Ravanini}. In the following we will 
be only interested in the limit $\beta\to\infty$, so that we will 
always use Eq. \reff{mixedgapeq}.}
to $\beta$ by \cite{Magnoli-Ravanini}
\be
\beta = \, {4 I_{L,T}(m_{L,T}^2)^2 \over 4 I_{L,T}(m_{L,T}^2) + 
    r ( m^2_{L,T} I_{L,T}(m^2_{L,T}) - 1)}\; ,
\label{mixedgapeq}
\ee
where 
\be
      I_{L,T}(m_{L,T}^2) \equiv\, {1\over LT} \sum_{n_1,n_2} 
                {1\over \hat{p}^2 + m^2_{L,T}}\; ,
\ee
with $p_1 = 2 n_1\pi/L$, $p_2 = 2 n_2\pi/T$. 

The isovector Green's function is given by 
\be
G_V(x-y;L,T)  =
      {1\over {I}_{L,T}(m^2_{L,T})} \, {1\over LT}
       \sum_{n_1,n_2} {e^{ip\cdot(x-y)} \over \hat{p}^2 + m^2_{L,T}}\; .
\ee
All other correlations are obtained using Eq. \reff{factorization}. In the
$N$-vector case we can use the gap equation to substitute 
${I}_{L,T}(m^2_{L,T})$ with $\beta$.

In this paper we will study the finite-size-scaling properties of 
various quantities. We define the vector and tensor
susceptibilities
\begin{eqnarray}
\chi_V(L,T) &=& \sum_x G_V(x;L,T), \\
\chi_T(L,T) &=& \sum_x G_T(x;L,T). 
\end{eqnarray}
Using the explicit expressions for the two-point functions we get
\begin{eqnarray}
\chi_V(L,T) &=& {1\over {\cal I}_{L,T}(m^2_{L,T}) m^2_{L,T}} \\
\chi_T(L,T) &=& {1\over ({\cal I}_{L,T}(m^2_{L,T}))^2} \, {1\over LT} 
            \sum_{n_1,n_2} {1\over (w(p) + m^2_{L,T})^2}.
\end{eqnarray}
We want also to define a quantity behaving as a correlation length.
In an infinite lattice there are essentially two possibilities.
One can define the exponential correlation length from the large-$x$
behaviour of a given two-point function\footnote{Here and in the
following we will indicate the infinite-volume limit of an observable
${\cal O}(L,T)$ with ${\cal O}(\infty)$.} $G(x;\infty)$: one considers 
a wall-wall correlation function
\be
G^{(w)} (y;\infty) = \sum_x G((x,y);\infty),
\ee
and then defines 
\be
\xi^{(exp)}(\infty) =\, - \lim_{y\to +\infty} 
   {y\over \log G^{(w)} (y;\infty) }\; .
\label{xiexpdef}
\ee
The mass gap $\mu(\infty)$ is the inverse of 
$\xi^{(exp)}(\infty)$. A second possibility is the second-moment correlation
length $\xi^{(m)}(\infty)$ that is defined by 
\be
\left[\xi^{(m)}(\infty) \right]^2 = {1\over4} 
{\sum_x |x|^2 G(x;\infty)\over \sum_x G(x;\infty)}\; .
\label{xi2def}
\ee
The factor $1/4$ has been introduced in order to have 
$\xi^{(m)}(\infty) =  \xi^{(exp)}(\infty)$ for Gaussian models. 

We must now give the definitions in finite volume. Of course the 
exponential correlation can only be defined in a strip. In this case
we can still use the definition \reff{xiexpdef}. For the 
second-moment correlation length we can use any definition that 
becomes equivalent to \reff{xi2def} in the limit $L,T\to\infty$. 
We will consider here three different definitions: given a two-point
function $G(x;L,T)$ and its Fourier transform $\hat{G}(p;L,T)$ we define
\begin{eqnarray}
\left( \xi^{(m,1)}(L,T)\right)^2 &=& {1\over 2 \hat{p}_{0x}^2}
     \left[ {\hat{G}(0;L,T)\over \hat{G}(p_{0x};L,T)} - 1 \right] +\, 
     {1\over 2 \hat{p}_{0y}^2} 
     \left[ {\hat{G}(0;L,T)\over \hat{G}(p_{0y};L,T)} - 1 \right] \; ,
\label{xi2def1}  \\
\left( \xi^{(m,2)} (L,T) \right)^2 &=& 
     {L^2\over 8\pi^2} \left(1 - 
       {\hat{G}(p_{0x};L,T)\over \hat{G}(0;L,T)}\right) + 
     {T^2\over 8\pi^2} \left(1 - 
       {\hat{G}(p_{0y};L,T)\over \hat{G}(0;L,T)}\right) \; ,
    \label{xi2def2} \\
\left( \xi^{(m,3)}(L,T)\right)^2 &=& {1\over 4 \hat{G}(0;L,T)} 
    \sum^{\lfloor L/2\rfloor}_{i = 1 - \lfloor (L+1)/2\rfloor} \, 
    \sum^{\lfloor T/2\rfloor}_{j = 1 - \lfloor (T+1)/2\rfloor}
    (i^2 + j^2) G((i,j);L,T), \nonumber \\ [-2mm]
{} 
\label{xi2def3}
\end{eqnarray}
where $p_{0x} = (2 \pi/L,0)$, $p_{0y} = (0,2 \pi/T)$ and 
$\lfloor x\rfloor$ is the largest integer smaller than or equal to $x$. 
The third definition evidently coincides with \reff{xi2def} for 
$L,T\to\infty$. To verify the correctness of the other two definitions
notice that Eq. \reff{xi2def} can be rewritten as 
\be
\left[ \xi^{(m)}(\infty) \right]^2 =\, - {1\over 4\ \hat{G}(0;\infty)} 
   \sum_\mu \left. {\partial^2 \over \partial p_\mu^2} \hat{G}(p;\infty)
   \right|_{p=0}\; .
\ee
Expanding in $1/L^2$ it is easy to verify that both $\xi^{(m,1)}(L,T)$
and $\xi^{(m,2)}(L,T)$ converge to \reff{xi2def} for $L,T\to\infty$. 
Essentially \reff{xi2def1} and \reff{xi2def2} are definitions in which 
one approximates the second derivative of $\hat{G}$  
with the difference at two nearby points. Thus these definitions 
converge to $\xi^{(m)}(\infty)$ as $1/L^2$ (notice that 
$G(x;L,T) \to G(x,\infty)$ exponentially). The third definition 
represents instead the ``best" approximation on a finite lattice since
$\xi^{(m,3)}(L,T)$ converges to $\xi^{(m)}(\infty)$ exponentially. 
This is indeed a general result that can be proved using the relation
\be
\sum^{\lfloor L/2\rfloor}_{i = 1 - \lfloor (L+1)/2\rfloor} 
    i^2 f(i)\, = \, 2\ \sum_{n=1}^{L-1} {(-1)^n\over \hat{q}^2}
    (\hat{f}(q) - \hat{f}(0))\ {\cal{P}}_L(q),
\label{sumi2f}
\ee
valid for every function $f$. Here $\hat{f}$ is the Fourier transform of 
$f$, $q=2\pi n/L$ and 
\be
  {\cal P}_L \left( {2 \pi n \over L}\right) =\ 
   \cases{1 & for $L$ even; \cr
       \cos \displaystyle{\pi n\over L} & for $L$ odd.}
\ee
If $\hat{f}(q)$ is meromorphic (as a function of the {\em complex} 
variable $q$) in the strip $0\le \hbox{\rm Re}\ q\le 2\pi$, 
periodic of period $2\pi$,
and with simple poles at $\overline{q}_i$, then we can evaluate this sum
to obtain 
\be
- \hat{f}''(0) - {L\over4} \sum_i 
      {R_i\over \sin^2(\overline{q}_i/2) \sin(L \overline{q}_i/2)}
    {\cal{P}}_L(\overline{q}_i),
\ee
where $R_i$ is the residue of $\hat{f}(q)$ at $\overline{q}_i$. 
Thus the convergence rate is $L\exp(-L \overline{q}_I/2)$ where 
$\overline{q}_I = \min_i |{\rm Im} (\overline{q}_i)|$. For the specific case 
of the isovector correlation length one expects the nearest singularities 
(for $\beta\to\infty$ at least) to be at $q = \pm i \mu(L)$ where 
$\mu(L)$ is the mass gap. Thus we expect a convergence rate of 
$L e^{-\mu(L) L/2}$. A general Green function will not be in general
a meromorphic function of $q$ as cuts will appear as well. We expect 
however that the definition 
\reff{xi2def3} will show the same exponential convergence rate.

Using Eq. \reff{sumi2f} we can rewrite Eq. \reff{xi2def3} as 
\begin{eqnarray}
\left(\xi^{(m,3)}(L,T)\right)^2 &=& 
    {1\over 2 \hat{G}(0;L,T)} \sum_{(n_1,n_2)\not=(0,0)} {1\over \hat{q}^2} 
    \left( \hat{G}(q;L,T) - \hat{G}(0;L,T)\right) \times
\nonumber \\ [2mm]
    && \quad \left( (-1)^{n_1} {\cal P}_L(q_1) \delta_{q_2 0} + 
           (-1)^{n_2} {\cal P}_T(q_2) \delta_{q_1 0} \right).
\end{eqnarray}
Let us now give explicit expressions for the isovector correlation length:
using the 
isovector two-point function \reff{isovector} we get on a finite lattice:
\begin{eqnarray}
\left(\xi^{(m,1)}_V(L,T)\right)^2 &=& {1\over2} 
   \left( {w(p_{0x})\over \hat{p}^2_{0x}} +
          {w(p_{0y})\over \hat{p}^2_{0y}} \right) 
   {1\over m^2_{L,T}}, \\
\left(\xi^{(m,2)}_V(L,T)\right)^2 &=& 
  {1\over 8\pi^2} \left( {w(p_{0x}) L^2 \over w(p_{0x}) + m^2} + 
         {w(p_{0y}) T^2 \over w(p_{0y}) + m^2} \right)\; , \\
\left(\xi^{(m,3)}_V(L,T)\right)^2 &=& -{1\over2} 
     \sum_{n_1=1}^{L-1} (-1)^{n_1}{w(q_1)\over \hat{q}_1^2} 
         {1\over w({q}_1)+ m_{L,T}^2} 
         {\cal P}_L (q_1) \nonumber \\
                          && - {1\over2} 
      \sum_{n_2=1}^{T-1} (-1)^{n_2} {w(q_2)\over \hat{q}_2^2} 
         {1\over w({q}_2) + m_{L,T}^2} {\cal P}_T (q_2) .
\end{eqnarray}
In infinite volume we have instead: 
\be
  \xi_V^{(m)} (\infty) \, =\, {1\over m_\infty}\; .
\ee
For the mass gap $\mu_V(L)$ and the exponential correlation length 
$\xi^{(exp)}_V(L)$ we must solve the equation
\be
     w(i\mu_V(L),0) + m^2_{L,\infty} = \, 0\; .
\ee
An explicit solution can be obtained only for the simplest $w(p)$. 
For the hamiltonians we have considered in this section we have
\be
\mu_V(L) \, =\, 
   \cases{2\ \hbox{\rm arcsh} 
            \left(\displaystyle{m_{L,\infty}\over2}\right) & 
           $\qquad$   for $H^{std}$ and $H^{diag}$; \cr
       2\ \hbox{\rm arcsh} \left\{ 
      \displaystyle{\sqrt{6}\over2} \left[1 - 
      \left(1 - \displaystyle{m^2_{L,\infty}\over3}
     \right)^{1/2} \right]^{1/2} \right\} & $\qquad $ for $H^{Sym}$. }
\ee
In our calculation we will only need the expression of $\mu_V(L)$
for $m_{L,\infty}\to 0$. In this limit we obtain 
\be
   \mu_V(L)\, =\, m_{L,\infty} 
    \left[ 1 + {1\over2} \left(\alpha_1 + \alpha_2 - {1\over12}\right)
    m^2_{L,\infty} + O(m^4_{L,\infty})\right].
\ee

Isotensor observables are defined using the tensor two-point
function \reff{isotensor}.
For the mass gap it is easy to verify that $\mu_T(L) = 2 \mu_V(L)$.

\section{$N$-vector model}

\subsection{The gap equation}

In this section we want to discuss the corrections to FSS for the hamiltonian
\reff{eq2.1}. 

Let us consider a fixed value of $\beta$. Let $m_\infty$ and $m_{L,T}$
be the mass parameters corresponding to $\beta$ in infinite volume
and in a box $L\times T$. It is immediate to obtain a relation 
between $m_\infty$ and $m_{L,T}$. Indeed from the gap equation we obtain 
\be
   {\cal I}_{L,T} (m^2_{L,T}) \, =\, 
   {\cal I}_{\infty} (m^2_{\infty}) .
\ee
Now let us consider the finite-size-scaling limit,
$m_\infty$, $m_{L,T}\to0$, $L,T\to\infty$ with 
$m_{L,T} L\equiv z$ and $T/L\equiv \rho$ fixed. Using the results
\reff{sommaasintotica} and \reff{intasintotico} we obtain
\be
{m^2_\infty\over m^2_{L,T}} \, =\, f_m (z;\rho) \left(1 +  
          {\Delta_{m,1}(z;\rho)\over L^2 }\log L
        + {\Delta_{m,2}(z;\rho)\over L^2} \right)
\label{m2sum2}
\ee
with corrections of order $O(\log^2 L/L^4)$, where 
\begin{eqnarray}
f_m(z;\rho) &\hskip -4pt =& \hskip -8pt 
         {32\over z^2} e^{-4\pi F_0(z;\rho)} ,
\label{fm}\\
\Delta_{m,1}(z;\rho) &\hskip -4pt =& \hskip -8pt 
        {1\over4} (12 \alpha_1 + 16 \alpha_2 -1) 
        \left( 32 e^{-4 \pi F_0(z;\rho)} - z^2\right) ,
\label{Deltam1}\\
\Delta_{m,2}(z;\rho) &\hskip -4pt =& \hskip -8pt
 16\pi (12 \alpha_1 + 16 \alpha_2 - 1) F_0(z;\rho)
         e^{-4 \pi F_0(z;\rho)}
\nonumber \\ [2mm]
&& \hskip -12pt - 4 (8\alpha_1 + 8\alpha_2 - 1) e^{-4 \pi F_0(z;\rho)}
    - {4 \pi}\left( {\cal F}_1(z;\rho) + 
      32 e^{-4 \pi F_0(z;\rho)} \Lambda_1\right).
\label{Deltam2}
\end{eqnarray}
The functions $F_0(z;\rho)$ and ${\cal F}_1(z;\rho)$ are defined in the 
appendix, Eqs. \reff{F0explicito} and \reff{F1storto}. The 
function $f_m(z;\rho)$ is the FSS function for the ratio 
$m_\infty^2/m^2_{L,T}$. As expected, it is universal (it does not
depend on the explicit form of the coupling $J(x)$) and depends 
on the modular parameter $\rho$. The corrections instead are 
{\em not} universal. However the dependence on $J(x)$ is very simple:
the only relevant quantities are $\alpha_1$ and $\alpha_2$ that are
connected to the small-$q^2$ behaviour of $w(q)$ and $\Lambda_1$
given by
\be
\Lambda_1 =\, \int {d^2p\over (2\pi)^2} 
     \left[ {1\over w(p)^2} - {1\over (\hat{p}^2)^2} +
     {2\over (\hat{p}^2)^3} \left( \alpha_1 \sum_\mu \hat{p}^4_\mu 
       + \alpha_2 (\hat{p}^2)^2 \right) \right]  \; .
\label{Lambda1testo}
\ee
The corrections behave in general as $\log L/L^2$, except when 
$12 \alpha_1 + 16 \alpha_2 - 1= 0$.
This cancellation happens for improved hamiltonians for which 
$\alpha_1 = {1\over12}$ and $\alpha_2 = 0$ and also
for many other hamiltonians that are not improved but nonetheless 
satisfy $12 \alpha_1 + 16 \alpha_2 - 1= 0$. 
To understand the relevance of this combination, let us introduce 
polar coordinates $q_x = q \cos \theta$, $q_y = q \sin \theta$. Then
\be
w(q) = q^2 + {1\over 16} (12 \alpha_1 + 16 \alpha_2 - 1) q^4 + 
    {1\over 48} (12 \alpha_1 - 1) q^4 \cos 4 \theta + O(q^6).
\ee
Thus, if $(12 \alpha_1 + 16 \alpha_2 - 1) = 0$, one cancels 
the first rotationally-invariant subleading operator, leaving 
a correction that is associated to a lattice operator that vanishes 
when averaged over the angle $\theta$. This last property 
is the reason why this quantity does not couple to the leading correction.
This fact is not unexpected. Indeed the leading correction to scaling
is usually associated to a rotationally-invariant operator (for a discussion
for the two-point function in infinite volume see Ref. \cite{funzione2pt}). 

For $\alpha_1 = {1\over12}$ and $\alpha_2 = 0$ the expression for 
$\Delta_{m,2}(z;\rho)$ simplifies drastically, becoming
\be
\Delta_{m,2}(z;\rho) = 4 \pi
   (32 e^{-4 \pi F_0(z;\rho)} - z^2) \left({1\over96\pi} - \Lambda_1\right).
\ee
Thus for improved hamiltonians there is the possibility of eliminating 
even the $1/L^2$ corrections choosing $J(x)$ so that 
\be
\Lambda_1 = {1\over 96 \pi}.
\label{conditionLambda1}
\ee
Notice that this condition is global, that is it does not only fix the 
small-$q$ behaviour of $w(q)$, but it depends on the behaviour of $w(q)$ 
over all the Brillouin zone.

A particular hamiltonian satisfying Eq. \reff{conditionLambda1} is
\begin{eqnarray}
\hskip -5pt
H^{Sym2}  = - N\sum_{x\mu} 
  \left[ \left( {4\over3} + 15 a\right) \bsigma_x\cdot\bsigma_{x+\mu} - \, 
         \left( {1\over12} + 6 a\right) \bsigma_x\cdot\bsigma_{x+2\mu} + \,
         a \bsigma_x\cdot\bsigma_{x+3\mu} \right]
\end{eqnarray}
where $a = 0.00836533968(1)$.
The function $w(q)$ is given by 
\be
w(q) = \hat{q}^2 + {1\over12} \sum_\mu \hat{q}_\mu^4 + 
   a \sum_\mu \hat{q}_\mu^6 .
\ee

For this hamiltonian the corrections to FSS behave as $\log L/L^4$. Of course
one could improve further: using a hamiltonian with $w(q) = q^2 + O(q^8)$ 
satisfying Eq. \reff{conditionLambda1} one could get rid also of the terms
$\log L/L^4$; however the cancellation of the terms $1/L^4$ will 
again require a global condition of the type \reff{conditionLambda1}.

It is interesting to understand our results in terms of perturbation
theory. Within the Symanzik improvement program the conditions 
$\alpha_1 = {1\over12}$ and $\alpha_2 = 0$ are required for tree-level
improvement: if the theory is tree-level improved , then the corrections
behave as $1/L^2$ instead of $\log L/L^2$. In Ref. \cite{CP_lettera}
it was shown that
the condition \reff{conditionLambda1} is necessary for improvement
at one loop. The simplifying feature of the model is that,
once the theory is one-loop improved, it is improved to all
orders of perturbation theory.  This explains why, if condition
\reff{conditionLambda1}  is satisfied, corrections to scaling behave 
as $\log L/L^4$. As we shall discuss in the following section, 
for a generic model, for instance for a mixed 
$O(N)$-$RP^{N-1}$ theory, we expect only the $1/L^2$ term to be canceled 
so that the corrections to scaling would still behave as 
$1/(L^2 \log L)$. 

We have performed various checks of the expressions 
\reff{fm}, \reff{Deltam1} and \reff{Deltam2}. First of all we have 
compared our results with previous work.   For the strip 
$f_m(z;\infty)$ was computed by L\"uscher \cite{Luscher_82}.  
In this case, as 
$\lim_{\rho\to\infty} M_{1,1}(z;\rho) = 0$, using the explicit expression
for $F_0(z;\rho)$, Eq. \reff{F0explicito}, and Eq. \reff{G0asintotico}, 
we get
\begin{eqnarray}
f_m(z;\infty)  &=& 
   {32 \over z^2} \exp\left[ - {2\pi\over z} - 2 \gamma_E + 
     \log {\pi^2\over2} - 2\, G_0 \left({z\over 2\pi}\right)\right] 
\nonumber \\
   &=& \exp\left[ -4 \sum_{n=1}^\infty K_0 (nz)\right]\; ,
\end{eqnarray}
that agrees with the result of Ref. \cite{Luscher_82}.

\begin{table}
\begin{center}
\begin{tabular}{|l|rr|rr|r|}
\hline
 $L$ &  $R_{exact}(L;2,1)$ & $R_{expan} (L;2,1)$ & 
 $D_{exact}(L;2,1)$ & $D_{expan} (L;2,1)$ & $\delta_2(L;2,1)$\\
\hline
4  &  0.23892847 &  0.24124682 &  0.0649479 &  0.0752812 &  $-$0.428  \\ 
6  &  0.23332246 &  0.23379910 &  0.0399609 &  0.0420854 &  $-$0.345  \\ 
8  &  0.23028713 &  0.23044079 &  0.0264318 &  0.0271167 &  $-$0.303  \\ 
10 &  0.22856980 &  0.22863479 &  0.0187774 &  0.0190671 &  $-$0.282  \\ 
12 &  0.22751364 &  0.22754600 &  0.0140699 &  0.0142141 &  $-$0.270  \\ 
14 &  0.22681775 &  0.22683573 &  0.0109682 &  0.0110483 &  $-$0.262  \\ 
16 &  0.22633412 &  0.22634493 &  0.0088126 &  0.0088607 &  $-$0.255  \\ 
20 &  0.22572118 &  0.22572580 &  0.0060806 &  0.0061012 &  $-$0.247  \\ 
32 &  0.22497047 &  0.22497124 &  0.0027345 &  0.0027380 &  $-$0.232  \\ 
64 &  0.22453987 &  0.22453993 &  0.0008153 &  0.0008155 &  $-$0.218  \\ 
128&  0.22441005 &  0.22441006 &  0.0002367 &  0.0002367 &  $-$0.208  \\
\hline
$\infty$ &       &  0.22435696 &            &            &  $-$0.148  \\
\hline
\end{tabular}
\end{center}
\caption{Values of $R_{exact}(L;z,\rho)$, 
$R_{expan}(L;z,\rho)$, $D_{exact}(L;z,\rho)$,
$D_{expan}(L;z,\rho)$ and $\delta_2(L;z,\rho)$ 
for the standard hamiltonian $H^{std}$, $\rho = 1$ and $z=2$.
 }
\label{tableFSS1}
\end{table}

We have furthermore performed a detailed numerical check for the standard 
hamiltonian $H^{std}$. Given $L$, $z$ and $\rho$ we have 
first computed $m^2_{L,T} = z^2/L^2$, then $\beta$ from the 
finite-volume gap equation $\beta = {\cal I}_{L,T}(m^2_{L,T})$ and 
finally $m^2_\infty$ from $\beta = {\cal I}_{\infty}(m^2_\infty)$: 
in this way we have obtained for each lattice size $L$ and $z$ the 
ratio $R_{exact}(L;z,\rho) \equiv m^2_\infty/m^2_{L,T}$. Then we computed 
\be
\delta_1(L;z,\rho)\, =\, 
  {L^4\over \log^2 L} \left[ R_{exact}(L;z,\rho) 
    - R_{expan}(L;z,\rho) \right],
\ee
where $R_{expan}(L;z,\rho)$ is the r.h.s of Eq. \reff{m2sum2}. In this way
we have tried to verify that indeed $\delta_1(L;z,\rho)$ at fixed $z$ goes to 
a constant for $L\to\infty$. 
Numerically we find that $\delta_1(L;z,\rho)$ varies slowly 
with $L$ and that the behaviour is compatible with the 
presence of $1/\log L$ and $1/\log^2 L$ corrections. 
A better check can be obtained if we include the term of order
$\log^2 L/L^4$ that can be easily computed 
\be
\Delta_{m,3}(z;\rho)\, =\, (12\alpha_1+16\alpha_2-1)^2
\left( 96 e^{-8\pi F_0(z;\rho)} - 4 z^2 e^{-4\pi F_0(z;\rho)} + 
    {z^4\over32}\right).
\ee
Then we consider
\be
\delta_2(L;z,\rho)\, =\, 
  {L^4\over \log L} \left[ 
        R_{exact}(L;z,\rho) - R_{expan}(L;z,\rho) - \,
    f_m(z;\rho) \Delta_{m,3}(z;\rho) {\log^2 L\over L^4} \right].
\ee
In this case we should be able to verify that
\be
\delta_2(L;z,\rho) \approx \delta_{20}(z;\rho) + 
      {\delta_{21}(z;\rho)\over \log L}
\ee
for large values of $L$.
The results for $\rho = 1$ and $z=2$ 
(this value of $z$ corresponds to the region where the corrections to FSS 
are larger) are shown in Table \ref{tableFSS1} where we also give 
the deviations from FSS, i.e. the quantity 
\be 
  D(L;z,\rho) = \left[ {R(L;z,\rho)\over f_m(z;\rho)} - \, 1\right]\; .
\label{defD}
\ee
A plot of $\delta_2(L;2,1)$ versus $1/\log L$ shows, as expected, a 
linear behaviour from which we can estimate 
$\delta_{20}(2;1) \approx -0.148$ and $\delta_{21}(2;1) \approx -0.291$.

Let us now consider the limits $z\to \infty$ and $z\to 0$.
Asymptotic expansions of the FSS functions 
can be obtained using the expansions of 
$F_0(z;\rho)$ and ${\cal F}_1(z;\rho)$ reported in 
sections \ref{secB.2.1} and \ref{secB.2.2}.
For large $z$ we have
\begin{eqnarray}
f_m(z;\rho) &=& 1 - 2 e^{-z} \sqrt{2\pi\over z} - 
                    2 e^{-\rho z} \sqrt{2\pi\over \rho z} + 
     O(z^{-3/2} e^{-z},z^{-3/2} e^{-\rho z}), \\
\Delta_{m,1}(z;\rho) &=& - {1\over2} (12 \alpha_1 + 16 \alpha_2 -1) z^2
     \left(e^{-z} \sqrt{2\pi\over z} +
           e^{-\rho z} \sqrt{2\pi\over \rho z} \right)  \nonumber \\
&& \qquad\qquad + O(z^{1/2} e^{-z},z^{1/2} e^{-\rho z}) ,
\label{Deltam1largez}\\[1mm]
\Delta_{m,2}(z;\rho) &=& {\pi\over6} (12 \alpha_1 + 12 \alpha_2 -1) z^2
     \left(e^{-z} \sqrt{z\over 2\pi} +
           e^{-\rho z} \sqrt{\rho z \over 2\pi} \right) \nonumber \\ 
&& + {z^2\over4} (12 \alpha_1 + 16 \alpha_2 - 1) 
    \, \log{z^2\over32}\,
  \left(e^{-z} \sqrt{2\pi\over z} +
           e^{-\rho z} \sqrt{2\pi\over \rho z} \right)  \nonumber \\ [2mm]
&& \qquad \qquad + 
     O(z^{3/2} e^{-z},z^{3/2} e^{-\rho z}) .
\label{Deltam2largez}
\end{eqnarray}
The result agrees with what is expected: for $z\to\infty$ the 
FSS function converges to 1 exponentially
\cite{Luscher_massa}. Also the corrections 
vanish in the same way and thus they are extremely tiny for large $z$. 

Let us now consider the perturbative limit (small $z$).
For finite values of $\rho$, for $z\ll 1$ and $z\ll 1/\rho$, we find 
\begin{eqnarray}
f_m(z;\rho) &=& {16\pi^2\over z^2} \eta(i\rho)^4 e^{-2\gamma_E} 
    \exp\left(-{4\pi\over \rho z^2}\right) \, 
    \left(1 - 4\pi z^2 F_{01}(\rho) + O(z^4)\right) ,
\label{fmsmallz}
\\
\Delta_{m,1}(z;\rho) &=& - {z^2\over4} (12 \alpha_1 + 16\alpha_2 - 1) +\, 
    O\left[e^{-4\pi/(\rho z^2)}\right] \; ,
\label{Deltam1smallz}
\\
\Delta_{m,2}(z;\rho) &=& - 4\pi (1 - 12 \alpha_1) F_{10}(\rho) - 
    {4\pi \alpha_2\over\rho} \nonumber \\
&& \hskip -10pt 
  - 4\pi z^2\left[ (1-12\alpha_1)F_{11}(\rho) + 2\alpha_2 F_{00}(\rho) 
    + {\alpha_1\over 8\pi} - \Lambda_1\right] + \, O(z^4) .
\label{Deltam2smallz}
\end{eqnarray}
Here $\eta(\tau)$ is Dedekind's $\eta$-function \cite{Chandra}
\be
\eta(\tau) = \, e^{\pi i \tau/12} \prod_{n=1}^\infty
   \left( 1 - e^{2 \pi i n \tau}\right),
\ee
and $F_{01}(\rho)$, $F_{10}(\rho)$ and $F_{11}(\rho)$ are defined 
in the appendix: see Eqs. \reff{F01def}, \reff{F10def} and \reff{F11def}.

For the strip the previous expansions are not valid. In this case, for small
$z$, we get
\begin{eqnarray}
f_m(z;\infty) &=& {16\pi^2\over z^2} e^{-2\gamma_E} e^{-2\pi/z}
     \left(1 + {z^2\over 4\pi^2} \zeta(3) + O(z^4)\right) ,
\label{fmsmallzstrip}
\\
\Delta_{m,1}(z;\infty) &=& - {z^2\over4} (12 \alpha_1 + 16\alpha_2 - 1) +\, 
    O(\exp(-2\pi/z) ),
\label{Deltam1smallzstrip}
\\
\Delta_{m,2}(z;\infty) &=& {\pi^2\over18} (12\alpha_1 - 1) - 
 {\pi z\over 4} (12 \alpha_1 + 12\alpha_2 - 1) + O(z^2) .
\label{Deltam2smallzstrip}
\end{eqnarray}
These formulae can also be used when $\rho \gg 1$. Indeed they 
approximate the FSS functions for $1/\rho \ll z \ll 1$.

It is interesting to obtain these expansions within perturbation theory (PT).
The idea is to start from the perturbative prediction for $m_\infty(\beta)$
\be
m_\infty^2 =\, 32 e^{4\pi \Lambda_0} e^{-4\pi \beta} 
     \left(1 + O(\beta e^{-4\pi \beta})\right),
\label{minfsumLTpert}
\ee
and the perturbative expansion of $m_{L,T}(\beta)$:
\be
    z \equiv L\ m_{L,T} = \sum_{n=1}^\infty {m_n(L,T)\over \beta^n}.
\ee
Then we use the last equation to express perturbatively $\beta$ in terms
of $z$ and finally we substitute the result in Eq.
\reff{minfsumLTpert}. In this way we obtain the expansions
\reff{fmsmallz}, \reff{Deltam1smallz} and 
\reff{Deltam2smallz} and the analogous ones on the strip. 
It must be noted that in this perturbative
expansion the combination $(12\alpha_1 + 16\alpha_2 -1)$ arises 
naturally: indeed it is the coefficient of the unique 
$\log L/L^2$ term that appears in the expansion. Thus 
$12\alpha_1 + 16\alpha_2 -1=0$ is the improvement condition of the 
renormalized perturbative expansion. 

To conclude the discussion we want to comment on the validity of PT: 
finite-volume $PT$ is valid in the limit $\beta\to\infty$ at 
$L$ fixed while the FSS limit we are interested in corresponds to 
$\beta\to\infty$, $L\to \infty$ at $z$ fixed. Thus our perturbative
derivation of the FSS scaling functions involves an {\em a priori} unjustified
extension of the validity of PT
\cite{Pat-Seil_comment,CEPS_O3_reply_to_Pat-Seil}. 
For the leading contribution this should
be correct (naively because the result is $L$-independent) but the 
situation is less clear for the corrections: in this case the explicit 
calculation shows that the extension is valid also for the 
$1/L^2$ corrections, but, as we shall see in the next section, this 
is a special feature of the large-$N$ $N$-vector model: 
in general the corrections to FSS computed in PT 
need a ``resummation" to correctly describe the FSS regime.

\begin{figure}
\vspace*{-1cm} \hspace*{-0cm}
\begin{center}
\epsfxsize = 0.9\textwidth
\leavevmode\epsffile{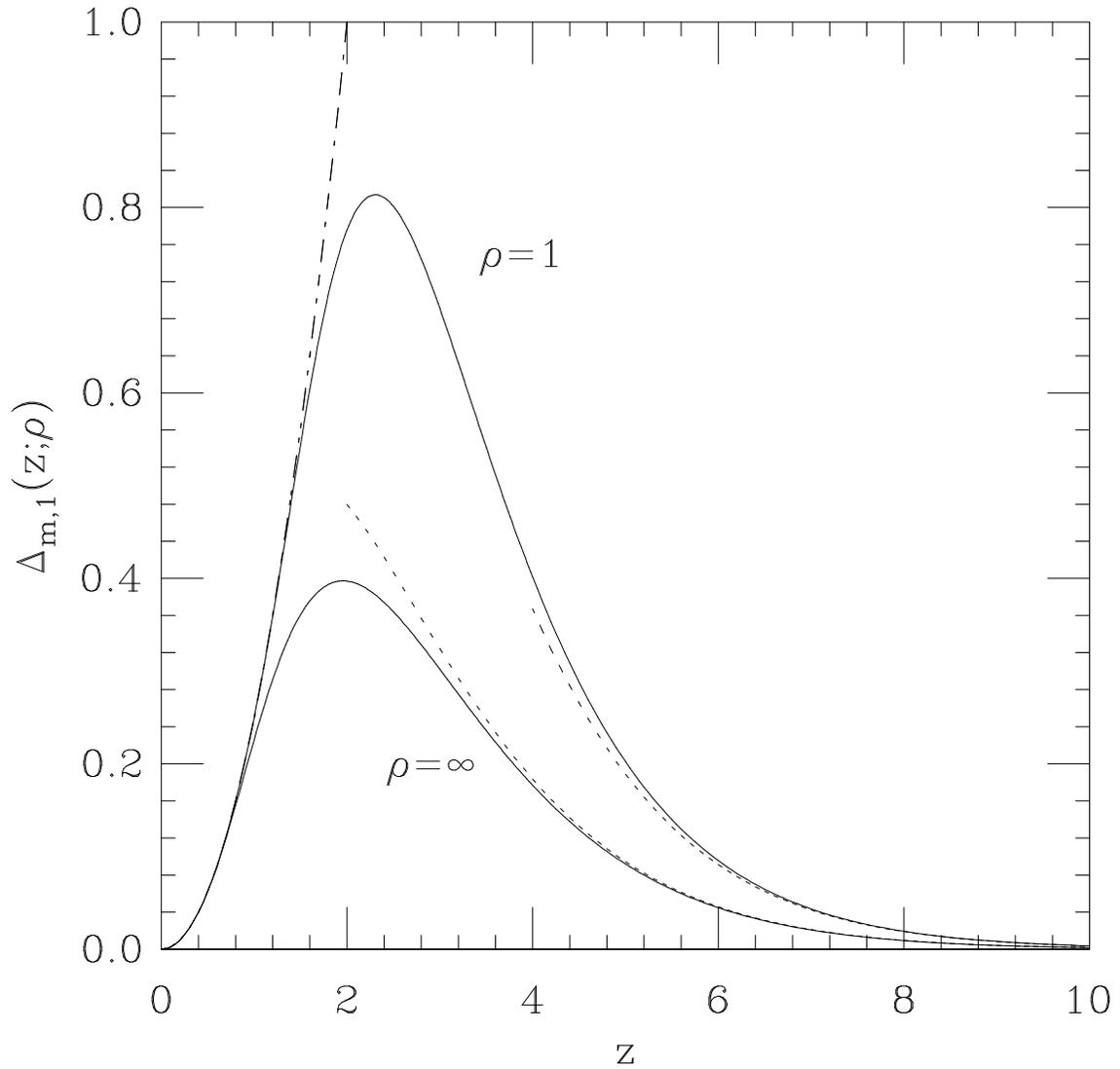}
\end{center}
\vspace*{-1cm}
\caption{$\Delta_{m,1}(z;\rho)$ for the standard hamiltonian $H^{std}$ for
$\rho=1$ and $\rho=\infty$. 
The dashed lines correspond to the asymptotic expansions    
\protect\reff{Deltam1largez} and \protect\reff{Deltam1smallz}.
$\Delta_{m,1}(z;\rho)$ for $H^{diag}$ 
is obtained by multiplying the vertical scale by $4/3$.
}
\label{fig1}
\end{figure}

\begin{figure}
\vspace*{-1cm} \hspace*{-0cm}
\begin{center}
\epsfxsize = 0.9\textwidth
\leavevmode\epsffile{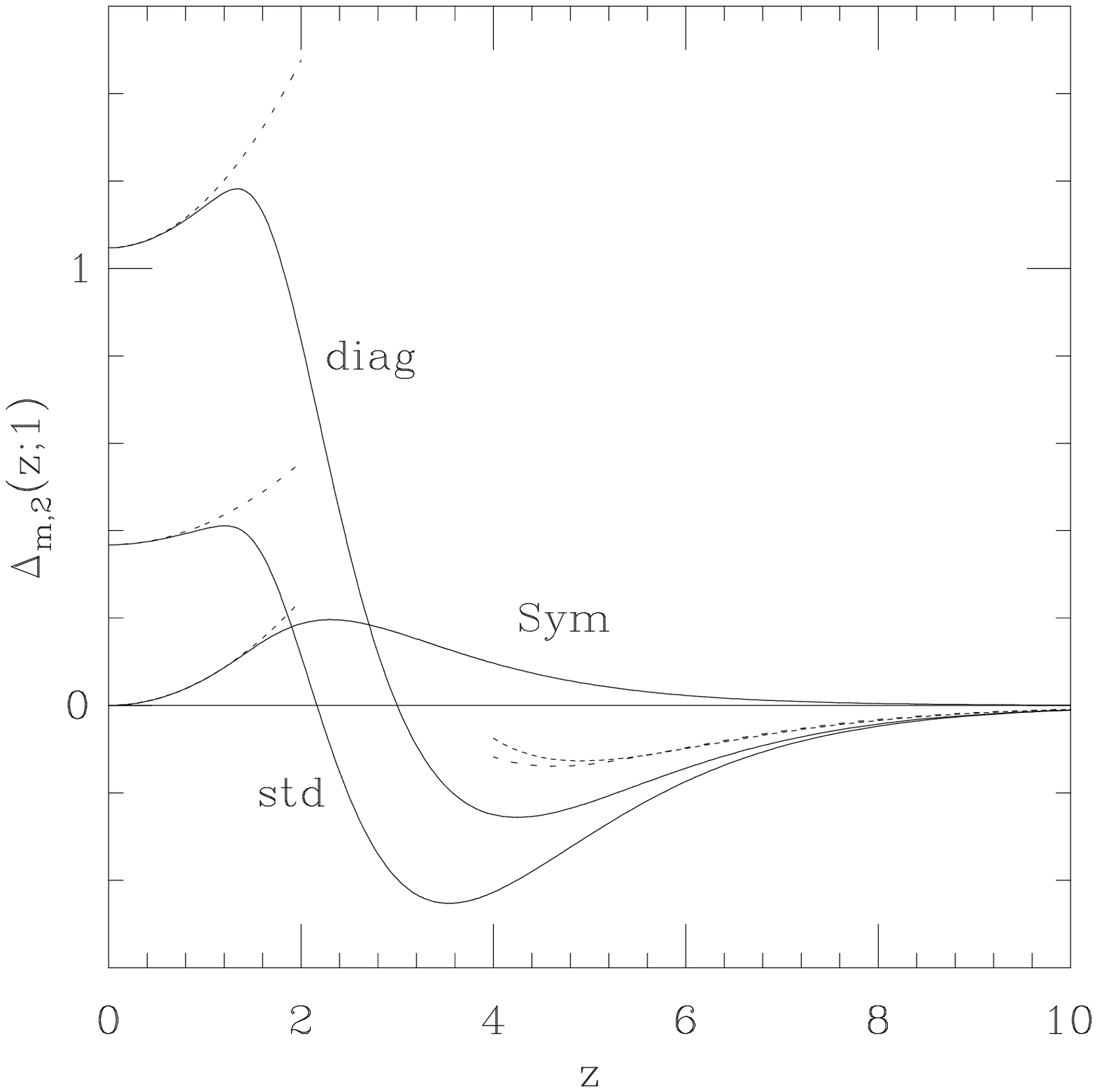}
\end{center}
\vspace*{-1cm}
\caption{$\Delta_{m,2}(z;1)$ for $H^{std}$ (``std"), $H^{diag}$
(``diag") and $H^{Sym}$ (``Sym"). The dashed lines are the 
asymptotic expansions \protect\reff{Deltam2largez} 
and \protect\reff{Deltam2smallz}.
}
\label{fig2}
\end{figure}

\begin{figure}
\vspace*{-1cm} \hspace*{-0cm}
\begin{center}
\epsfxsize = 0.9\textwidth
\leavevmode\epsffile{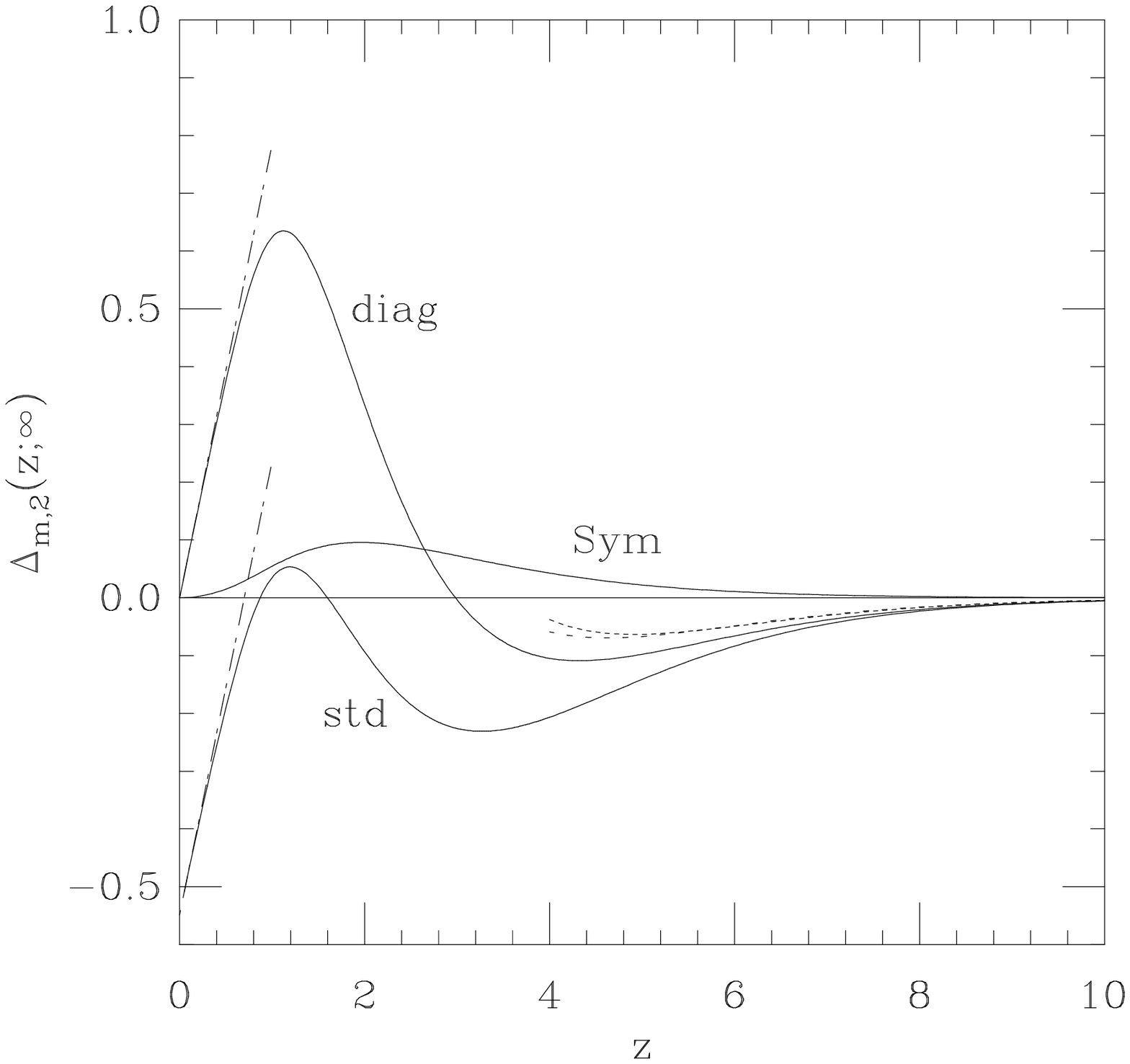}
\end{center}
\vspace*{-1cm}
\caption{$\Delta_{m,2}(z;\infty)$ for $H^{std}$ (``std"), $H^{diag}$
(``diag") and $H^{Sym}$ (``Sym"). The dashed lines are the 
asymptotic expansions \protect\reff{Deltam2largez} 
and \protect\reff{Deltam2smallzstrip}.
}
\label{fig3}
\end{figure}

The functions $\Delta_{m,1} (z;\rho)$ and $\Delta_{m,2}(z;\rho)$ are 
reported in the figures \ref{fig1}, \ref{fig2}, and \ref{fig3} 
for the torus with $\rho=1$ and for the strip $\rho=\infty$
for the various hamiltonians we have introduced.

From these plots one can immediately recognize a few basic facts: 
the region where the corrections to FSS are larger corresponds 
to $1\ltapprox z\ltapprox 4$ (the same has been found 
numerically in Monte Carlo simulations of $H^{std}$ with $N=3$
\cite{o3_letter}). 
In this region, for $H^{std}$ and $H^{diag}$ and small 
values of $L$, say $L=10$, the $\log L/L^2$ term gives 
a contribution which is $2 - 4$ times larger than the 
$1/L^2$ term and the corrections are positive. 
For these two hamiltonians the corrections become negative 
for large values of $z$ (this can be easily checked 
from the asymptotic expansions \reff{Deltam1largez} and 
\reff{Deltam2largez}). They are also negative for $H^{std}$ 
in the small-$z$ region on the strip or on the torus for large values
of $\rho$. Numerically we find
that $H^{diag}$ is the hamiltonian with the largest 
corrections, while $H^{Sym}$ is the ``best" one, as expected. 

We have finally checked that our expansion \reff{m2sum2} describes 
well the corrections to FSS even for small values of $L$. 
In table \ref{tableFSS2} we give $D_{exact}(L;z,\rho)$ and 
$D_{expan}(L;z,\rho)$
for $H^{diag}$ and $H^{Sym}$ for $\rho=1$ and $z=2$. For the 
first hamiltonian there is good agreement even at $L=4$, while for 
the latter there is a somewhat larger discrepancy, probably due to the 
larger spatial extent of the Symanzik hamiltonian.
We have also computed $D_{exact}(L;z,\rho)$ for the same values 
of $\rho$ and $z$ for $H^{Sym2}$: for $L=4$ (resp. 10) we get 
$D_{exact}(L;2,1) = 0.0005743$ (resp. $0.00001853$). The corrections
are estremely small (at $L=4$ they are 100 times smaller than 
those present for $H^{std}$): improvement really works!

\begin{table}
\begin{center}
\begin{tabular}{|l|rr|rr|}
\hline
     & \multicolumn{2}{c|}{$H^{diag}$} &
      \multicolumn{2}{c|}{$H^{Sym}$}   \\
\hline
 $L$ &  
 $D_{exact}(L;2,1)$ & $D_{expan}(L;2,1)$ & 
 $D_{exact}(L;2,1)$ & $D_{expan}(L;2,1)$  \\
\hline
4  &  0.1363230 &  0.1435180 &  0.0039380 &  0.0116907   \\  
6  &  0.0736526 &  0.0752402 &  0.0035162 &  0.0051959   \\  
8  &  0.0463553 &  0.0468952 &  0.0023457 &  0.0029227   \\  
10 &  0.0320544 &  0.0322879 &  0.0016198 &  0.0018705   \\  
12 &  0.0235981 &  0.0237157 &  0.0011666 &  0.0012990   \\  
14 &  0.0181629 &  0.0182287 &  0.0008802 &  0.0009543   \\  
16 &  0.0144512 &  0.0144909 &  0.0006858 &  0.0007307   \\  
20 &  0.0098297 &  0.0098468 &  0.0004483 &  0.0004676   \\  
\hline
\end{tabular}
\end{center}
\caption{Deviations from FSS for $H^{diag}$ and $H^{Sym}$:
here $\rho = 1$ and $z=2$.
 }
\label{tableFSS2}
\end{table}
\subsection{Observables}

Let us now compute the FSS curves and the corresponding corrections for the 
observables we have introduced in Sec. \ref{sec2}. We will
first consider the quantities that are obtained from the 
isovector correlation function, then we will discuss 
isotensor observables. 

\subsubsection{Isovector sector}

From Eq. \reff{m2sum2} it is immediate to obtain the finite-size-scaling
curves and their leading corrections for the various observables. The
susceptibility $\chi_V$ does not require any additional calculation
since
\be
{\chi_V (L,T)\over \chi_V(\infty)} = {m_\infty^2\over m^2_{L,T}} \; .
\ee
For the second-moment correlation lengths, neglecting terms of 
order $\log^2 L/L^4$, we obtain: 
\begin{eqnarray}
\left( {\xi^{(m,1)}_V(L,T) \over \xi^{(m)}_V(\infty)} \right)^2 
    &=&   {m^2_\infty\over m^2_{L,T}} 
          \left(1 + {2\pi^2\over L^2} (\alpha_1 + \alpha_2) 
          {1+\rho^2\over \rho^2}\right) \; ,
\label{xim1FSS}\\
\left( {\xi^{(m,2)}_V(L,T) \over \xi^{(m)}_V(\infty)} \right)^2 
    &=&  {z^2\over2} A_1(z;\rho) {m^2_\infty\over m^2_{L,T}} 
      \left(1 + {\pi^2\over 3L^2} (12 \alpha_1 + 12 \alpha_2 - 1)
       {A_2(z;\rho)\over A_1(z;\rho)}\right), \nonumber \\ [-1mm]
{} \label{xim2FSS} \\ [-1mm]
\left( {\xi^{(m,3)}_V(L,T) \over \xi^{(m)}_V(\infty)} \right)^2 
    &=& {m^2_\infty\over m^2_{L,T}} 
   \left[ 1 - {z\over4} \left( {1\over \sinh z/2} + 
           {\rho \over \sinh \rho z/2}\right) \right.
\nonumber \\
   && \qquad \left. + {1\over L^2} \left( Q_{p(L)}(z) + {1\over \rho^2} 
                          Q_{p(T)}(\rho z)\right) \right]\; ,
\label{xim3FSS}
\end{eqnarray}
where $p(M)$ is the parity of $M$ ($M=L,T$) and 
\begin{eqnarray}
A_1(z;\rho) &=& {4\pi^2 (1 + \rho^2) + 2 \rho^2 z^2 \over 
                (4 \pi^2 + z^2) (4 \pi^2 + \rho^2 z^2)}\; , \\
A_2(z;\rho) &=& {z^2\over (4 \pi^2 + z^2)^2} + 
                {\rho^2 z^2 \over (4 \pi^2 + \rho^2 z^2)^2} \; ,
\\
Q_{even}(z) &=& (12 \alpha_1 + 12 \alpha_2 - 1)
       {z^4\over192} {\cosh z/2\over \sinh^2 z/2} -
      (4 \alpha_1 + 4 \alpha_2 - 1)
       {z^3\over32} {1\over \sinh z/2} , \\
Q_{odd}(z) &=& (12 \alpha_1 + 12 \alpha_2 - 1)
       {z^4\over192} {\cosh z/2\over \sinh^2 z/2} - 
      (\alpha_1 + \alpha_2)
       {z^3\over8} {1\over \sinh z/2} .
\end{eqnarray}
Let us now consider the asymptotic limit $z\to\infty$. 
In the FSS limit it is easy to obtain
\begin{eqnarray}
\hskip -25pt
\left( {\xi^{(m,1)}_V(L,T) \over \xi^{(m)}_V(\infty)} \right)^2_{FSS} 
 &=& 1 - 2 e^{-z} \sqrt{{2\pi\over z}} - 
         2 e^{-\rho z} \sqrt{{2 \pi \over \rho z}} + 
         O(z^{-3/2} e^{-z}, z^{-3/2} e^{-\rho z})  ,
\label{xim1largez}
\\
\hskip -25pt
\left( {\xi^{(m,2)}_V(L,T) \over \xi^{(m)}_V(\infty)} \right)^2_{FSS} 
  &=& 1 - {2\pi^2 (1+\rho^2)\over \rho^2 z^2} + O(z^{-4}) ,
\label{xim2largez}
\\
\hskip -25pt
\left( {\xi^{(m,3)}_V(L,T) \over \xi^{(m)}_V(\infty)} \right)^2_{FSS} 
  &=& 1 - {z\over2} e^{-z/2} - {\rho z\over2} e^{-\rho z/2} 
    + O(z^{-1/2} e^{-z}, z^{-1/2} e^{-\rho z}) .
\label{xim3largez}
\end{eqnarray}
From these expansions one immediately sees that the FSS function
for $\xi^{(m,2)}_V$ goes to 1 only as a power as 
$z\to\infty$. The approach is very slow and indeed it reaches 
1 at the 1\% level only for $z\approx 60$. This is extremely 
inconvenient for Monte Carlo applications: indeed 
in order  to determine numerically the FSS curve one has to perform runs 
up to the value of $z$ where the FSS curve becomes 
1 within error bars: in this case runs with $z\approx 60$ are 
required, which means that simulations on very large lattices are needed.
The origin of these power corrections can be identified in the 
definition that approximates the infinite-volume $\xi^{(m)}_V(\infty)$
with corrections of order $1/L^2$: the $1/L^2$ terms give rise to the 
corrections of order $1/z^2$. The first definition should suffer
from the same problem because also in this case $\xi^{(m,1)}_V(L,T)$ 
converges to $\xi^{(m)}_V(\infty)$ with corrections of order $1/L^2$.
Instead Eq. \reff{xim1largez} shows corrections of order 
$O(e^{-z}/\sqrt{z})$. This is a peculiarity due to the particular 
form of $G_V(x)$ ($G_V(x)$ is a free-field two-point function). 
However for different Green's functions 
terms of order $1/z^2$ are expected and indeed they are present 
for $\xi^{(m,1)}_T$, cf. Eq. \reff{fxiTlargez}. As expected the FSS function 
for $\xi^{(m,3)}_V$ converges to 1 with corrections of order 
$z e^{-z/2}$: in this case the FSS function is 1 at the 1\% level 
already at $z\approx 15$. 

The large-$z$ behaviour of the FSS-functions can be easily computed not
only in the large-$N$ limit, but for all values of $N$. The
basic observation is that $G_V(x;L,T)$ converges to 
$G_V(x;\infty)$ with corrections of order $L^p e^{- \mu_V(\infty) L}$. 
Therefore, in order to compute the large-$z$ expansion, one can simply
replace $G_V(x;L,T)$ with $G_V(x;\infty)$. 
The function $G_V(x;\infty)$ is well known in the critical limit.
Indeed, if $\hat{G}_V(p;\infty)$ is the corresponding Fourier 
transform, then, in the limit $p\to 0$, $\xi^{(m)}_V(\infty)\to\infty$
with $p \xi^{(m)}_V(\infty)\equiv Q$ fixed, we have
\cite{Fisher-Burford,funzione2pt,CPRV_lettera}
\be
{\hat{G}_V(0;\infty) \over \hat{G}_V(p;\infty)} = D(Q).
\ee
The function $D(Q)$ can be expanded in the limit $Q\to 0$ in powers of
$Q^2$: 
\be
D(Q) = \, \sum_{n=0}^\infty b_n Q^{2n},
\ee
with $b_0 = b_1 = 1$. This expansion converges up to the three-particle
cut, i.e. for $|Q| < 3 s_m$ where $s_m$ is defined by
\be
s_m = \lim_{\beta\to\infty} \mu_V(\infty) \xi^{(m)}_V(\infty);
\ee
$s_m$ is the ratio between the second-moment
and the exponential correlation length. Moreover $D(Q)$ has a zero in
correspondence to the one-particle poles, $Q=\pm i s_m$. In a neighbourhood
of these points, we have
\be
D(Q) = s_z \left({Q^2\over s_m^2} + 1\right).
\ee
Using these results it is straightforward to compute the FSS scaling curves
in terms of $y = L/\xi^{(m)}_V(\infty)$ in the limit $y\to\infty$. 
Disregarding terms of order $y^p e^{-y}, y^p e^{-\rho y}$ we obtain
\begin{eqnarray}
\hskip -25pt
\left( {\xi^{(m,1)}_V(L,T) \over \xi^{(m)}_V(\infty)} \right)^2_{FSS} 
 &=& 1 + {1\over2} \sum_{n=1}^\infty 
    b_{n+1} \left[\left( {2\pi\over y}\right)^{2n} + 
                   \left( {2\pi\over \rho y}\right)^{2n} \right]\; ,
\label{xim1largezNgen}
\\
\hskip -25pt
\left( {\xi^{(m,2)}_V(L,T) \over \xi^{(m)}_V(\infty)} \right)^2_{FSS} 
  &=& {1\over2} \left[ 
   {\sum_{n=0}^\infty b_{n+1} (2\pi/y)^{2n} \over 
    \sum_{n=0}^\infty b_{n} (2\pi/y)^{2n} }  + 
   (y\to \rho y) \right]\; ,
\label{xim2largezNgen}
\\
\hskip -25pt
\left( {\xi^{(m,3)}_V(L,T) \over \xi^{(m)}_V(\infty)} \right)^2_{FSS} 
  &=& 1 - {y\over 2 s_z s_m} \left[
   e^{-s_m y/2} + \rho  e^{-s_m\rho y/2} \right]\; .
\label{xim3largezNgen}
\end{eqnarray}
In the large-$N$ limit $b_n=0$ for $n\ge 2$, $s_m=s_z=1$, and 
$y=z$ in the FSS limit, so that one recovers our previous results,
Eqs. \reff{xim1largez}, \reff{xim2largez}, \reff{xim3largez}. 
For generic values of $N$ numerical estimates of the various constants 
are reported in Ref. \cite{CPRV_lettera}. The deviations from the large-$N$
values are extremely small: for $N=3$ one finds from a strong-coupling 
analysis \cite{CPRV_lettera}
$b_2 = -1.2 (2) \cdot 10^{-3}$, $s_m = 0.9994(1)$, $s_z = 1.0013 (2)$, 
while a precise Monte Carlo simulation gives 
\cite{Meyer_unp} $s_m = 0.9992(6)$. Using Eqs. 
\reff{xim1largezNgen}, \reff{xim2largezNgen}, \reff{xim3largezNgen}, 
it is evident that the first definition is always the most convenient one
except for extremely large values of $y$ ($y \gtapprox 20$ for $N=3$), 
where the deviations are extremely tiny.
This is in agreement with the observation of Ref. \cite{Alles-Symanzik}:
they found numerically that, for $7\ltapprox y \ltapprox 10$,
$\xi^{(m,3)}_V$ had finite-size 
corrections larger than $\xi^{(m,1)}_V$. Using their data we can 
check the large-$y$ behaviour of the FSS function of  $\xi^{(m,3)}_V$.
We find that the data of Ref. \cite{Alles-Symanzik} --- they belong to the 
range $7\ltapprox y \ltapprox 10$ --- are well described by the 
formula
\be
\left( {\xi^{(m,3)}_V(L,T) \over \xi^{(m,1)}_V(L,T)} \right)^2 = 
1 + \alpha {L\over \xi^{(m,1)}_V(L,T)} \, 
\exp \left[ - {L\over 2\xi^{(m,1)}_V(L,T)}\right], 
\ee
where
\be
 \alpha = \cases{1.023 \pm 0.012 & for $N=3$, \cr
                 1.001 \pm 0.007 & for $N=8$,}
\ee
in good agreement with our previous results.

\begin{figure}
\vspace*{-1cm} \hspace*{-0cm}
\begin{center}
\epsfxsize = 0.9\textwidth
\leavevmode\epsffile{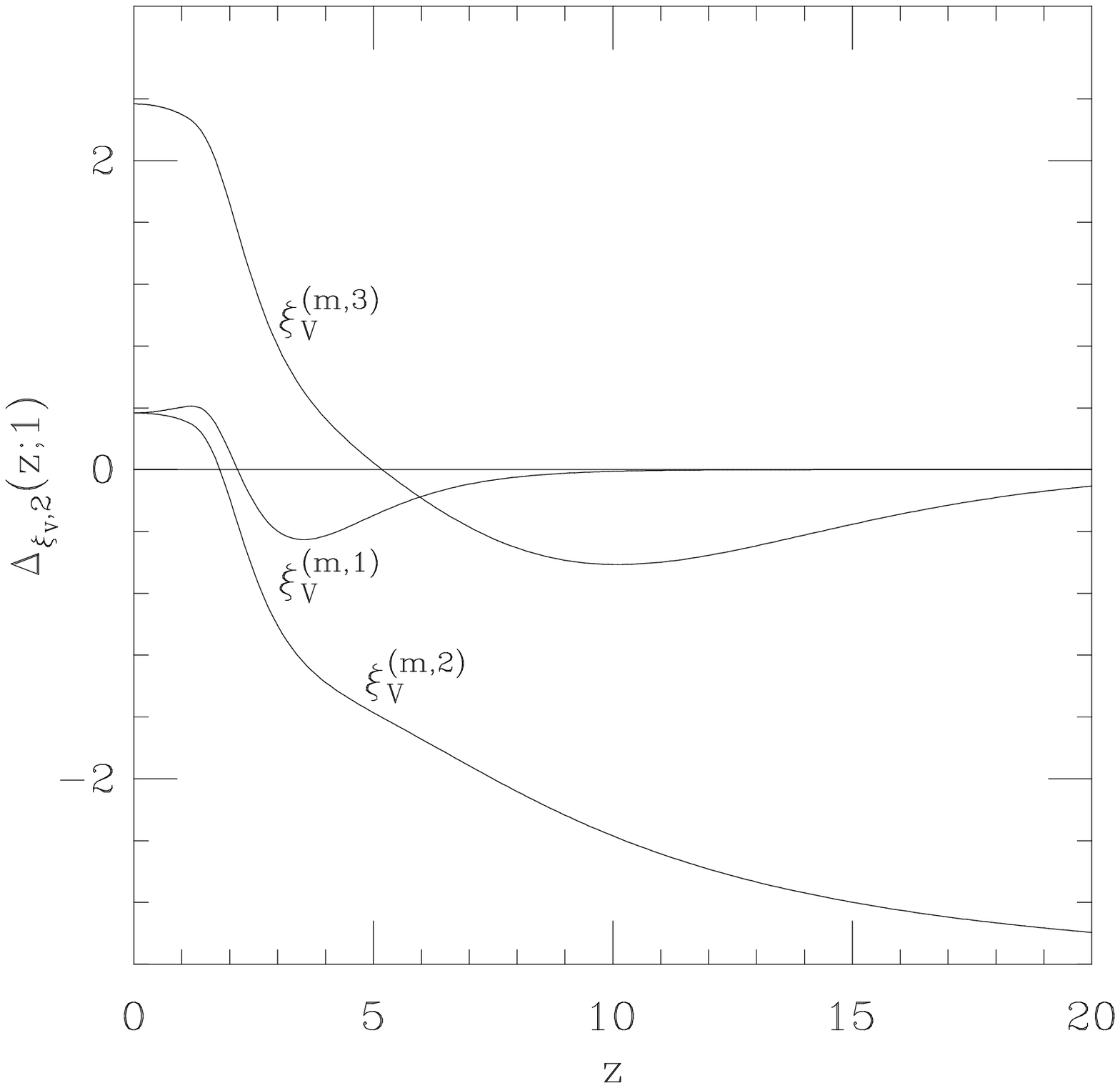}
\end{center}
\vspace*{-1cm}
\caption{$\Delta_{\xi_V,2}(z;1)$ for $H^{std}$ 
for the three different definitions 
of second-moment correlation length.
}
\label{fig4}
\end{figure}

\begin{figure}
\vspace*{-1cm} \hspace*{-0cm}
\begin{center}
\epsfxsize = 0.9\textwidth
\leavevmode\epsffile{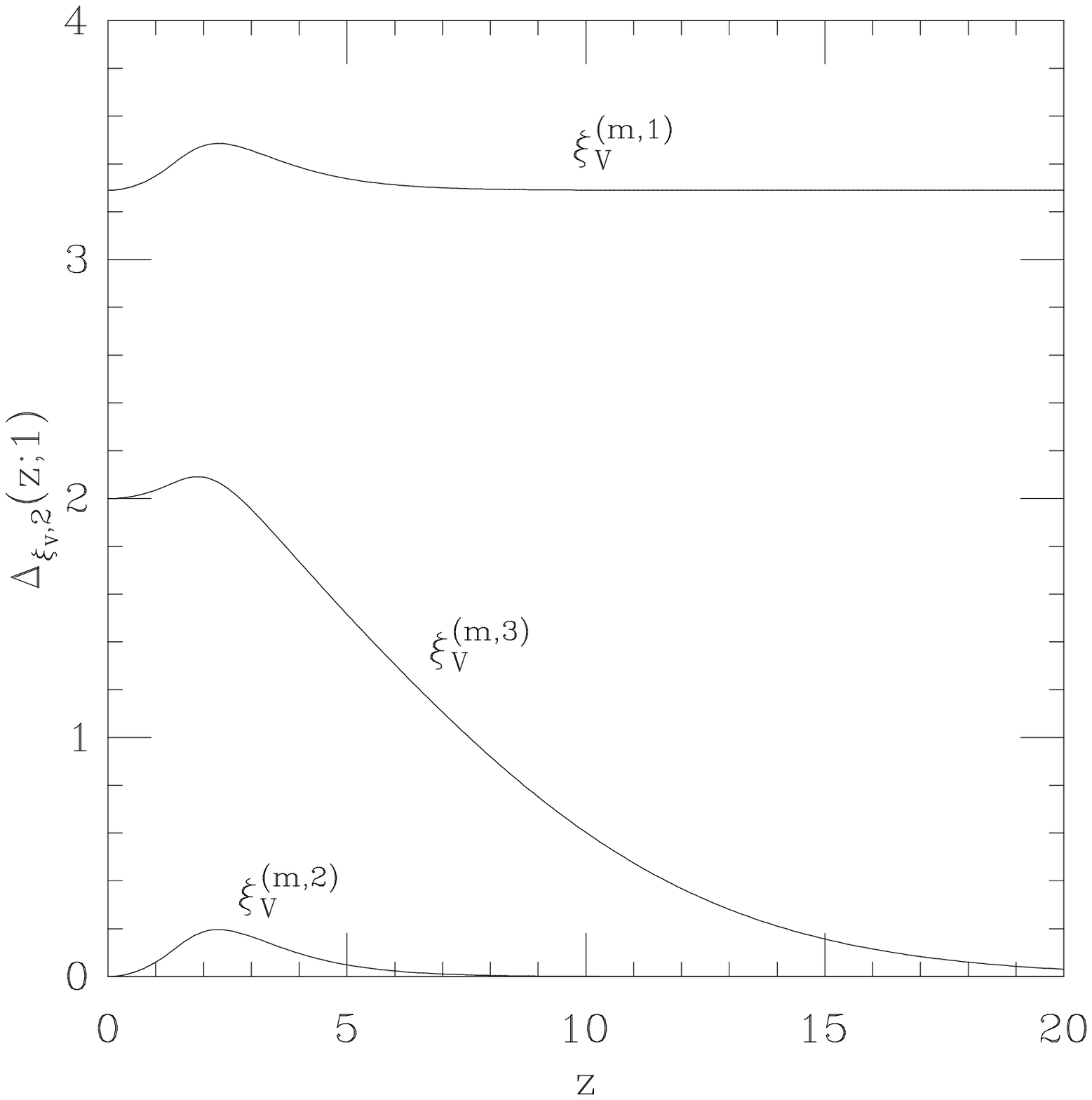}
\end{center}
\vspace*{-1cm}
\caption{$\Delta_{\xi_V,2}(z;1)$ for 
$H^{Sym}$  for the three different definitions 
of second-moment correlation length.
}
\label{fig5}
\end{figure}

Let us now consider the corrections to scaling. The term proportional to
$\log  L/L^2$ is identical in all cases to $\Delta_{m,1}(z;\rho)$,
cf. Eq. \reff{Deltam1}. The contribution proportional to 
$1/L^2$ depend instead on the definition of $\xi$. 
In Figs. \ref{fig4} and \ref{fig5} 
we report the deviations from FSS for the three definitions
for the standard and the Symanzik hamiltonians
($\Delta_{\xi,1}$ and $\Delta_{\xi,2}$ are defined in analogy with 
Eq. \reff{m2sum2}).
Notice that for $\xi^{(m,1)}_V$ and $\xi^{(m,3)}_V$ 
the corrections proportional to $1/L^2$ do not vanish even 
when $\alpha_1 = {1\over12}$, $\alpha_2 = 0$ and 
Eq. \reff{conditionLambda1} are satisfied. This is expected since 
the second-moment correlation length is an off-shell quantity. Therefore 
the definition of the correlation length must be improved, as well as 
the hamiltonian. For instance, if one uses $\xi^{(m,1)}_V$ and the 
Symanzik hamiltonian one does not see any improvement:
this definition has large corrections to scaling, and the behaviour 
is worse for the Symanzik hamiltonian than for the standard one.
In this case there is a simple remedy to
the problem: modify the definition in such a way that $\xi(L,T) \approx 
\xi_\infty + O(L^{-4}, T^{-4}, L^{-2} T^{-2})$. Analogously one could 
proceed for $\xi^{(m,3)}_V$. The second definition is automatically
improved but this is a peculiarity of the large-$N$ limit.

Let us finally discuss the mass gap $\mu_V(L)$ and the exponential
correlation length $\xi^{(exp)}_V = 1/\mu_V(L)$. We have computed 
the FSS functions expressing them in terms of $\mu_V(L)$ itself,
i.e. using as variable $x\equiv \mu_V(L) L$ instead of $z$. 
We get
\be
\left({\mu_V(\infty)\over \mu_V(L)}\right)^2 = 
   f_{m}(x;\infty) \left( 
   1 + {\Delta_{\mu,1}(x;\infty)\over L^2}\log L + 
       {\Delta_{\mu,2}(x;\infty)\over L^2} \right)\; ,
\ee
where $\Delta_{\mu,1}(x;\infty) = \Delta_{m,1}(x;\infty)$ and
\begin{eqnarray}
\Delta_{\mu,2}(x;\infty) &=& \Delta_{m,2}(x;\infty) + 
       {\pi x^3\over6} (12 \alpha_1 + 12 \alpha_2 -1) 
    {\partial F_0\over \partial x}(x;\infty) 
\nonumber \\
&& \quad + {8\over3} (12 \alpha_1 + 12 \alpha_2 -1) 
  e^{-4 \pi F_0(x;\infty)}\; .
\end{eqnarray}
Thus only the $1/L^2$ term differs from the expansion of 
$m^2_\infty/m^2_{L,T}$. The asymptotic behaviour for large $x$ is analogous 
to \reff{Deltam2largez} while for  $x\to 0$ we have
\be
\Delta_{\mu,2}(x;\infty) = 
   {\pi^2\over18} (12 \alpha_1 -1) - 
   {\pi x\over3} (12 \alpha_1 + 12 \alpha_2 - 1) + O(x^2).
\ee
Notice that, since this quantity is defined on-shell, it is improved 
once the hamiltonian is improved. 

\subsubsection{Isotensor sector}

Let us now consider the isotensor observables. The calculation of the 
FSS  function for the isotensor susceptibility $\chi_T$ is 
straightforward. We obtain
\be
{\chi_T(L,T)\over \chi_T(\infty)} \,=\, 
   f_{\chi_T}(z;\rho) \left(
    1 + {\Delta_{\chi_T,1}(z;\rho)\over L^2} \log L +
    {\Delta_{\chi_T,2}(z;\rho)\over L^2} \right),
\ee
with corrections of order $\log^2 L/L^4$ where 
\begin{eqnarray}
\hskip -20pt
f_{\chi_T}(z;\rho) &=& 
   - {64 \pi\over z} {\partial F_0\over \partial z}(z;\rho)
   e^{-4\pi F_0(z;\rho)} ,
\\
\hskip -20pt
\Delta_{\chi_T,1}(z;\rho) &=& 
    {1\over4} (12\alpha_1 + 16\alpha_2 -1) 
    \left[ 64 e^{-4 \pi F_0(z;\rho)} - z^2 + 
   {z\over 2\pi} \left({\partial F_0\over \partial z}(z;\rho)\right)^{-1}
    \right] \; ,
\\
\hskip -20pt
\Delta_{\chi_T,2}(z;\rho) &=&
   32 \pi (12\alpha_1 + 16\alpha_2 -1) F_0(z;\rho) e^{-4 \pi F_0(z;\rho)}
 \nonumber \\ [2mm]
  && - 4 (28 \alpha_1 + 32 \alpha_2 -3) e^{-4 \pi F_0(z;\rho)} 
     - 4 \pi {\cal F}_1(z;\rho) 
 \nonumber \\
  && + {\partial{\cal F}_1/\partial z (z;\rho)\over 
        \partial F_0/\partial z(z;\rho)} - 
     256 \pi \Lambda_1 e^{-4 \pi F_0(z;\rho)}\; .
\end{eqnarray}
As before the $\log L/L^2$ corrections cancel 
if $12\alpha_1 + 16\alpha_2 -1=0$. The function $\Delta_{\chi_T,2}(z;\rho)$
simplifies considerably if $\alpha_1 = \smfrac{1}{12}$ and 
$\alpha_2 = 0$ . In this case 
\be
\Delta_{\chi_T,2}(z;\rho)\, =\, 
   \left( {1\over 96\pi} - \Lambda_1\right)
   \left[ 256 \pi e^{-4 \pi F_0(z;\rho)} -4 \pi z^2 + 
    2z \left({\partial F_0\over \partial z}(z;\rho)\right)^{-1} \right]\; .
\ee
Therefore, if Eq. \reff{conditionLambda1} is satisfied, 
$\chi_T$ has only corrections 
of order $\log L/L^4$.
It is straightforward to compute the expansions of the various FSS functions
in the limit $z\to\infty$. Using the results of Sect. 
\ref{secB.2.1} and \ref{secB.2.2} we obtain
\begin{eqnarray}
f_{\chi_T}(z;\rho) &=& 1 + \sqrt{2 \pi z}\ e^{-z} +
                           \sqrt{2 \pi \rho z}\ e^{-\rho z} +
     O(z^{-1/2} e^{-z},z^{-1/2} e^{-\rho z}) ,
\label{fchiTlargez}\\
\Delta_{\chi_T,1}(z;\rho) &=& {z^2\over4} (12 \alpha_1 + 16 \alpha_2 -1)
     \left(\sqrt{2\pi z} \ e^{-z} +
           \sqrt{2\pi\rho z} \  e^{-\rho z} \right) \nonumber \\
&& \qquad\qquad + O(z^{3/2} e^{-z},z^{3/2} e^{-\rho z}) ,
\label{DeltachiT1largez}\\
\Delta_{\chi_T,2}(z;\rho) &=& - {\pi z^4\over 12} 
   (12 \alpha_1 + 12 \alpha_2 -1) 
     \left( {e^{-z} \over \sqrt{2\pi z}} +
           {\rho^2 e^{-\rho z} \over \sqrt{2\pi \rho z}} \right) \nonumber \\ 
&& - {z^2\over8} (12 \alpha_1 + 16 \alpha_2 - 1) 
    \, \log{z^2\over32}\,
  \left(e^{-z} \sqrt{2\pi z} +
           e^{-\rho z} \sqrt{2\pi\rho z} \right)  \nonumber \\ [2mm]
&& \qquad \qquad + 
     O(z^{5/2} e^{-z},z^{5/2} e^{-\rho z}) .
\label{DeltachiT2largez}
\end{eqnarray}
As expected, $f_{\chi_T}(z;\rho)$ behaves as $z^p e^{-z}$, but 
$p$ differs from the value it assumes for other observables 
(see e.g. the large-$z$ behaviour of $\chi_V$). Indeed, while the 
exponential behaviour is completely general, the power $p$ 
depends on the observable.

\begin{figure}
\vspace*{-1cm} \hspace*{-0cm}
\begin{center}
\epsfxsize = 0.9\textwidth
\leavevmode\epsffile{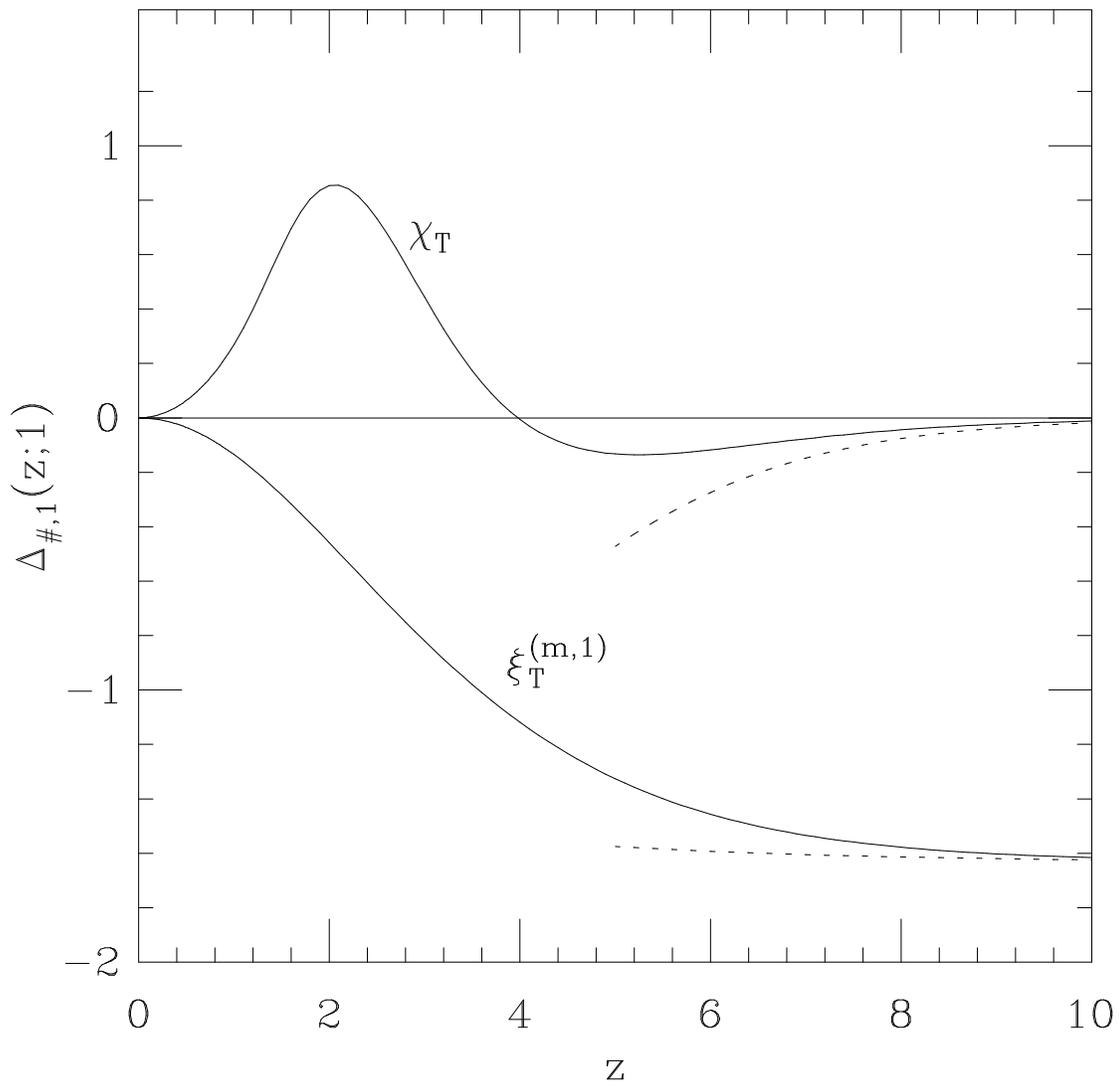}
\end{center}
\vspace*{-1cm}
\caption{$\Delta_{\chi_T,1}(z;1)$ and $\Delta_{\xi_T,1}(z;1)$ for 
$H^{std}$. The dashed lines are the large-$z$ asymptotic expansions,
Eqs. \protect\reff{DeltachiT1largez} and \protect\reff{DeltaxiT1largez}.
}
\label{fig6}
\end{figure}

\begin{figure}
\vspace*{-1cm} \hspace*{-0cm}
\begin{center}
\epsfxsize = 0.9\textwidth
\leavevmode\epsffile{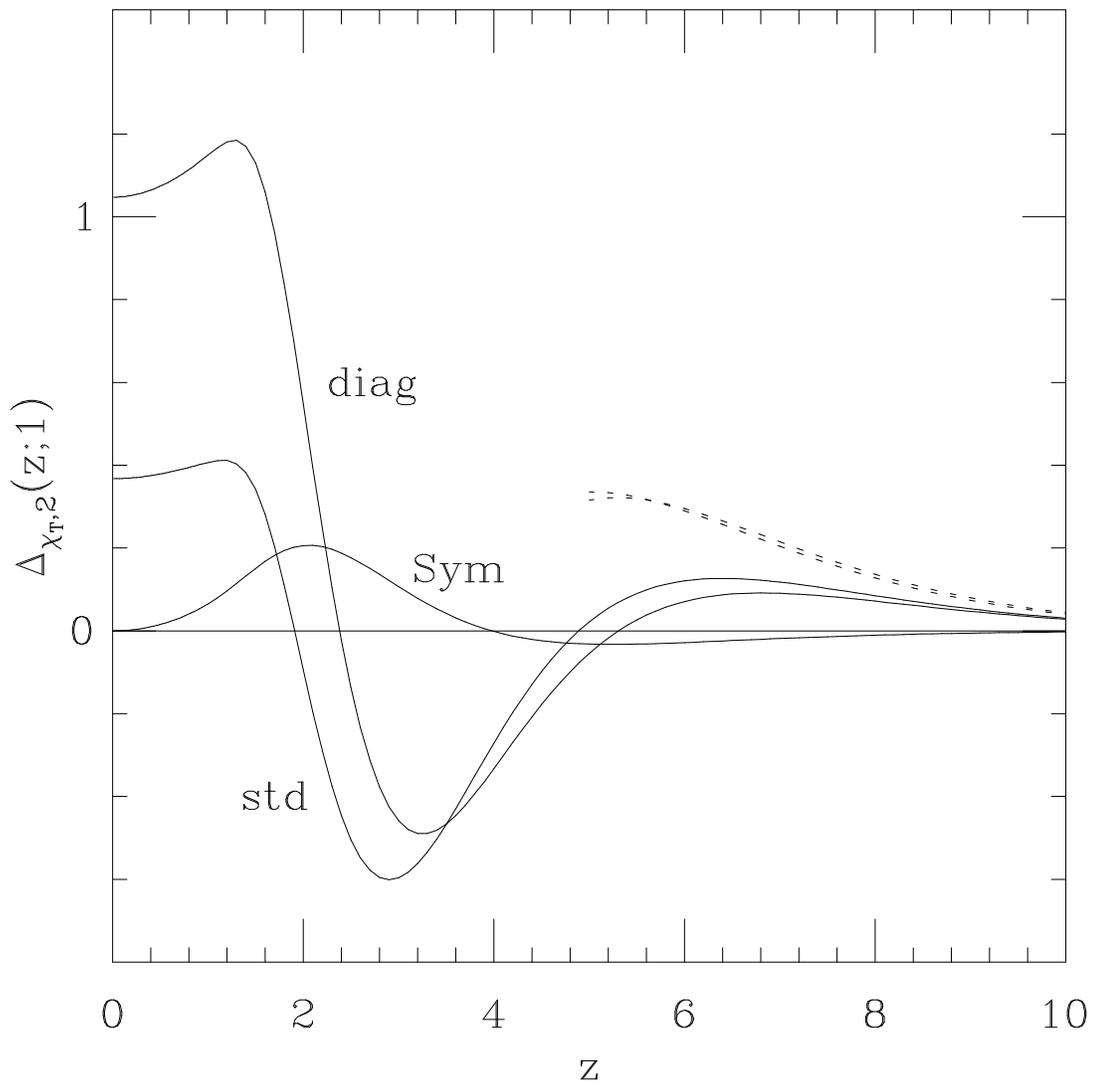}
\end{center}
\vspace*{-1cm}
\caption{$\Delta_{\chi_T,2}(z;1)$  for
$H^{std}$, $H^{Sym}$ and $H^{diag}$. 
The dashed lines are the large-$z$ asymptotic expansions,
Eq. \protect\reff{DeltachiT2largez}. 
}
\label{fig7}
\end{figure}

In Figs. \ref{fig6} and \ref{fig7} we report the graphs of 
$\Delta_{\chi_T,1}(z;\rho)$ and $\Delta_{\chi_T,2}(z;\rho)$ for $\rho=1$ 
and different hamiltonians. The behaviour is very similar to the 
behaviour of $m^2_{L,T}/m^2_\infty$. The FSS corrections are quite small.
The Symanzik hamiltonian shows the best behaviour, while the diagonal 
hamiltonian is the one with the largest deviations from FSS.

Finally let us compute the FSS curve for the tensor correlation length.
We will restrict the discussion to the standard action; the generalization
to generic hamiltonians is straightforward but the final expressions 
are cumbersome. Moreover we will restrict our discussion to 
$\xi^{(m,1)}_T$, that is to the definition used in numerical simulations.
We obtain
\be
\left({\xi_T^{(m,1)}(L,T)\over \xi_T^{(m)}(\infty)}\right)^2 
\, =\, f_{\xi_T}(z;\rho) \left(
    1 + {\Delta_{\xi_T,1}(z;\rho)\over L^2} \log L +
    {\Delta_{\xi_T,2}(z;\rho)\over L^2} \right), 
\ee
where 
\begin{eqnarray}
\hskip -20pt
f_{\xi_T}(z;\rho)  &=& 192\ \Xi_1(z;\rho) e^{-4 \pi F_0(z;\rho)} ,
\\
\hskip -20pt
\Delta_{\xi_T,1}(z;\rho) &=& 
    \Delta_{m,1}(z;\rho) +{\Xi_2(z;\rho)\over \Xi_1(z;\rho)} + 
    8 e^{-4 \pi F_0(z;\rho)} ,
\\
\hskip -20pt
\Delta_{\xi_T,2}(z;\rho) &=& 
    \Delta_{m,2}(z;\rho) +{\Xi_3(z;\rho)\over \Xi_1(z;\rho)} -
    4 e^{-4 \pi F_0(z;\rho)} + 
    16 \pi F_0(z;\rho) e^{-4 \pi F_0(z;\rho)} ,
\end{eqnarray}
and 
\begin{eqnarray}
\hskip -20pt
\Xi_1(z;\rho) &=& 
   {1\over 8\pi^2} \left[
  -1 - \rho^2  - {1\over 2z}{\partial F_0\over \partial z}(z;\rho)
    \left( {1\over F_3(z;\rho)} + {1\over F_3(\rho z;1/\rho)}
    \right) \right]\; ,
\\
\hskip -20pt
\Xi_2(z;\rho) &=& 
   {1\over 128\pi^3} \left[
     {1\over F_3(z;\rho)} + {1\over F_3(\rho z;1/\rho)} \right.
\nonumber \\
&& \quad \left. 
    + {1\over 2z}{\partial F_0\over \partial z}(z;\rho)
    \left( {1\over F_3^2(z;\rho)} + {1\over \rho^2 F_3^2(\rho z;1/\rho)}
    \right) \right]\; ,
\\
\hskip -20pt
\Xi_3(z;\rho) &=& 
   {1\over 8\pi^2} \left[
   - {1\over 2z}{\partial F_1\over \partial z}(z;\rho)
    \left( {1\over F_3(z;\rho)} + {1\over F_3(\rho z;1/\rho)}
    \right)   \right.
\nonumber \\ 
\hskip -5pt
&& - {2 \pi^2\over 3} - {\pi^2\over 6 z} {\partial F_0\over \partial z}(z;\rho)
\left( {1\over F_3(z;\rho)} + {1\over \rho^2 F_3(\rho z;1/\rho)}
    \right)
\nonumber \\
\hskip -5pt
&& \left. + {1\over 2z}{\partial F_0\over \partial z}(z;\rho) 
   \left({F_4(z;\rho)\over F_3^2(z;\rho)} + 
         {F_4(\rho z;1/\rho)\over \rho^2 F_3^2(\rho z;1/\rho)} + 
      {1\over 16\pi} {\log \rho\over \rho^2 F_3^2(\rho z;1/\rho)}
   \right)\right] .
\end{eqnarray}
Using Eqs. \reff{F0largez}, \reff{F1largez}, \reff{F3largez}, and
\reff{F4largez} it is straightforward to compute the large-$z$
behaviour of the various FSS functions. We obtain
\medskip 
\begin{eqnarray}
\hskip -10pt
f_{\xi_T} (z;\rho) &=& 1 - {\pi^2\over 15 z^2} {1 + \rho^2\over \rho^2} + 
     {2\pi^4 \over 63 z^4} {1 + \rho^4\over \rho^4} + O(z^{-6}) ,
\label{fxiTlargez}
\\
\hskip -10pt
\Delta_{\xi_T,1} (z;\rho) &=&  
    - {\pi^2\over12} {1 + \rho^2\over \rho^2} + 
    {\pi^4\over 180 \rho^4 z^2} (3 - 2\rho^2 + 3 \rho^4)
    + O(z^{-4}), 
\label{DeltaxiT1largez}
\\
\hskip -10pt
\Delta_{\xi_T,2} (z;\rho) &=&
    {\pi^2\over24\rho^2} (1 + \rho^2) \left( \log{z^2\over32} + 4\right) 
\nonumber \\ 
 && \hskip -30pt
  - {\pi^4\over 360\rho^4 z^2} (3 - 2 \rho^2 + 3 \rho^4) \log{z^2\over32} -
    {\pi^4\over 90\rho^4 z^2} (1 - \rho^2)^2 + 
   O(\log z/z^4).
\label{DeltaxiT2largez}
\end{eqnarray}
As expected $f_{\xi_T}(z;\rho)$ approaches one as $1/z^2$ and thus it 
reaches the asymptotic value for $z\to\infty$ within 1\% only for 
$z\approx 9$. Notice moreover that $\Delta_{\xi_T,2} (z;\rho)$ 
diverges logarithmically as $z\to\infty$. This fact signals the 
non-uniformity of the expansion in $z$. This is not unexpected. 
Indeed, for each fixed $z$, we expect the expansion to be reliable only
if $m_{L,T} \ll 1$, i.e. if the correlation length is much larger than
a lattice spacing. Therefore we expect the expansion to be  valid
only if $z\ll L$. If $1 \ll z \ll L$, $\log z/L^2$ is a small number and thus 
the expansion is completely under control. 

\begin{figure}
\vspace*{-1cm} \hspace*{-0cm}
\begin{center}
\epsfxsize = 0.9\textwidth
\leavevmode\epsffile{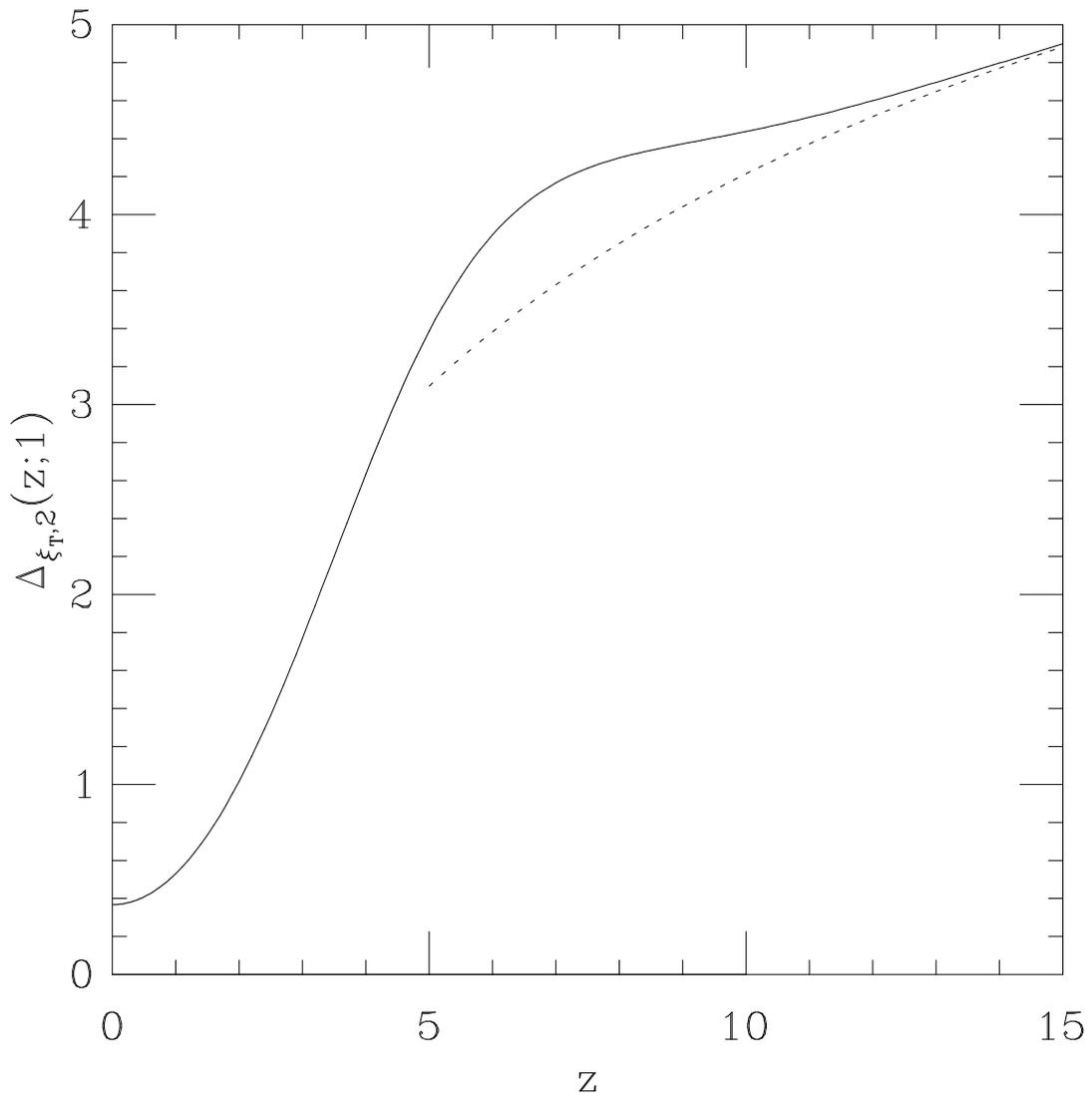}
\end{center}
\vspace*{-1cm}
\caption{$\Delta_{\xi_T,2}(z;1)$  for
$H^{std}$. The dashed line is the large-$z$ asymptotic expansion,
Eq. \protect\reff{DeltaxiT2largez}.
}
\label{fig8}
\end{figure}

The functions $\Delta_{\xi_T,1} (z;\rho)$ and
$\Delta_{\xi_T,2} (z;\rho)$ are reported for $\rho = 1$ in 
Figs. \ref{fig6} and \ref{fig8}. From these plots, 
comparing with the analogous graphs for other observables,
one can immediately see
that the corrections to FSS for $\xi^{(m,1)}_T$ are quite large. 
This is particularly evident in the large-$z$ region, where the 
$\log L/L^2$ term goes to a constant (for the isovector correlation length
this term vanishes exponentially), while the $1/L^2$ term diverges 
as $\log z$.

\section{Mixed $O(N)$-$RP^{N-1}$ model}

In this section we compute the FSS corrections for the hamiltonian
\reff{mixedmodel}. As before we want to obtain a relation between
$m_\infty^2$ and $m^2_{L,T}$ at fixed $\beta$. Using now the gap
equation \reff{mixedgapeq} we have
\be
 {4 I_{L,T}(m_{L,T}^2)^2 \over 4 I_{L,T}(m_{L,T}^2) + 
    r ( m^2_{L,T} I_{L,T}(m^2_{L,T}) - 1)} = 
 {4 I_{\infty}(m_{\infty}^2)^2 \over 4 I_{\infty}(m_{\infty}^2) + 
    r ( m^2_{\infty} I_{\infty}(m^2_{\infty}) - 1)} \; .
\ee
We will now discuss the FSS limit in which  $L,T\to\infty$, $\beta\to\infty$, 
$m_{L,T},m_\infty\to0$ with $m_{L,T} L\equiv z$ and $T/L\equiv \rho$ fixed.
At leading order we can disregard the terms 
$m^2_{\infty} I_{\infty}(m^2_{\infty})$ and 
$m^2_{L,T} I_{L,T}(m^2_{L,T}) $ in the denominators obtaining
\be
 {I_{L,T}(m_{L,T}^2)^2 \over 4 I_{L,T}(m_{L,T}^2) - r} = 
 {I_{\infty}(m_{\infty}^2)^2 \over 4 I_{\infty}(m_{\infty}^2) - r}\; ,
\ee
that implies $I_{L,T}(m_{L,T}^2) = I_{\infty}(m_{\infty}^2)$. Thus, at leading
order, the relation between $m_{L,T}$ and $m_{\infty}$ is identical to the one 
we have discussed in the previous section. Consequently the FSS functions 
for the hamiltonian \reff{mixedmodel} are identical to 
the FSS functions of models with hamiltonian \reff{eq2.1}, as expected on the 
basis of universality\footnote{Under suitable
assumptions (absence of vortices) that are verified in the large-$N$ limit,
one can prove that the FSS functions for the $RP^{N-1}$ model with 
periodic boundary conditions are equal to the $N$-vector FSS functions
with fluctuating periodic/antiperiodic boundary conditions 
\protect\cite{Hasenbusch}. 
In the large-$N$ limit the antiperiodic contribution vanishes,
hence $RP^{\infty}$ has the same FSS functions of the $N$-vector model.}. 
The corrections will however be different. Writing
\be
I_\infty (m_{\infty}^2) = I_{L,T}(m_{L,T}^2) + 
    \delta\ m_{L,T}^2\; ,
\ee
a simple computation gives
\be
\delta = {r\over2}\left( {m_{\infty}^2\over m_{L,T}^2} - 1\right)\
    {I_{L,T}(m_{L,T}^2)^2 \over 2 I_{L,T}(m_{L,T}^2) - r}\; .
\ee
Solving for $m_{\infty}^2/m_{L,T}^2$ we get finally 
\begin{eqnarray}
{m_{\infty}^2\over m_{L,T}^2} &=&
  f_m(z;\rho)\left(
      1 + {(\Delta_{m,1}(z;\rho)+\Delta_{m,1}^r (z;\rho))\over L^2}\log L +
          {(\Delta_{m,2}(z;\rho)+\Delta_{m,2}^r (z;\rho))\over L^2} +\right.
\nonumber \\
  && + \left. {\widehat{\Delta_{m}^r} (L;z;\rho)\over L^2} \right)\; ,
\label{ratiommixed}
\end{eqnarray}
where $f_m(z;\rho)$, $\Delta_{m,1}(z;\rho)$ and
$\Delta_{m,2}(z;\rho)$ are defined in Eqs. \reff{fm},
\reff{Deltam1} and \reff{Deltam2}, with $\alpha_1 = \alpha_2 = 0$ and
\begin{eqnarray}
\Delta_{m,1}^r (z;\rho) &=& 
  {r\over2} \left(z^2 - 32 e^{-4\pi F_0(z;\rho)}\right),
\\
\Delta_{m,2}^r (z;\rho) &=& 
  {\pi r\over2} \left(2 F_0(z;\rho) + r\right) 
    \left(z^2 - 32 e^{-4\pi F_0(z;\rho)}\right),
\\
\widehat{\Delta_{m}^r} (L;z;\rho) &=& 
   {\pi^2 r^3\over2} \left(z^2 - 32 e^{-4\pi F_0(z;\rho)}\right) \, 
   {1\over \log L + 2\pi F_0(z;\rho) - \pi r}\; .
\label{hatdelta}
\end{eqnarray}
The result \reff{ratiommixed} is quite different from Eq. \reff{m2sum2}.
Indeed, while before the corrections had a very simple dependence 
on $\log L$, now the corrections involve $\widehat{\Delta_{m}^r}
(L;z;\rho)$ that is not a simple polynomial in $\log L$. 
Notice that, for large 
$L$ at fixed $z$, $\widehat{\Delta_{m}^r} (L;z;\rho)$ behaves as 
$1/\log L$. Therefore, in the FSS limit, 
the corrections still behave as $\log L/L^2$.

We should make a second remark about $\widehat{\Delta_{m}^r}
(L;z;\rho)$. It is easy to convince oneself from the 
asymptotic expansions, Eqs. \reff{F0largez}, \reff{F0smallz}, and 
\reff{F0smallzstrip}, that $F_0(z;\rho)$ assumes any real value. 
Therefore, for each value of $L$, there is a value $z_c$ such that 
the denominator in Eq. \reff{hatdelta} vanishes, and therefore
$\widehat{\Delta_{m}^r} (L;z;\rho)$ diverges. When $L\to\infty$, 
$z_c\to\infty$; more precisely, using Eq. \reff{F0largez}, we have 
$z_c \approx e^{-\pi r} \sqrt{32} L$. This singularity is a signal of the 
fact that the expansion is not uniform in $z$. For each $z$ the 
expansion is valid only when $L\gg z$, i.e. for $m_{L,T} \ll 1$. 
In other words the expansion makes sense only when the correlation length
is much larger than a lattice spacing.
\begin{table}
\begin{center}
\begin{tabular}{|l|rr|rr|}
\hline
 $L$ &  $R_{exact}(L;2,1)$ & $R_{expan} (L;2,1)$ &
 $D_{exact}(L;2,1)$ & $D_{expan} (L;2,1)$ \\
\hline
8 &  0.35709886 & 0.34256990 & 0.5916549 & 0.5268967 \\
10 & 0.29398737 & 0.29002279 & 0.3103555 & 0.2926846 \\
12 & 0.26809044 & 0.26656021 & 0.1949281 & 0.1881076 \\
14 & 0.25463070 & 0.25391316 & 0.1349356 & 0.1317374 \\
16 & 0.24665643 & 0.24627448 & 0.0993928 & 0.0976904 \\
32 & 0.22943124 & 0.22941299 & 0.0226170 & 0.0225356 \\
64 & 0.22561220 & 0.22561112 & 0.0055948 & 0.0055900 \\
128& 0.22467784 & 0.22467777 & 0.0014302 & 0.0014300 \\ \hline
$\infty$ &      & 0.22435696 & & \\
\hline
\end{tabular}
\end{center}
\caption{Values of $R_{exact}(L;z,\rho)$,
$R_{expan}(L;z,\rho)$, $D_{exact}(L;z,\rho)$, and
$D_{expan}(L;z,\rho)$ 
for the $RP^\infty$ model ($r=1$), $\rho = 1$ and $z=2$.
 }
\label{tableFSS3}
\end{table}

\begin{figure}
\vspace*{-1cm} \hspace*{-0cm}
\begin{center}
\epsfxsize = 0.9\textwidth
\leavevmode\epsffile{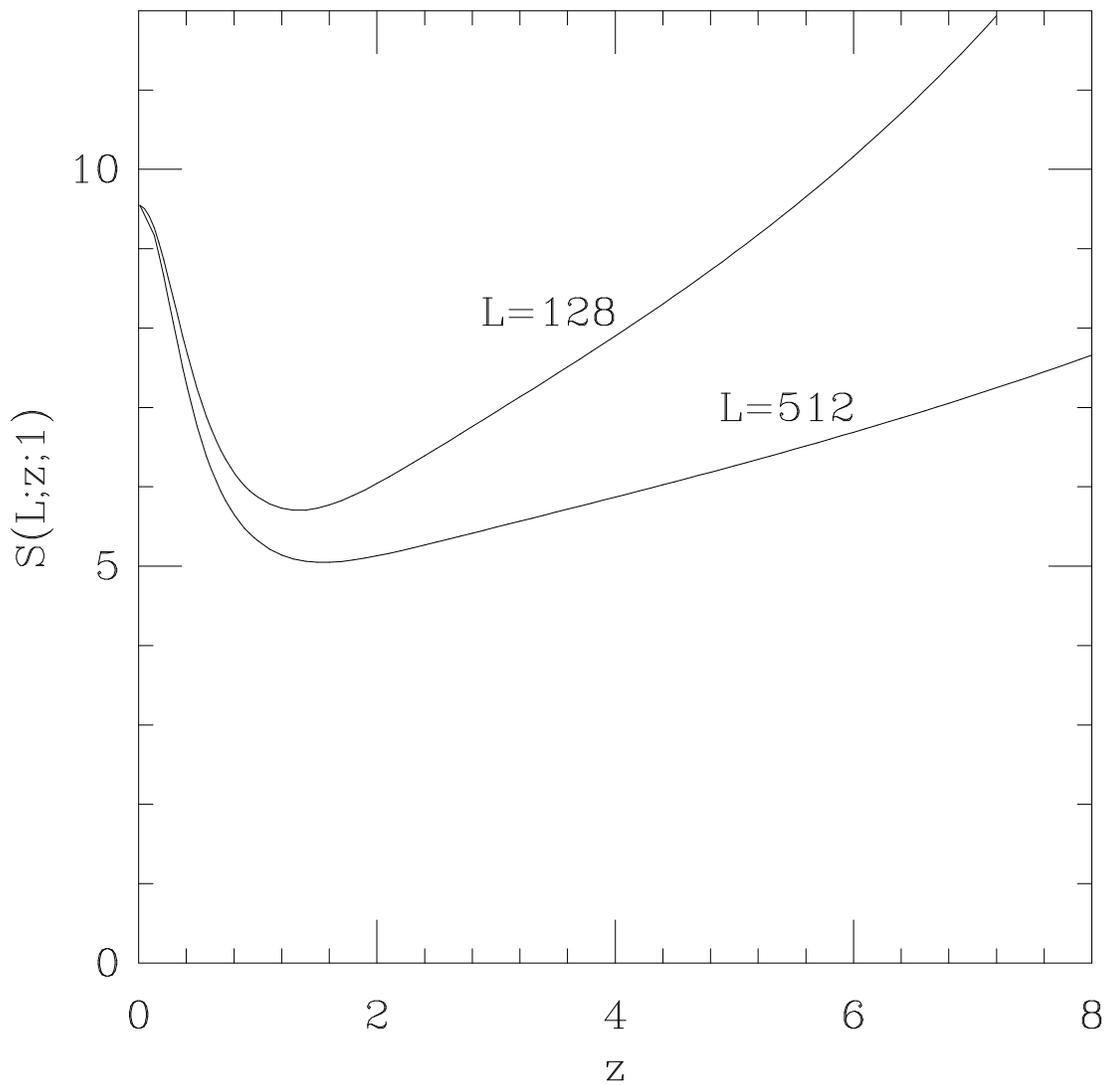}
\end{center}
\vspace*{-1cm}
\caption{$S(L;z;1)$ for $L=128$ and $L=512$. For $L\to\infty$, 
$S(L;z;1)$ converges to 3 for all $z\not=0$.
}
\label{fig9}
\end{figure}

The corrections are larger for the
mixed model than for the vector model. 
For instance $\Delta^r_{m,1}(z;\rho)/\Delta_{m,1}(z;\rho) =
2r$ so that the logarithmic correction in the $RP^\infty$ model ($r=1$)
is three times larger than the corresponding one in the $N$-vector
model ($r=0$). For the values of $L$ that are used in Monte Carlo simulations,
say $8\le L \le 128$, however all terms in Eq. \reff{ratiommixed} 
contribute to the FSS corrections. In table \ref{tableFSS3} 
we report for $r=1$, $\rho = 1$ and $z=2$, the same quantities reported in 
table \ref{tableFSS1}. Comparing the two tables we see that the 
corrections to FSS for $RP^\infty$ are seven times larger than those 
for the $N$-vector model in the same range of values of $L$. Only for 
$L\approx50$ (resp. $L\approx150$) 
the corrections are smaller than 1\% (resp.  0.1\%). 
To compare the corrections for the $RP^{\infty}$ and the $N$-vector model
with the standard hamiltonian for all values of $z$, in Fig.
\ref{fig9} we report
\be
S(L;z;\rho) = {D_{expan}(L;z,\rho)_{r=1}\over D_{expan}(L;z,\rho)_{r=0}}
\ee
for $L=128,512$ and $\rho = 1$ ($D(L;z,\rho)$ is defined in Eq. \reff{defD}). 
For these values of $L$, corrections are 5-10 times larger in 
the $RP^{\infty}$ model.

It is interesting to understand the origin of $\widehat{\Delta_{m}^r}
(L;z;\rho)$ in terms of perturbation theory. First of all, 
for $z$ small, (for simplicity we consider the strip case, analogous
results are valid for generic $\rho$) we can expand,
cf. Eq. \reff{F0smallzstrip},
\be
\widehat{\Delta_{m}^r} (L;z;\rho) = 
   {\pi^2 r^3\over2} {z^3\over \pi + z\log L} 
   \left[ 1 - {\pi z (2 \overline{F}_{00} - r) \over \pi + z\log L} + 
    O(z^2) \right]. 
\ee
This is not yet a perturbative expansion in $z$ due to the presence of
the term $z\log L$ in the denominators. This term cannot be expanded in
$z$ since we are considering an asymptotic expansion at fixed $z$ with
$L\to\infty$. However, if we ignore this problem and expand the denominators
in powers of $z$, we obtain 
\be
\widehat{\Delta_{m}^r} (L;z;\rho)_{pert} = 
{\pi r^3\over 2} z^3 \left( 
  1 - {z\over \pi} \log L - z (2 \overline{F}_{00} - r) + 
   O(z^2 \log^2 L) \right). 
\label{widehatm3pert}
\ee
In general $\widehat{\Delta_m^r} (L;z;\rho)$ has
a polynomial expansion in $z$ with coefficients that are
polynomials in $\log L$: 
\be
\widehat{\Delta_{m}^r} (L;z;\rho)_{pert} =
z^3 \sum_{n=0}^\infty P_n(\log L) z^n,
\label{widehatm3pert2}
\ee
where $P_n(x)$ is a polynomial of degree $n$. 
This expansion is clearly incorrect in the FSS limit $L\to \infty$ at fixed
$z$. However Eq. \reff{widehatm3pert2} correctly describes the theory
in the limit $z\to 0$ at fixed $L$. This is the limit 
in which PT works correctly \cite{Pat-Seil_comment,CEPS_O3_reply_to_Pat-Seil},
and indeed the expansion \reff{widehatm3pert2} can be directly obtained 
with a perturbative calculation.
Therefore our results show that in order to correctly compute the 
corrections to FSS one needs to resum the perturbative expansion. 
This reflects the fact that the perturbative limit $z\to 0$ with 
$L$ large and fixed does not commute with the FSS limit 
$L\to\infty$ with $z$ fixed and small. It should also be noticed that
the infinite series of logarithms appearing in Eq. \reff{widehatm3pert2}
resums 
to give corrections of order $1/(L^2\log L)$. 
This is a result far from obvious: in general one expects 
series of the form \reff{widehatm3pert2} to give powers of $L$,
i.e. to resum to $L^{p(z)}$, where $p(z)$ is some function of $z$.
For $N=\infty$ no power is generated, but we have no proof 
that this will be true for generic values of $N$. The only argument we
have against the appearance of power corrections is based on a naive 
application of renormalization-group ideas. The corrections to FSS are
due to the irrelevant operators of the theory. Since the 
$N$-vector model is asymptotically free, operators have canonical
scaling dimensions with logarithmic corrections. Therefore we always
expect a behaviour of the form $(\log L)^p/L^2$. 

\begin{figure}
\vspace*{-1cm} \hspace*{-0cm}
\begin{center}
\epsfxsize = 0.9\textwidth
\leavevmode\epsffile{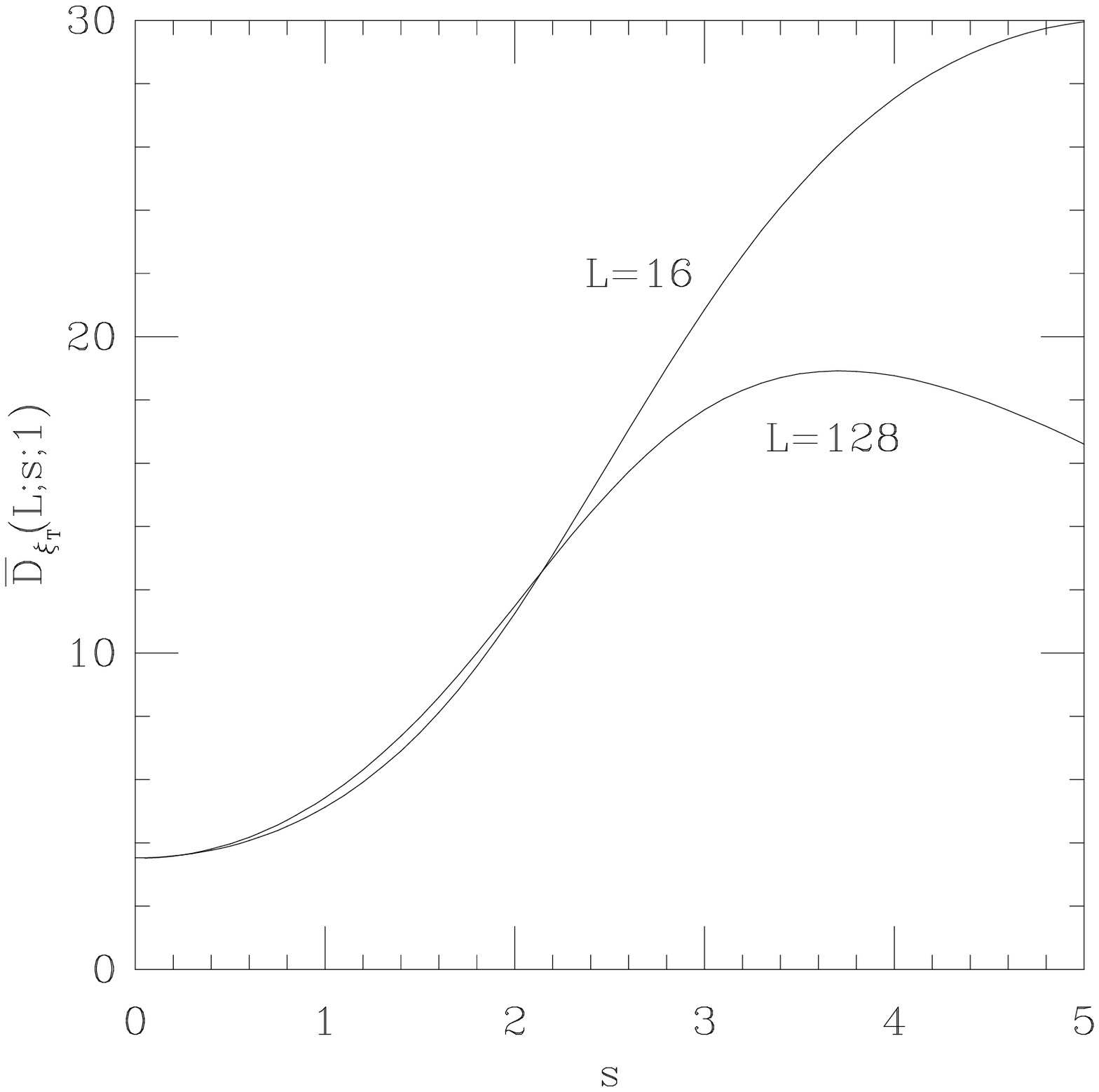}
\end{center}
\vspace*{-1cm}
\caption{$\overline{D}_{\xi_T}(L;s;1)$ for the $RP^{\infty}$ model 
for two different values of $L$: $L=16$ and $L=128$.
}
\label{fig10}
\end{figure}

\begin{figure}
\vspace*{-1cm} \hspace*{-0cm}
\begin{center}
\epsfxsize = 0.9\textwidth
\leavevmode\epsffile{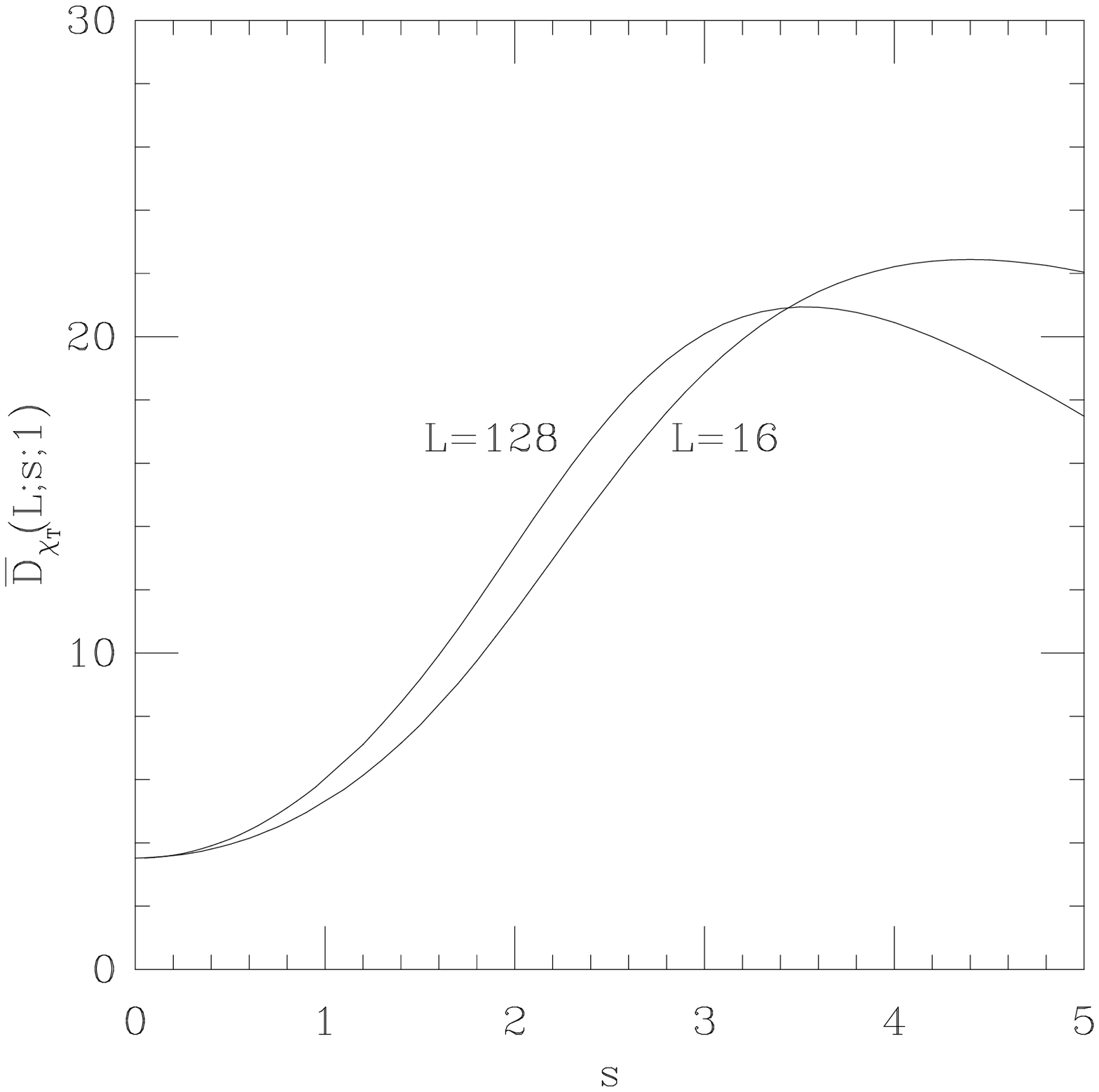}
\end{center}
\vspace*{-1cm}
\caption{$\overline{D}_{\chi_T}(L;s;1)$ for the $RP^{\infty}$ model 
for two different values of $L$: $L=16$ and $L=128$.
}
\label{fig11}
\end{figure}

Using the results of the previous section it is easy to obtain the FSS
functions and their leading corrections for the various observables.
For the isovector second-moment correlation lengths the expressions 
\reff{xim1FSS}, \reff{xim2FSS} and \reff{xim3FSS} still hold with
$m_\infty^2/m_{L,T}^2$ given by Eq.
\reff{ratiommixed}. For the susceptibility $\chi_V$ we have 
\be
{\chi_V(L,T)\over \chi_V(\infty)} \, =\,
{m_\infty^2\over m_{L,T}^2} \left[
  1 + {r\over 4z^2 L^2} \left(32 e^{-4\pi F_0(z;\rho)} -z^2\right) 
    {\log L + 2 \pi F_0(z;\rho)\over \log L + 2 \pi F_0(z;\rho) - \pi r}
\right]\; ,
\ee
while for $\chi_T$ we have
\begin{eqnarray}
{\chi_T(L,T)\over \chi_T(\infty)} &=& 
\left( {\chi_T(L,T)\over \chi_T(\infty)} \right)_{r=0} \left[
      1 + {\Delta_{m,1}^r (z;\rho)\over L^2}\log L +
          {\Delta_{m,2}^r (z;\rho)\over L^2} +
          {\widehat{\Delta_{m}^r} (L;z;\rho)\over L^2} \right. 
\nonumber \\
&& \left. 
  + {r\over 2z^2 L^2} \left(32 e^{-4\pi F_0(z;\rho)} -z^2\right) 
    {\log L + 2 \pi F_0(z;\rho)\over \log L + 2 \pi F_0(z;\rho) - \pi r}
\right]\; .
\end {eqnarray}
Finally for $\xi_T^{(m,1)}$ we have
\be
\left({\xi^{(m,1)}_T(L,T)\over \xi^{(m,1)}_T(\infty)} \right)^2 =\
\left( {\xi^{(m,1)}_T(L,T)\over \xi^{(m,1)}_T(\infty)} \right)_{r=0}^2 \left[
      1 + {\Delta_{m,1}^r (z;\rho)\over L^2}\log L +
          {\Delta_{m,2}^r (z;\rho)\over L^2} +
          {\widehat{\Delta_{m}^r} (L;z;\rho)\over L^2} \right].
\ee
To conclude our discussion we come back again to the $RP^\infty$
model. In this case the FSS functions are usually reported in terms 
of $s = L/\xi_T^{(m,1)}(L,T)$. Indeed, because of the gauge symmetry, 
one cannot define observables in the isovector sector. 
For any observable $\cal O$, we define the FSS deviations
\be
\overline{D}_{\cal O}(L;s;\rho) = L^2\left[
   {{\cal O}(L,T)\over {\cal O}(\infty)} 
   {1\over \overline{f}_{\cal O} (s;\rho)} - 1\right].
\ee
where $\overline{f}_{\cal O} (s;\rho)$ is the FSS function of ${\cal O}$
expressed in terms of $s$. In Figs. \ref{fig10} and \ref{fig11} we
report $\overline{D}_{\xi_T}(L;z;1)$ and 
$\overline{D}_{\chi_T} (L;z;1)$ for $L=16$ and $L=128$. The corrections
are extremely large if one compares them with the analogous results 
for the $N$-vector model. This is especially true in the large-$s$
region. Moreover the corrections are positive. These results 
are in qualitative agreement with the results 
of Ref. \cite{Hasenbusch}. 
   
\section{Conclusions}

In this paper we have investigated the corrections to FSS in the 
large-$N$ limit for a vast class of models and we have studied
their relation with the improvement program of Symanzik.

In the large-$N$ limit we find that the corrections behave as 
$f(z,L) \log L/L^2$ where $f(z,L)$ can be expanded in powers of 
$1/\log L$ and is such that $f(z,\infty)$ is finite for all values of 
$z$. Thus, for large values of $L$, corrections behave as 
$\log L/L^2$. Tree-level improved hamiltonians have corrections
to FSS behaving as $1/L^2$: the effect of the improvement 
is the cancellation of a logarithm. Subsequent perturbative 
improvement should give hamiltonians with corrections of order 
$1/(L^2 \log^l L)$ ($l=1$ for one-loop improvement and so on).

We have shown explicitly that the FSS limit and the perturbative limit
commute in the calculations of the FSS functions but {\em not}
for the calculation of the next-to-leading term.
Corrections to FSS cannot be computed in perturbation theory
unless an infinite series of logarithms is resummed.

Finally we have investigated if there  is any sign of large correction
in the $RP^\infty$ model. We find that this model shows deviations 
from FSS that are much larger than those of the $N$-vector model.
We find qualitative agreement with the results of Ref.
\cite{Hasenbusch}.

\section*{Acknowledgments}
We thank Ferenc Niedermayer, Paolo Rossi, Juan Ruiz-Lorenzo, 
Alan Sokal, and Ettore Vicari
for many useful comments.

\appendix

\section{Definitions} \label{secA}

\subsection{Functions 
 $G_k(\alpha)$, $H_k(\alpha)$ and ${\cal H}_k (\alpha)$} \label{secA.1}

Let us define the following functions:
\begin{eqnarray}
G_{k}(\alpha) &=&  \sum_{n=1}^{\infty} \left[
    (n^2 + \alpha^2)^{k-1/2} - \sum_{m=0}^{k} 
     {k-1/2 \choose m} \alpha^{2 m} n^{2 k - 2 m - 1}\right] \; ,
\label{Gkdef}
\\
H_k(\alpha) &=&  
     \sum_{n=1}^{\infty} {1\over (n^2+\alpha^2)^{k+1/2}} \; ,
\\
{\cal H}_k(\alpha) &=& \sum_{n=1}^{\infty} 
            {1\over (1-4n^2) (n^2+\alpha^2)^{k+1/2}}\; .
\label{Hkdef}
\end{eqnarray}
The first and the third ones are defined for integers $k\ge 0$ while 
the second one is defined for $k\ge 1$: for these values of $k$
the sums converge for all values of $\alpha$.
The functions $G_k(\alpha)$ and $H_k(\alpha)$ are known in 
statistical mechanics under the name of remnant functions
(see App. D of Ref. \cite{Barber-Fisher_73} and 
Ref. \cite{BF-Arch-Rat-Mech-Anal}).

We want now to discuss their 
behaviour for $\alpha\to 0$ and $\alpha\to\infty$. We will focus
on those values of $k$ that appear in our final results, i.e.
to $k=0,1$.

The expansion for $\alpha\to 0$ is trivial. We obtain
\begin{eqnarray}
G_0(\alpha) &=& \sum_{k=1}^\infty 
     (-1)^k {2k \choose k} \zeta(2 k +1 ) 
      \left( {\alpha\over2} \right)^{2 k} ,\\
G_1(\alpha) &=& 2 \sum_{k=1}^\infty {(-1)^k\over k}
     {2k \choose k} \zeta(2 k+ 1) 
     \left( {\alpha\over2}\right)^{2 k+ 2} ,\\
H_1(\alpha) &=& \sum_{k=0}^\infty (-1)^k (2 k+1) 
     {2k \choose k} \zeta(2k+3) 
     \left( {\alpha\over2}\right)^{2 k}  ,\\
{\cal H}_0 (\alpha) &=& \sum_{k=0}^\infty (-1)^k {2k \choose k}
    \alpha^{2k} \left[1 - 2 \log 2 + \sum_{n=1}^k {\zeta(2n+1)\over 4^n}\right]
   \; , \\
{\cal H}_1 (\alpha) &=& 4 \sum_{k=0}^\infty (-1)^k (2k+1) {2k \choose k}
    \alpha^{2k} \left[1 - 2 \log 2 + \sum_{n=1}^{k+1}
      {\zeta(2n+1)\over 4^n}\right]\; ,
\end{eqnarray}
where we have used  
\be
\sum_{n=1}^\infty {1\over (1-4n^2) n^{2q+1}} \, =\, 4^q (1 - 2 \log 2) +\, 
     4^q \sum_{s=1}^q {\zeta(2s+1)\over 4^s}\; .
\ee
All series converge for $|\alpha|< 1$.

We want now to derive the asymptotic expansions for $\alpha\to \infty$.
In order to obtain them let us derive a different representation for the 
functions $G_k(\alpha)$ and $H_k(\alpha)$.

Let us first consider $G_0(\alpha)$. We rewrite it  as 
\be
G_0(\alpha) = \lim_{\epsilon\to0+} 
   \left[ - \zeta(1 + \epsilon) + 
       {1\over \Gamma (1/2 + \epsilon/2) } \int^\infty_0 
       {dx \over x^{(1-\epsilon)/2}} e^{-x \alpha^2} 
       \sum_{n=1}^\infty e^{-x n^2}\right]\; .
\ee
Using the Poisson resummation formula \cite{Morse} we obtain 
\begin{eqnarray}
G_0(\alpha) &=& \lim_{\epsilon\to 0+} \left\{
   - \zeta(1 + \epsilon) + {\alpha^{-\epsilon}\over\Gamma(1/2+\epsilon/2)}
      \left( {\sqrt{\pi}\over2} \Gamma\left({\epsilon\over2}\right) 
           - {1\over 2 \alpha} \Gamma\left({1+\epsilon\over2}\right)
      \right)\right. \nonumber \\
     && \qquad \left. + {\sqrt{\pi}\over\Gamma(1/2 + \epsilon/2)}
        \int_0^\infty {dx\over x^{1-\epsilon/2}} e^{- x\alpha^2} 
        \sum_{n=1}^\infty \exp\left( - {\pi^2 n^2\over x}\right)
        \right\}\; .
\end{eqnarray}
Taking the limit we get the representation
\cite{BF-Arch-Rat-Mech-Anal}
\be
G_0(\alpha) = - \log {\alpha\over2} - \gamma_E - {1\over 2\alpha} 
    + 2 \sum_{n=1}^\infty K_0 (2 \pi n \alpha) ,
\label{G0asintotico}
\ee
where $K_0$ is a modified Bessel function \cite{GR}. 
The corresponding representations for $G_k(\alpha)$ and $H_k(\alpha)$,
$k\ge 1$ can be obtained by integration and derivation of the 
previous relation. We obtain
\cite{BF-Arch-Rat-Mech-Anal}
\begin{eqnarray}
G_1(\alpha) &=& {1\over 12} + {\alpha^2\over2} 
    \left( - \log {\alpha\over2} + { 1\over2} - \gamma_E - {1\over\alpha}
    \right) - {\alpha\over\pi} \sum_{n=1}^\infty 
                    {1\over n} K_1(2 \pi n \alpha),     \\
H_1(\alpha) &=& {1\over \alpha^2} - {1\over 2 \alpha^3} + 
       {4 \pi\over \alpha} \sum_{n=1}^\infty n K_1(2 \pi n \alpha).
\end{eqnarray}
This representation of the functions $G_k(\alpha)$ and 
$H_k(\alpha)$ allows an immediate derivation of the asymptotic
expansion for large values of $\alpha$ since \cite{GR}
\be
K_n (x) = \sqrt{{\pi\over 2x}} e^{-x} \left[
    1 + {1\over 2x} \left(n^2 - {1\over4}\right) + O(x^{-2})\right] 
   \;\; .
\ee
Let us now consider the functions ${\cal H}_k(\alpha)$. We rewrite 
${\cal H}_k(\alpha)$ as 
\be
{\cal H}_k(\alpha) \, =\, - {1\over2} {1\over \alpha^{2k+1}} + 
   {1\over2} {1\over \Gamma(k+1/2)} 
   \int^\infty_0 dx\, x^{k-1/2} e^{-x\alpha^2} g(x),
\label{calHderivation1}
\ee
where 
\be
g(x) = \, \sum_{n=-\infty}^{+\infty} {e^{-x n^2}\over {1-4 n^2}}\; .
\ee
Now in the integral appearing in Eq. \reff{calHderivation1} the relevant 
region for large $\alpha^2$ corresponds to small values of $x$. 
Therefore we need the small-$x$ expansion of $g(x)$. First of all notice that
\be
g(0) = 1 + 2 \sum_{n=1}^{+\infty} {1\over 1-4n^2} = 0.
\label{gforx0}
\ee
Then one immediately verifies that $g(x)$ satisfies an equation of the form
\be
g'(x) + {1\over4} g(x) - 
{1\over 4} \sum_{n=-\infty}^{+\infty} e^{-x n^2} = 0,
\ee
so that, using Eq. \reff{gforx0}, we get
\be
g(x) = {1\over4} e^{-x/4} \int_0^x dy\ e^{y/4} 
     \sum_{n=-\infty}^{+\infty} e^{-y n^2}.
\ee
Using the Poisson resummation formula \cite{Morse} we can rewrite it as 
\be
g(x) = {\sqrt{\pi}\over4} e^{-x/4} \int_0^x {dy\over \sqrt{y}}\ e^{y/4}
     \sum_{n=-\infty}^{+\infty} e^{-\pi^2 n^2/y} .
\ee
In the small-$x$ region only the term with $n=0$ is relevant so that
\begin{eqnarray}
g(x) &\approx& {\sqrt{\pi}\over4} e^{-x/4} \int_0^x {dy\over \sqrt{y}}\ e^{y/4}
\nonumber
\\
&\approx& {\sqrt{\pi x}\over2} 
   \left(1 - {x\over6} + {x^2\over 60} - {x^3\over 840} + 
     {x^4\over 15120} + O(x^5)\right)\; .
\end{eqnarray}
We obtain eventually for large values of $\alpha$
\begin{eqnarray}
{\cal H}_k(\alpha) &=& - {1\over 2} {1\over \alpha^{2k+1}} + \, 
     {\sqrt{\pi}\over 4\ \Gamma(k+1/2)} {1\over \alpha^{2k+2}} \times
\nonumber \\
&&\left[ k! - {(k+1)!\over 6\alpha^2} + 
    {(k+2)!\over 60\alpha^4} -
    {(k+3)!\over 840\alpha^6} +
    {(k+4)!\over 15120\alpha^8} \right] +O(\alpha^{-2k-12}).
 \nonumber \\ [-2mm]
{}
\end{eqnarray}

\subsection{The functions $M_{pq}(z;\rho)$, 
   $N_{pq}(\rho)$, ${\cal M}_{pq}(z;\rho)$ 
   and ${\cal N}_{pq}(\rho)$} \label{secA.2}

A second set of functions appear in our calculations. We define
\begin{eqnarray}
\hskip -60pt
M_{pq}(z;\rho) & =& 
    (2 \pi)^p \sum_{n=-\infty}^\infty
     {1\over (4 \pi^2 n^2 + z^2)^{p/2}}
     {1\over [ \exp(\rho \sqrt{4 \pi^2 n^2 +z^2}) - 1]^q},   \\
\hskip -60pt
{\cal M}_{pq}(z;\rho) & =& 
     (2 \pi)^p \sum_{n=-\infty}^\infty
     {1\over (1-4n^2) (4 \pi^2 n^2 + z^2)^{p/2}}
     {1\over [ \exp(\rho \sqrt{4 \pi^2 n^2 +z^2}) - 1]^q},   \\
\hskip -60pt
N_{pq}(\rho) & =& 
       \sum_{n=1}^\infty {1\over n^p}
     {1\over [ \exp(2 \pi \rho n) - 1]^q}, \\
\hskip -60pt
{\cal N}_{pq}(\rho) & =& 
      \sum_{n=1}^\infty {1\over (1-4n^2) n^p}
     {1\over [ \exp(2 \pi \rho n) - 1]^q},
\end{eqnarray}
where $q$ is a positive integer.
We want to compute here the asymptotic expansion of $M_{pq}(z;\rho)$ and 
${\cal M}_{pq}(z;\rho)$
for $z\to 0$ and $z\to\infty$ at fixed $\rho$. The first expansion is
straightforward and we get for $M_{pq}(z;\rho)$
\begin{eqnarray}
M_{pq}(z;\rho) &=& \left( {2\pi\over z}\right)^p {1\over (\rho z)^q}
     \sum_{n_1,\ldots,n_q = 0}^\infty
     {B_{n_1} \ldots B_{n_q}\over n_1!\ldots n_q!}
     (\rho z)^{n_1 + \ldots + n_q} + 2 N_{pq}(\rho) \nonumber \\
     && + 2 z^2 \left[ - {p\over 8\pi^2} N_{p+2,q}(\rho) -
         {q \rho\over 4 \pi}
         \left( N_{p+1,q}(\rho) + N_{p+1,q+1}(\rho)\right)    \right]
   + O(z^4). \nonumber \\ [-2mm]
{}
\end{eqnarray}
For ${\cal M}_{pq}(z;\rho)$ the expansion is analogous 
with the substitution of 
$N_{pq}$ with ${\cal N}_{pq}$.

For large $z$ let us consider first $M_{pq}(z;\rho)$. We obtain 
\be
M_{pq}(z;\rho) = \, \sum_{n=-\infty}^\infty
       {(2\pi)^p \over (4 \pi^2 n^2 + z^2)^{p/2}}
       \exp\left[ - q \rho \sqrt{4 \pi^2 n^2 + z^2}\right] +
        O(e^{-2 \rho q z} z^{1/2-p}).
\ee
This last sum can be evaluated using the 
Poisson resummation formula \cite{Morse}. Define
\be
\hat{f}(\omega) =
   \int^\infty_{-\infty} dt\,  e^{i\omega t}
       {(2\pi)^p \over (4 \pi^2 t^2 + z^2)^{p/2}}
       \exp\left[ - q \rho \sqrt{4 \pi^2 t^2 + z^2}\right]\; .
\ee
Then
\be
M_{pq} (z;\rho) \approx \sum_{n=-\infty}^\infty \hat{f}(2 \pi n).
\ee
For $z\to\infty$ we have
\be
\hat{f}(\omega) = {2\pi\over \sqrt{z}} (q \rho z)^{1-p}
     (\omega^2 + 4 \pi^2 q^2 \rho^2)^{(2p-3)/4}
     \exp\left[- {z\over 2\pi} \sqrt{\omega^2 + 4 \pi^2 q^2 \rho^2}\right]
      \left( 1 + O(z^{-1})\right).
\ee
Therefore
\be
M_{pq}(z;\rho) = {1\over \sqrt{q\rho}}
            \left( {z\over 2\pi}\right)^{1/2 -p} e^{-q \rho z}
            \left( 1 + O(z^{-1})\right).
\ee
For ${\cal M}_{pq}(z;\rho)$ we will not need the explicit asymptotic 
behaviour. It is however easy to convince oneself that, for large $z$, 
${\cal M}_{pq}(z;\rho)$ goes to zero faster than $z^{-p} e^{-q\rho z}$.

To conclude this subsection let us derive a set of relations
for the functions $N_{pq}(\rho)$.
First of all let us notice that $N_{1,1}(\rho)$ can be related to
Dedekind's $\eta$-function \cite{Chandra}
\be
\eta(\tau) = \, e^{\pi i \tau/12} \prod_{n=1}^\infty
   \left( 1 - e^{2 \pi i n \tau}\right).
\ee
Indeed
\begin{eqnarray}
N_{1,1}(\rho) &=& \sum_{n=1}^\infty \sum_{k=1}^\infty
          {1\over n} e^{-2 \pi n k \rho} =\,
           -\log \left[ \prod_{k=1}^\infty
                  \left(1 - e^{-2 \pi k \rho}\right)\right]
    \nonumber \\
            & = & - {\pi \rho\over 12} - \log \eta (i\rho).
\end{eqnarray}
Following the same steps it is possible to prove that, for $p>0$,
\be
N_{-p,1} (\rho) \, =\, {(-1)^p\over (2\pi)^p} 
    {d^p\over  d\rho^p} N_{p,1}(\rho).
\label{Nmenop1}
\ee
This relation, together with
\be
 {d\over d\rho} N_{pq}(\rho)\, =\, 
   -2 \pi q \left[ N_{p-1,q}(\rho) + N_{p-1,q+1} (\rho)\right]\; ,
\label{derivataNpq}
\ee
allows to prove that all $N_{pq}(\rho)$ with $p\le -1$ can be expressed
in terms of $N_{0,q'}(\rho)$.
From Eqs. \reff{Nmenop1} and \reff{derivataNpq} we obtain the following 
relations we will use in the following:
\begin{eqnarray}
\hskip -50pt
&& N_{-1,1}(\rho)\, =\, N_{0,1}(\rho) + N_{0,2}(\rho)  =\, 
    {1\over 24} + {1\over 2\pi} {d\over d\rho} \log \eta(i\rho) ,
\label{Nlogeta1} \\
\hskip -50pt
&& N_{-2,1}(\rho) + N_{-2,2}(\rho) \, =\, 
    N_{-1,1}(\rho) + 3 N_{-1,2}(\rho) + 2 N_{-1,3}(\rho)\, =\, 
     - {1\over 4\pi^2} {d^2\over d\rho^2} \log \eta(i\rho).
 \nonumber \\ [-2mm]
{}
\label{Nlogeta2}
\end{eqnarray}
Finally, for $p\ge 0$,
let us derive a relation between $N_{4 p + 3,1}(\rho)$
and $N_{4 p + 3,1}(1/\rho)$ that will allow us to compute explicitly
$N_{4 p + 3,1}(1)$.
Let us start from
\be
\sum_{n=-\infty}^\infty {1\over n^2 + \alpha^2}=\,
         {\pi\over \alpha} +
         {2 \pi\over\alpha} {1\over e^{2\pi\alpha}-1}.
\ee
It follows
\begin{eqnarray}
&& \sum_{m,n=1}^\infty \left( {1\over n^{4 p + 2}} +
      {1\over (\rho m)^{4 p + 2}} \right) \,
      {1\over n^2 + \rho^2 m^2} = \nonumber \\
&& \qquad - {1\over2} \zeta(4 p + 4) + {\pi\over 2\rho} \zeta(4 p + 3)
    + \, {\pi\over \rho} N_{4 p + 3,1} (1/\rho) \nonumber \\
&& \qquad + {1\over \rho^{4p + 2}} \left[
    - {1\over 2\rho^2} \zeta(4p+4) + {\pi\over2\rho} \zeta(4p+3) +
      {\pi\over\rho} N_{4p+3,1}(\rho)\right]\; .
\label{identityA.5.1}
\end{eqnarray}
However, for integer $p\ge 0$, we can also compute the first sum as
\begin{eqnarray}
&& \sum_{m,n=1}^\infty {1\over (nm\rho)^{4 p+ 2}}
   {n^{4 p + 2} + (\rho m)^{4 p + 2}\over n^2 + (\rho m)^{2}}
   \nonumber \\
&& \qquad = \, \sum_{k=0}^{2p} (-1)^k
    \sum_{m,n=1}^\infty {1\over (mn\rho)^{4p+2}} (\rho m)^{2k}
        n^{4p - 2k}
   \nonumber \\
&& \qquad = \, \sum_{k=0}^{2p} (-1)^k {1\over \rho^{4p-2k+2}}
            \zeta(4p-2k+2)\zeta(2k+2).
\label{identityA.5.2}
\end{eqnarray}
Comparing Eqs. \reff{identityA.5.1} and \reff{identityA.5.2} we obtain
a relation between $N_{4p+3,1}(\rho)$ and $N_{4p+3,1}(1/\rho)$. For
$\rho =1$ we obtain
\be
N_{4p+3,1}(1) =\, {1\over2\pi} \zeta(4p+4) -
  {1\over2} \zeta(4p+3) + {1\over2\pi}
  \sum_{k=0}^{2p} (-1)^k \zeta(4p-2k+2) \zeta(2k+2).
\ee 
Particular cases are 
\begin{eqnarray}
N_{3,1} (1) &=& {7\pi^3\over 360} - {1\over2} \zeta(3) \, \approx\, 
      0.001871373\; , \\
N_{7,1} (1) &=& {19\pi^7\over 113400} - {1\over2} \zeta(7) \, \approx\, 
      0.001870964\; . 
\end{eqnarray}
Analogous relations can be gotten starting from the more general
sum
\be
\sum_{m,n=1}^\infty \left( {1\over n^{4p+2}} +
         {1\over (\rho m)^{4p+2}}\right) {1\over (n^2 + \rho^2 m^2)^q}.
\ee
We leave the derivation to the reader. We will not need these relations here.

\section{Asymptotic Expansions of Lattice Sums}   \label{secB}

In this appendix we will study generic sums involving the lattice 
propagator for a Gaussian theory with arbitrary interactions. 
We will present an algorithmic procedure that allows to derive
systematically the expansion in powers of $1/L$ of sums of this type.
The results will be expressed in terms of the functions we have
introduced in App. \ref{secA}. The method presented in this Appendix
applies to a square lattice $L\times T$ with periodic boundary 
conditions but generalizes easily to other types of boundary conditions.
It can also be used to study lattice sums in more than two dimensions.
In Sect. \ref{secB.1} we compute some preliminary one-dimensional
sums; the general procedure is presented in Sect. \ref{secB.2}.

\subsection{One-dimensional sums}  \label{secB.1}

\subsubsection{The Euler-Mac Laurin formula}  \label{secB.1.1}

The basic tool we will use is the Euler-Mac Laurin formula
\cite{Abramowitz}. In its 
general form it is given by
\begin{eqnarray}
\sum_{k=n}^{m-1} f(k)
         &=& \int^m_n dx f(x) - {1\over2} (f(m) - f(n)) \nonumber \\
          && + \sum_{k=1}^{N} {B_{2k}\over (2k)!}\,
              \left( f^{(2k-1)}(m) - f^{(2k-1)}(n)\right) \nonumber \\
          && + {1\over (2 N+1)!} \int_n^m dx\, f^{(2N+1)} (x) 
               B_{2 N +1} (x - \lfloor x\rfloor),
\label{EulerMacLaurin1}
\end{eqnarray}
where $B_k$ are the Bernoulli numbers and $B_k(x)$ are the 
Bernoulli polynomials defined by \cite{GR,Abramowitz}
\be
B_n(x) = \sum_{k=0}^n  {n \choose k}  B_k x^{n-k}\;\; .
\ee
In the following we will be interested in sums of the form 
\be
{1\over L} \sum_{n=0}^{\alpha L - 1} f(p),
\label{sumA1}
\ee
where $p = 2 \pi n/L$. We will try to compute the  asymptotic expansion
of the sum \reff{sumA1}
for $L\to \infty$ with $\alpha$ fixed. It is easy to obtain from
Eq. \reff{EulerMacLaurin1} the following formula:
\begin{eqnarray}
{1\over L} \sum_{n=0}^{\alpha L -1} f(p) &=& 
    \int_0^{2 \pi \alpha} {dp\over 2\pi} f(p) - {1\over 2L} 
           (f(2 \pi \alpha) - f(0)) 
      \nonumber \\
    && + {1\over 2\pi} \sum_{k=1}^N {B_{2 k}\over (2 k)!} 
         \left( {2\pi\over L}\right)^{2 k} 
         \left( f^{(2 k-1)}(2 \pi \alpha) - f^{(2 k -1)}(0)\right) 
      \nonumber \\
    && + {1\over (2 N+1)!} \left( {2\pi\over L}\right)^{2 N + 1} \,
       \int^{2\pi \alpha}_0 {dp\over 2\pi} 
       f^{(2 N + 1)}(p) \hat{B}_{2 N + 1} (p),
\label{EulerMacLaurin}
\end{eqnarray}
with
\be
      \hat{B}_n(p) = B_n \left( {Lp\over 2 \pi} - 
        \left\lfloor {Lp\over 2\pi}\right\rfloor\right)\;\;\; .
\ee
Thus, as long as the last integral in Eq. \reff{EulerMacLaurin} is finite,
i.e. $f^{(2 N + 1)} (p)$ is integrable in the interval $[0,2\pi \alpha]$, 
the previous formula gives the asymptotic expansion of the sum 
\reff{sumA1} in powers of $1/L$ up to order $1/L^{2 N}$. 
An important case corresponds to $\alpha=1$ and $f(p)$ periodic of period
$2\pi$, i.e. $f(p+2\pi) = f(p)$. In this case all the 
$1/L$ corrections vanish and we obtain
\be
{1\over L} \sum_{n=0}^{L -1} f(p) = 
    \int_0^{2 \pi} {dp\over 2\pi} f(p) +\, O(L^{-2 N -1}).
\ee
It is easy to prove that a similar result holds for generic 
$d$-dimensional sums. If $f(p_1,\ldots,p_d)$ is a periodic function in all 
variables of period $2\pi$ and 
${\partial_1^{n_1}\ldots \partial_d^{n_d} f(p)}$ is integrable in 
$[0,2\pi]^d$ for all $n_1,\ldots,n_d$ such that 
$n_1 + \ldots + n_d \le 2 N + 1$, then, for $L\to\infty$ with
$L_1/L$, $L_2/L \ldots , L_d/L$ fixed, we have
\be
{1\over L_1\ldots L_d} \sum_{n_1=0}^{L_1-1} 
   \sum_{n_2=0}^{L_2-1} \ldots \sum_{n_d=0}^{L_d-1} 
   f(p_1,\ldots, p_d) =\,
    \int_{[0,2\pi]^d} {d^d p\over (2 \pi)^d} f(p) + \,
    O( L^{-2 N - 1}),
\label{EulerMacLaurindim}
\ee
where in the l.h.s.  $p_i = 2 \pi n_i/L_i$.

\subsubsection{Asymptotic expansions of $\sum n^{-p}$} \label{secB.1.2}

Here we will discuss the asymptotic expansion of sums of the form
\be
    \sum_{n=1}^{L} {1\over n^p},  \label{sumunosunp}
\ee
for $L\to\infty$. When $p$ is a negative integer it is easy to perform
the summation exactly. Indeed ($q=-p$) 
\be
\sum_{n=1}^L n^q \, =\, {d^q\over d\alpha^q}
     \left. \left[ {1 - e^{\alpha(L+1)} \over 1 - e^\alpha}\right]
     \right|_{\alpha=0}\; .
\ee
The simplest cases are 
\begin{eqnarray}
\sum_{n=1}^L n &=& {1\over 2} L ( L + 1), \\
\sum_{n=1}^L n^2 &=& {1\over6} L(L+1)(2 L + 1), \\
\sum_{n=1}^L n^3 &=& {1\over4} L^2 (L+1)^2. 
\end{eqnarray}
Let us now consider the sum \reff{sumunosunp} with $p > 1$. Rewriting 
it as
\be
\sum_{n=1}^L {1\over n^p} \, =\, \zeta(p) -\, 
   \sum_{n=L+1}^\infty {1\over n^p},
\ee
where $\zeta(n)$ is Riemann zeta function, 
and using the Euler-Mac Laurin formula for the second sum,
we get the asymptotic expansion
\be
\sum_{n=1}^L {1\over n^p} \, =\, \zeta(p) -\,
    {1\over \Gamma(p)} \sum_{n=0}^\infty {B_n\over n!} 
    \Gamma (n+p-1) L^{1-n-p}.
\ee
Finally, taking in the previous formula the limit $p\to 1$, we have
the asymptotic expansion
\be
\sum_{n=1}^L {1\over n} = \, 
   \log L +\, \gamma_E + {1\over 2L} - 
    \sum_{n=1}^\infty {B_{2n}\over 2n} {1\over L^{2n}},
\ee
where $\gamma_E \approx 0.577215665$ is the Euler constant.

\subsubsection{Asymptotic expansions of 
   $\sum (n^2 + \alpha^2)^{k+1/2}$   }   \label{secB.1.3}

In this section we consider sums of the form
\be
   \sum_{n=1}^{L-1} (n^2 + \alpha^2)^{k+1/2}\; .
\ee
Again we want to compute the asymptotic expansion for $L\to\infty$ with
$\alpha$ fixed. Let us first consider the case $k\ge - 1$. In this case
we rewrite the sum as 
\begin{eqnarray}
&& \sum_{n=1}^{L-1} \left[
    (n^2 + \alpha^2)^{k+1/2} - \sum_{m=0}^{k+1} 
     {k+1/2 \choose m} \alpha^{2 m} n^{2 k - 2 m + 1}\right] 
 \nonumber \\
&& \quad + \sum_{m=0}^{k+1} \left[ {k+1/2 \choose m} \alpha^{2m} 
    \sum_{n=1}^{L-1} n^{2 k - 2 m + 1}\right]\; .
\label{sumA2}
\end{eqnarray}
We have already explained how to compute the last sums in the previous
subsection. We will now discuss the first sum that we rewrite as 
\be
G_{k+1}(\alpha)  - 
  \sum_{n=L}^{\infty} \left[
    (n^2 + \alpha^2)^{k+1/2} - \sum_{m=0}^{k+1} 
     {k+1/2 \choose m} \alpha^{2 m} n^{2 k - 2 m + 1}\right] \; ,
\label{eqB.18}
\ee
where $G_k(\alpha)$ is defined in Eq. \reff{Gkdef}. 
The last term appearing in Eq. \reff{eqB.18}
can be easily computed using the 
Euler-Mac Laurin formula \reff{EulerMacLaurin1}.

In the following we will need the previous sums for $k=-1,0$. Explicitly
we have
\begin{eqnarray}
\sum_{n=1}^{L-1} \sqrt{n^2 + \alpha^2} &=& 
    {1\over2} L(L-1) + G_1(\alpha) + 
    {\alpha^2\over2} \left( \log L + \gamma_E\right) - 
    {\alpha^2\over 4 L} +\, O(L^{-2}),    \\
\sum_{n=1}^{L-1} {1\over \sqrt{n^2 + \alpha^2}} &=&
     \log L + \gamma_E + G_0(\alpha) - {1\over 2L} - {1\over 12 L^2}
     + {\alpha^2\over4} \left({1\over L^2} + {1\over L^3}\right) 
     + \, O(L^{-4}).   \nonumber \\ [-4mm]
&& {}
\end{eqnarray}
For $k<-1$ the computation is straightforward as no subtraction is 
needed in this case. For $k=1$ we have
\be
\sum_{n=1}^{L-1} {1\over (n^2 + \alpha^2)^{3/2}}\,
    = \, H_1(\alpha) - {1\over 2 L^2} - {1\over 2 L^3} + 
    O(L^{-4}),
\ee
where $H_1(\alpha)$ is defined in Eq. \reff{Hkdef}.

\subsubsection{Computation of 
     $\sum (\hat{p}^2 + \alpha^2)^{-q}$} \label{secB.1.4}

In this section we will compute exactly sums of the form 
\be
\sum_{n=0}^{L-1} {1\over (\hat{p}^2 + \alpha^2)^q},
\ee
for integer values of $q$. As usual, $\hat{p} = 2 \sin(p/2)$. 

If $q$ is negative the summation is trivial as (here $k\ge 1$)
\be
   \sum_{n=0}^{L-1} \hat{p}^{2k} =\, {2k \choose k} L\; .
\ee
Let us now discuss the case $q\ge 1$. Consider first $q=1$. Then
\be
\sum_{n=0}^{L-1} {1\over \hat{p}^2 + \alpha^2} =\,
   {1\over 2} \sum_{n=0}^{L-1} {1\over \omega - 
         \cos(2 \pi n /L)},
\ee
where $\omega = 1 + \alpha^2/2$. Then notice that 
\be
\lim_{R\to\infty} \int_{D_R} dz {\cot \pi z\over \omega - \cos (2 \pi z/L)}
  \, =\, 0,
\ee
where $D_R$ is the rectangle in the complex $z$-plane bounded by the lines
$z=-1/2$, $z= L-1/2$, $z=\pm i R$. Using the residue theorem we get
\be
\sum_{n=0}^{L-1} {1\over \omega - \cos (2 \pi n /L)} \, =\, 
     {L\over \sqrt{\omega^2 - 1}}\, 
     \coth \left( {L\over2} \hbox{\rm arcch}\,\omega\right).
\ee
We thus end up with 
\be
\sum_{n=0}^{L-1} {1\over \hat{p}^2 +\alpha^2}\, =\, 
   {L\over \alpha \sqrt{4 + \alpha^2}} \,
   \coth \left[ L\,\hbox{\rm arcsh}\, 
   \left( {\alpha\over2} \right)\right]\; .
\label{A30}
\ee
Higher values of $q$ can be handled by taking derivatives with respect to
$\alpha^2$ of the previous formula.

\subsubsection{Asymptotic expansion of 
    $\sum (\hat{p}^2 + m^2)^{k+1/2}$} \label{secB.1.5}

Let us now consider sums of the form
\be
\sum_{n=0}^{L-1} (\hat{p}^2 + m^2)^{k+1/2},
\ee
where $\hat{p} = 2 \sin (p/2)$, $p=2 \pi n/L$. We want to study these
sums in the finite-size-scaling limit, i.e. for $L\to \infty$,
$m\to 0$, with $m L\equiv z$ fixed. To compute these 
asymptotic expansions we proceed in the following way. 

Assuming $L$ even (the final result will not depend on
this assumption) we rewrite
\be
\sum_{n=0}^{L-1} \left( \hat{p}^2 + m^2 \right)^{k+1/2}\, =\, 
    2 \sum_{n=1}^{L/2-1} 
      \left( \hat{p}^2 + m^2 \right)^{k+1/2} +\, 
      (4 + m^2)^{k+1/2} + m^{2k+1}.
\label{A3.1}
\ee
Then let us consider the expansion of $(\hat{p}^2 + z^2/L^2)^{k+1/2}$
in powers of $1/L^2$: it can be written in the form 
\be
\left(\hat{p}^2 + {z^2\over L^2}\right)^{k+1/2} \, =\,
    \left( p^2 + {z^2\over L^2}\right)^{k+1/2} \,
    \sum_{h=0}^\infty {a_h(n^2,z^2)\over L^{4h} }
     {1\over (p^2 + z^2/L^2)^h},
\label{A3.2}
\ee
where $a_h(n^2,z^2)$ is a polynomial in $n^2$ and $z^2$. Let us indicate 
with $R_{k,q}(p,z;L)$ the sum of the first $q$ terms in \reff{A3.2}. Then we
rewrite
\begin{eqnarray}
\sum_{n=0}^{L-1} \left( \hat{p}^2 + m^2 \right)^{k+1/2} 
       &=& 2 \sum_{n=1}^{L/2-1} 
    \left[ \left( \hat{p}^2 + {z^2\over L^2}\right)^{k+1/2} -
        R_{k,q}(p,z;L)\right] 
    \nonumber \\
   && + 2  \sum_{n=1}^{L/2-1} R_{k,q} (p,z;L) +\, (4 + m^2)^{k+1/2}
      + m^{2k+1}.
\end{eqnarray}
We must then choose $q$. To fix its value we must decide the order
in $1/L$ to which we want to compute the expansion. Then we fix 
$q$ so that we can use the Euler-Mac Laurin formula for the first 
sum. It is trivial to reduce the second sum to a sum of terms
of the form studied in the previous paragraph.

We will now illustrate the method by computing the asymptotic
expansion of
\be
\sum_{n=0}^{L-1} {1\over \sqrt{\hat{p}^2 + m^2}},
\ee
including terms of order $1/L^2$. Since
\be
{1\over \sqrt{\hat{p}^2 + z^2/L^2}}\, =\, 
{1\over \sqrt{p^2 + z^2/L^2}} \,
  \left( 1 + {1\over 24} {p^4\over p^2 + z^2/L^2} +\, O(L^{-4})\right),
\ee
we rewrite
\begin{eqnarray}
\sum_{n=0}^{L-1} {1\over \sqrt{\hat{p}^2 + m^2}} &=& 
   2 \sum_{n=0}^{L/2-1} \left[ 
   {1\over \sqrt{\hat{p}^2 + m^2}} - 
   {1\over \sqrt{p^2 + m^2}} - {1\over24} {p^4\over (p^2 + m^2)^{3/2}}
   \right] +\, {1\over m} + 
   \nonumber \\
   && 2 \sum_{n=1}^{L/2-1} {1\over \sqrt{p^2 + m^2}} + 
     {1\over12} \sum_{n=1}^{L/2-1} {p^4\over (p^2 + m^2)^{3/2}} + 
      {1\over \sqrt{4 + m^2}}.
\end{eqnarray} 
The first sum can be computed up to order $1/L^2$ using the 
Euler-Mac Laurin formula. We obtain
\begin{eqnarray}
\hskip -20truept
   && \sum_{n=0}^{L/2-1} \left[ 
   {1\over \sqrt{\hat{p}^2 + m^2}} - 
   {1\over \sqrt{p^2 + m^2}} - {1\over24} {p^4\over (p^2 + m^2)^{3/2}}
   \right] = 
   \nonumber \\
   && \quad 
   L \left[ - {1\over 2\pi} \log {\pi\over4} - {\pi\over96} -
     {z^2\over16 \pi L^2} \left(-{1\over12} - {1\over2} 
     \log {\pi\over4} + {2\over \pi^2}\right)\right]
   \nonumber \\
   && \quad -{1\over4} + {1\over 2\pi} + {\pi\over48} - 
      {z^2\over 2L^2}\left(- {1\over16} + {1\over 2 \pi^3} + 
      {1\over 16\pi}\right) + {\pi\over 6 L}
      \left( {1\over \pi^2} - {1\over 24}\right) +\, O(L^{-3}) .
   \nonumber \\  [-4mm]
   && {}
\end{eqnarray}
The two remaining sums can be computed using the results of the 
previous subsection. We obtain finally
\begin{eqnarray}
\hskip -20truept
&& \sum_{n=0}^{L-1} {1\over \sqrt{\hat{p}^2 + m^2}}\, =\, 
   {L\over \pi}\left[ \log L + \gamma_E - \log {\pi\over2} + 
     G_0 (z/2 \pi)  \right] +\, {L\over z} 
   \nonumber \\
&& \qquad + {\pi\over 6 L} \left( G_1(z/2\pi) - {1\over 12}\right) 
   + {z^4\over 96\pi^3 L} H_1(z/2 \pi) 
   \nonumber \\
&& \qquad - {z^2 \over 16\pi L} \left( \log L + \gamma_E - 
    \log {\pi\over2} + {4\over3} G_0 (z/2\pi) - {1\over 6}\right) +\,
    O(L^{-3}).
\end{eqnarray}
We will also need the expansion of \reff{A3.1} for $k=0$ up to 
$O(L^{-3})$. Using the same method we obtain
\begin{eqnarray}
\hskip -20pt
&& \sum_{n=0}^{L-1} \sqrt{\hat{p}^2 + m^2} \,=\,
   {4 L\over \pi} - {\pi\over 3L} + {4 \pi\over L} G_1(z/2 \pi) 
   + {z\over L}
   \nonumber \\
&& \qquad + {z^2\over 2\pi L} \left( \log L + \gamma_E - 
     \log {\pi\over 2}\right) +\, O(L^{-3}).
\end{eqnarray}

\subsection{Two-dimensional sums} \label{secB.2}

In this Section we present our procedure to expand in powers of $1/L$
general sums with Gaussian propagators in the FSS limit. 
A general theorem for massless propagators was proved in Ref. 
\cite{LW-theorem}. Here we will improve their result showing 
that only even powers of $1/L$ appear in the expansion and providing 
an algorithmic method to compute the various coefficients.

\subsubsection{Asymptotic expansions of
$\sum \hat{p}_x^{2h} \hat{p}_y^{2k} (\hat{p}^2 + m^2)^{-q} $}\label{secB.2.1}

In this section we present a general procedure to derive asymptotic 
expansions of sums of the form
\be
{1\over LT} \sum_{n_x,n_y} {\hat{p}_x^{2h} \hat{p}_y^{2k}\over
      (\hat{p}^2 + m^2)^q}, 
\label{eqB.38}
\ee
where $p_x = 2 \pi n_x/L$, $p_y = 2 \pi n_y/T$, the sum extends over
$0\le n_x < L$, $0\le n_y < T$, in the finite-size-scaling limit,
i.e. for $L,T\to\infty$, $m^2\to 0$ with $T/L\equiv \rho$ and
$mL\equiv z$ fixed.

First of all let us notice that rewriting
$\hat{p}^{2k}_y = [ (\hat{p}^2 + m^2) - p_x^{2} - m^2]^k$ we can limit
ourselves to consider sums with $k=0$, i.e. sums of the form
\be
{1\over LT} \sum_{n_x,n_y} {\hat{p}_x^{2h} \over
           (\hat{p}+m^2)^q}.
\ee
The summation over $n_y$ can be performed exactly using the results
of section \ref{secB.1.4}. 
It is easy to see that the result will be a sum of terms
of the form 
\be
{1\over L} \sum_{n_x=0}^{L-1}
   (\hat{p}^2_x + m^2)^{a/2} (4 + \hat{p}^2_x + m^2)^{b/2}
   \left\{ \exp\left[ {2 T}\, \hbox{\rm arcsh}\ \left(
       {1\over2}\sqrt{\hat{p}^2_x + m^2}\right)\right] - 1 \right\}^{-c},
\label{eqB.40}
\ee
for integers $a$, $b$ and $c\ge 0$. If $c$ is strictly positive
it is simple to obtain an asymptotic expansion in powers
of $1/L^2$. Indeed $\hbox{\rm arcsh}\ x = 0$ if and only if
$x=0$. Therefore, for $L,T\to\infty$ the terms that contribute are those
for which $\hat{p} \approx 0$. Thus rewriting the previous sum as
\be
{1\over L} \sum_{n_x = \lfloor L/2\rfloor}^{1 - \lfloor (L+1)/2\rfloor}
    g\left( {2\pi n\over L}, {z^2\over L^2}, \rho L\right),
\ee
we expand the function $g$ in powers of $L$ at $n,z,\rho$ fixed:
\be
g\left( {2\pi n\over L}, {z^2\over L^2}, \rho L\right)\, =\
L^\alpha  \sum_{m=0}^\infty {1\over L^{2m}} \hat{g}_m (n,z^2,\rho).
\ee
The expansion of Eq. \reff{eqB.40} is simply given by
\be
L^{\alpha-1} \sum_{m=0}^\infty {1\over L^{2m}}
   \left[ \sum_{n=-\infty}^{+\infty} \hat{g}_m (n,z^2,\rho)\right].
\ee
Let us now consider the case $c=0$. If also $b=0$ we have discussed
the asymptotic expansion in section \ref{secB.1.5}. Suppose now $b\not=0$.
Then define
\be
R_{kl} = 2^k \sum_{n=0}^l {k/2 \choose n} 
          \left( {\hat{p}_x^2 + m^2\over 4} \right)^n,
\ee
and rewrite
\begin{eqnarray}
&& {1\over L} \sum_{n_x=0}^{L-1} 
     \left(\hat{p}^2_x + m^2\right)^{a/2} 
     \left(4 + \hat{p}^2_x + m^2\right)^{b/2} = \nonumber \\
&& \quad 
   {1\over L} \sum_{n_x=0}^{L-1} 
     \left(\hat{p}^2_x + m^2\right)^{a/2} \,
     \left[ \left( 4 + \hat{p}^2_x + m^2\right)^{b/2} - R_{bl}\right] 
   \nonumber \\
&& \qquad + {1\over L} \sum_{n_x=0}^{L-1} 
      \left(\hat{p}^2_x + m^2\right)^{a/2} R_{bl}.
\end{eqnarray}
Then choose $l$ so that one can apply the Euler-Mac Laurin formula
to the first sum: as the function is periodic of period $2\pi$, 
as we observed at the end of section \ref{secB.1.1} 
(see formula \reff{EulerMacLaurindim}) we can 
simply replace the sum with the corresponding integral. We thus obtain
\begin{eqnarray}
&& {1\over L} \sum_{n_x=0}^{L-1} 
     \left(\hat{p}^2_x + m^2\right)^{a/2} 
     \left(4 + \hat{p}^2_x + m^2\right)^{b/2} = \nonumber \\
&& \quad 
   \int_0^{2\pi} {dp\over 2\pi}\, 
     \left(\hat{p}^2_x + m^2\right)^{a/2} \,
     \left[ \left( 4 + \hat{p}^2_x + m^2\right)^{b/2} - R_{bl}\right] 
   \nonumber \\
&& \qquad + {1\over L} \sum_{n_x=0}^{L-1} 
      \left(\hat{p}^2_x + m^2\right)^{a/2} R_{bl}.
\end{eqnarray}
The computation of the remaining sums 
have been discussed in Section \ref{secB.1.5}.

To illustrate the method let us consider a specific case, the sum
\be
   I_{L,T}(m^2) \equiv \, {1\over LT} 
              \sum_{n_x,n_y} {1\over \hat{p}^2 + m^2}.
\ee
Using Eq. \reff{A30} we can perform the summation over $n_y$ obtaining
\begin{eqnarray}
&& {1\over L} \sum_{n_x=0}^{L-1} 
  {1\over \sqrt{\hat{p}_x^2 + m^2} \sqrt{4 + \hat{p}_x^2 + m^2} } 
   \nonumber \\
&& +\, {2\over L}\sum_{n_x=0}^{L-1} {1\over \sqrt{\hat{p}_x^2 + m^2} 
         \sqrt{4 + \hat{p}^2_x + m^2} }\, 
     \left\{ \exp\left[ {2T}\, \hbox{\rm arcsh}\ 
     \left( {1\over2} \sqrt{\hat{p}^2_x + m^2}\right)\right] - 1 
     \right\}^{-1}. 
    \nonumber \\ [-2mm]
{}
\end{eqnarray}
The asymptotic expansion of the second sum is immediately computed: we get
\begin{eqnarray}
&& {1\over L} \sum_{n_x=0}^{L-1} {1\over \sqrt{\hat{p}_x^2 + m^2} 
         \sqrt{4 + \hat{p}^2_x + m^2} }\, 
     \left\{ \exp\left[ {2 T}\, \hbox{\rm arcsh}\ 
     \left({1\over2} \sqrt{\hat{p}^2_x + m^2}\right)\right] - 1 
     \right\}^{-1}
  \nonumber \\
&& = {1\over 4\pi} M_{1,1} (z;\rho) + 
     {\pi\over 24 L^2} \left[ \left( {z\over 2\pi}\right)^4 M_{3,1}(z;\rho)
       - 2 \left( {z\over 2\pi}\right)^2  M_{1,1}(z;\rho) - 
         2 M_{-1,1}(z;\rho) \right]
    \nonumber \\
&& \qquad + {\pi^2\rho\over 12 L^2} 
   \left[ 2 M_{-2,1}(z;\rho) - 
          2 \left( {z\over 2\pi}\right)^2 M_{0,1}(z;\rho) + 
          \left( {z\over 2\pi}\right)^4 M_{2,1}(z;\rho) \right. 
    \nonumber \\
&& \qquad\qquad \left. + 2 M_{-2,2}(z;\rho) - 
           2 \left( {z\over2\pi}\right)^2 M_{0,2}(z;\rho) + 
           \left( {z\over 2\pi}\right)^4 M_{2,2} (z;\rho)
           \right] +\, O(L^{-4}). \nonumber \\ [-4mm]
&& {}
\label{Iexpsecondsum}
\end{eqnarray}
Let us now consider the first sum. We want to compute its asymptotic 
expansion including terms of order $1/L^2$. We rewrite it as 
\begin{eqnarray}
&& {1\over L} \sum_{n_x=0}^{L-1} {1\over \sqrt{\hat{p}_x^2 + m^2}} \,
      {1\over \sqrt{4 + \hat{p}^2_x + m^2} } \, =\, 
   \nonumber \\
&& \qquad = 
   {1\over L} \sum_{n_x=0}^{L-1} 
   {1\over \sqrt{\hat{p}_x^2 + m^2}} \, 
   \left[ {1\over \sqrt{4 + \hat{p}^2_x + m^2} } - {1\over2}
          +{1\over 16} (\hat{p}^2_x + m^2) \right] 
    \nonumber \\
&& \qquad \qquad + {1\over 2L} \sum_{n_x=0}^{L-1} 
       {1\over \sqrt{\hat{p}^2_x + m^2} }\, -\, 
      {1\over 16 L} \sum_{n_x=0}^{L-1} \sqrt{\hat{p}^2_x + m^2}.
\end{eqnarray}
The last two sums have been discussed in \ref{secB.1.5}. The first one, up to
terms of $O(L^{-4})$, can be replaced by the corresponding integral.
Expanding the integrand in powers of $m^2$ we get
\begin{eqnarray}
&&  {1\over L} \sum_{n_x=0}^{L-1} 
   {1\over \sqrt{\hat{p}_x^2 + m^2} }\, 
   \left[ {1\over \sqrt{4 + \hat{p}^2_x + m^2} } - {1\over2}
          +{1\over 16} (\hat{p}^2_x + m^2) \right] 
    \nonumber \\
&& \qquad = \, {1\over 4\pi} (1 - \log 2) + \
          {z^2\over 64\pi L^2} (1 + 2 \log 2) +\, O(L^{-4}\log L).
\end{eqnarray}
It follows
\begin{eqnarray}
\hskip -10pt
&& {1\over L} \sum_{n_x=0}^{L-1} {1\over \sqrt{\hat{p}^2_x + m^2}}\, 
     {1\over \sqrt{4 + \hat{p}^2_x + m^2} } \, =
   \nonumber \\
&& \quad {1\over 2z} + {1\over 2\pi} \left[ 
      \log L + \gamma_E - \log \pi + {1\over 2} \log 2 + 
      G_0 \left({z\over 2\pi}\right)\right] \nonumber \\
&& \quad \quad + {\pi\over 6 L^2} 
        \left[ {1\over12} - G_1\left( {z\over2\pi} \right) \right]
        - {z\over 16 L^2} + 
          {z^4\over 192\pi^3 L^2} H_1\left({z\over 2\pi}\right)
        \nonumber \\
&& \quad \quad - {z^2\over 16 \pi L^2} 
      \left[ \log L+ \gamma_E - \log \pi + {1\over2}\log 2 + 
       {2\over3} G_0 \left( {z\over2\pi}\right) - {1\over3} \right] 
        \, +\, O(L^{-4}\log L). \nonumber \\ [-2mm]
{}
\end{eqnarray}
Using Eq. \reff{Iexpsecondsum} and the previous expression we 
obtain the following result:
\be
I_{L,T}(m^2) =\, {1\over2\pi} \log L +\, 
          F_0(z;\rho) - {z^2\over 16 \pi L^2} \log L + 
         {1\over L^2} F_1(z;\rho) +\, O(L^{-4}\log L),
\label{ILTexpansion}
\ee
where 
\be
F_0(z;\rho) =\, {1\over 2z} + {1\over 2 \pi}
      \left(\gamma_E - {1\over2} \log {\pi^2\over2} + 
            G_0 \left({z\over2\pi}\right)\right) +\,
      {1\over 2\pi} M_{1,1} (z;\rho),
\label{F0explicito}
\ee
and 
\begin{eqnarray}
F_1(z;\rho) &=& {\pi\over6}\left({1\over12} - G_1\left({z\over2\pi}\right)
           \right) - {z\over16} +
           {z^4\over 192\pi^3} H_1\left({z\over2\pi}\right) 
\nonumber \\
&& - {z^2\over 16 \pi} \left[ \gamma_E - {1\over2} \log {\pi^2\over2}
     + {2\over3} G_0\left({z\over2\pi}\right) - {1\over3}\right] 
\nonumber \\
&& + {\pi\over12} 
     \left[ \left( {z\over 2\pi}\right)^4 M_{3,1}(z;\rho)
       - 2 \left( {z\over 2\pi}\right)^2  M_{1,1}(z;\rho) - 
         2 M_{-1,1}(z;\rho) \right]
    \nonumber \\
&& + {\pi^2\rho\over 6 } 
   \left[ 2 M_{-2,1}(z;\rho) - 
          2 \left( {z\over 2\pi}\right)^2 M_{0,1}(z;\rho) + 
          \left( {z\over 2\pi}\right)^4 M_{2,1}(z;\rho) \right. 
    \nonumber \\
&& \quad \left. + 2 M_{-2,2}(z;\rho) - 
           2 \left( {z\over2\pi}\right)^2 M_{0,2}(z;\rho) + 
           \left( {z\over 2\pi}\right)^4 M_{2,2} (z;\rho)
           \right]\; .
\label{F1explicito}
\end{eqnarray}
Beside this sum we will also need 
\begin{eqnarray}
&& {1\over LT} \sum_{n_x,n_y} {\sum_\mu \hat{p}_\mu^4\over 
    (\hat{p}^2 + m^2)^2} 
\nonumber \\
&& \qquad =\, 1 - {1\over \pi} \,
+ {1\over L^2}\left[ {\pi\over6} 
      - 2 \pi G_1\left({z\over2\pi}\right) - {3 z\over 4} +
       {z^4\over 16\pi^3} H_1 \left({z\over2\pi}\right)
         \right. \nonumber \\
&& \qquad\qquad \left. - 
       {3 z^2\over 4 \pi} \left( \log L - \log \pi + {1\over2} \log 2
       + \gamma_E - {1\over 6} 
       + {2\over3} G_0\left({z\over2\pi}\right) \right) \right] 
    \nonumber \\
&& \qquad - {\pi\over L^2}
    \left[ 2 M_{-1,1} (z;\rho) + 
    2 \left({z\over2\pi}\right)^2 M_{1,1} (z;\rho) -
    \left({z\over2\pi}\right)^4 M_{3,1} (z;\rho)
    \right] \nonumber \\
&& \qquad + {2 \pi^2\rho \over L^2}
    \left[ 2 M_{-2,1} (z;\rho) - 
    2 \left({z\over2\pi}\right)^2 M_{0,1} (z;\rho) +
    \left({z\over2\pi}\right)^4 M_{2,1} (z;\rho)
    \right. \nonumber \\
&& \qquad \qquad \left.
    + 2 M_{-2,2} (z;\rho) - 
    2 \left({z\over2\pi}\right)^2 M_{0,2} (z;\rho) +
    \left({z\over2\pi}\right)^4 M_{2,2} (z;\rho)
    \right] +\, O(L^{-4} \log L) \nonumber \\
&& = 1 - {1\over \pi} - {3 z^2\over 4\pi L^2} \log L +\, 
     {1\over L^2} \left(12 F_1(z;\rho) -{z^2\over 8\pi}\right)
     + O(L^{-4} \log L).
\label{I2LT2exp}
\end{eqnarray}

Finally we want to report the asymptotic expansions of $F_0(z;\rho)$ and 
$F_1(z;\rho)$ for $z\to 0$ and $z\to +\infty$. They are obtained 
using the asymptotic expansions of the functions
$G_k(z)$, $H_1(z)$ and $M_{pq}(z;\rho)$ reported in sections 
\ref{secA.1} and \ref{secA.2}. For $z\to +\infty$,  we obtain
\begin{eqnarray}
F_0(z;\rho) &=& - {1\over 4\pi} \log {z^2\over32} + 
   {e^{-z}\over\sqrt{2 \pi z}} \left(1 + O(z^{-1})\right) +
   {e^{- \rho z}\over\sqrt{2 \pi \rho z}} \left(1 + O(z^{-1})\right) ,
\label{F0largez}
\\
F_1(z;\rho) &=& {z^2\over 32\pi} \left(\log {z^2\over32} +  1 \right) +
   {z^3\over24}\ {e^{-z}\over\sqrt{2 \pi z}}  \left(1 + O(z^{-1})\right) +
   {\rho z^3\over24}{e^{- \rho z}\over\sqrt{2 \pi \rho z}} 
    \left(1 + O(z^{-1})\right) . 
\nonumber \\ [-2mm]
{} \label{F1largez}
\end{eqnarray}
Let us now consider the perturbative limit. If $\rho\not=\infty$, we 
get for $z\ll 1$, $z\ll 1/\rho$: 
\begin{eqnarray}
F_0(z;\rho) &=& {1\over \rho z^2} +\ F_{00}(\rho) + z^2 F_{01}(\rho) + O(z^4),
\label{F0smallz}
\\
F_1(z;\rho) &=&  F_{10}(\rho) + z^2 F_{11}(\rho) + O(z^4),
\end{eqnarray}
where
\begin{eqnarray}
F_{00}(\rho) &=& 
   {1\over2\pi} \left (\gamma_E - \log \pi + {1\over2} \log 2\right ) - 
   {1\over \pi} \log \eta(i\rho), \label{F00def} \\
F_{01}(\rho) &=& - {1\over 16\pi^3} \zeta(3) - {\rho^3\over 720} 
  - {1\over8\pi^3} N_{3,1}(\rho) - 
   {\rho\over 4\pi^2} (N_{2,1}(\rho) + N_{2,2}(\rho) ), 
\label{F01def} \\
F_{10}(\rho) &=& {\pi\over 72} - {1\over 12 \rho} - {\pi\over 3} N_{-1,1}(\rho)
      + {2 \pi^2 \rho\over 3} (N_{-2,1}(\rho) + N_{-2,2}(\rho) ),
\label{F10def} \\
F_{11}(\rho) &=& - {1\over 16\pi}\left(\gamma_E - \log \pi + {1\over2} \log 2 
      -{1\over3}\right) - {\rho\over 288} + {1\over 8\pi} \log \eta(i\rho) 
\nonumber \\
&& 
    + {\rho\over 12} (N_{0,1}(\rho) + N_{0,2}(\rho) ) - 
     {\pi \rho^2\over6} (N_{-1,1}(\rho) + 3 N_{-1,2}(\rho) + 2 N_{-1,3}(\rho))
\nonumber \\
&=& - {1\over 8} F_{00}(\rho) - {\rho\over 4\pi} F_{10}(\rho).
\label{F11def}
\end{eqnarray}
In the last formula we have used Eqs. \reff{Nlogeta1} and \reff{Nlogeta2} 
that also show that $F_{10}(\rho)$ can be expressed in terms of
derivatives of $\log \eta(i\rho)$.
For $\rho = 1$ we obtain the following numerical values:
\begin{eqnarray}
F_{00}(1) &\approx & \hphantom{-}  0.04876563317014130,  \\
F_{01}(1) &\approx &  -0.00386694659073721,  \\
F_{10}(1) &\approx &  -0.02924119479519021,  \\
F_{11}(1) &\approx &  -0.00376876379948390.  
\end{eqnarray}
For the strip ($\rho=\infty$) the previous expansions are not valid.
In this case we write
\begin{eqnarray}
F_0(z;\infty) &=& {1\over 2z} + \overline{F}_{00} + z^2 \overline{F}_{01} + 
    O(z^4), 
\label{F0smallzstrip} \\
F_1(z;\infty) &=& {\pi\over 72} - {z\over16} + z^2 \overline{F}_{11} +
    O(z^4), 
\end{eqnarray}
with
\begin{eqnarray}
\overline{F}_{00} &=& {1\over 2\pi} 
    \left (\gamma_E - \log \pi + {1\over2}\log 2 \right ) ,
\label{Fbar00}\\
\overline{F}_{01} &=& - {1\over 16 \pi^3} \zeta(3), \\
\overline{F}_{11} &=& - {1\over 16\pi} 
    \left(\gamma_E - \log \pi + {1\over2}\log 2 - {1\over3} \right ).
\end{eqnarray}
Finally let us comment on the duality property of the functions 
$F_0(z;\rho)$ and $F_1(z;\rho)$. The sum $I_{L,T}(m^2)$ is clearly 
symmetric in $L,T$ and thus it is a function $\Phi(m L,T/L,L)$ such that 
\be
   \Phi(m L, T/L, L) = \Phi(m T, L/T, T),
\ee
i.e. $\Phi(z,\rho,L) = \Phi(\rho z,1/\rho,\rho L)$. This implies for 
the functions $F_0(z;\rho)$ and $F_1(z;\rho)$ the following relations:
\begin{eqnarray}
F_0(z;\rho) &=& {1\over 2\pi} \log \rho +\, F_0 (\rho z;1/\rho), \\
F_1(z;\rho) &=& -{z^2\over 16\pi} \log \rho +\, {1\over \rho^2}
         F_1 (\rho z;1/\rho). 
\end{eqnarray}
These equations provide a non trivial check for the correctness of our 
asymptotic expansions and moreover imply the following relations 
on the expansion coefficients for $z\to 0$ 
\begin{eqnarray}
F_{00}(\rho) &=& {1\over 2\pi} \log \rho + F_{00}(1/\rho), \\
F_{01}(\rho) &=& \rho^2 F_{01}(1/\rho), \label{F01duality} \\
F_{10}(\rho) &=& {1\over \rho^2} F_{10}(1/\rho), \\
F_{11}(\rho) &=& - {1\over 16 \pi} \log \rho + F_{11}(1/\rho). 
\end{eqnarray}
The duality relation for 
$F_{00}(\rho)$, $F_{10}(\rho)$ and $F_{11}(\rho)$
can be obtained directly from the 
inversion property of Dedekind's $\eta$-function
\cite{Chandra}
\be
 \eta(-1/\tau)^2 = - i\tau \eta(\tau)^2,
\ee
where, in our case, we would identify $\tau = i\rho$. To prove directly
Eq. \reff{F01duality} one should use the relation obtained comparing 
Eq. \reff{identityA.5.1} with Eq. \reff{identityA.5.2} for $p=0$.

\subsubsection{Asymptotic expansion of 
                $\sum (w(p) + m^2)^{-1}$} \label{secB.2.2}

We want to compute here the asymptotic expansion, including 
terms of order $O(L^{-2})$, of the sum
\be
   {\cal I}_{L,T}(m^2) \, =\, {1\over LT}\
    \sum_{n_x,n_y} {1\over w(p) + m^2}
\ee
in the FSS limit. Generic sums of the type \reff{eqB.38} can be 
easily computed with the same technique.
We assume that,
for $-\pi \le p_i \le \pi$, $w(p)$ vanishes only for $p=0$ and 
that in a  neighbourhood of the origin $w(p)$ has 
an expansion of the form \reff{wqespansione}.
Then we rewrite
\begin{eqnarray}
{\cal I}_{L,T}(m^2) &=& 
    {1\over LT} \sum_p \left[ {1\over w(p) + m^2} - {1\over \hat{p}^2 + m^2}
      + {\alpha_1 \sum_\mu \hat{p}^4_\mu + \alpha_2 (\hat{p}^2)^2
           \over (\hat{p}^2 + m^2)^2 }\right]
   \nonumber \\
&& \qquad + {1\over LT} \sum_p {1\over \hat{p}^2 + m^2} -
   \, {1\over LT} \sum_p 
   {\alpha_1 \sum_\mu \hat{p}^4_\mu + \alpha_2 (\hat{p}^2)^2\over 
       (\hat{p}^2 + m^2 )^2}.
\end{eqnarray}
Since we want to compute ${\cal I}_{L,T}(m^2)$ up to $O(L^{-2})$ 
we can substitute 
the first sum with the corresponding integral 
(cf. Eq. \reff{EulerMacLaurindim}). Then, expanding
the integrand in powers of $m^2$, we obtain
\begin{eqnarray}
{\cal I}_{L,T}(m^2) &=& 
   \Lambda_0 + \alpha_1 \left(1 - {1\over \pi}\right) + \alpha_2 - 
          {z^2\over L^2} \Lambda_1 \nonumber \\
   && + I_{L,T}(m^2) 
      - {1\over LT} \sum_p 
        {\sum_\mu \alpha_1 \hat{p}^4_\mu + \alpha_2 (\hat{p}^2)^2
          \over (\hat{p}^2 + m^2)^2},
\end{eqnarray}
where we have introduced
\begin{eqnarray}
\Lambda_0 &=& \int {d^2p\over (2\pi)^2} 
     \left( {1\over w(p)} - {1\over \hat{p}^2} \right) ,
\label{Lambda0}\\
\Lambda_1 &=& \int {d^2p\over (2\pi)^2} 
     \left( {1\over w(p)^2} - {1\over (\hat{p}^2)^2} +
     {2\over (\hat{p}^2)^3} \left( \alpha_1 \sum_\mu \hat{p}^4_\mu 
       + \alpha_2 (\hat{p}^2)^2 \right) \right) ,
\label{Lambda1}
\end{eqnarray}
and we have used the result (see Appendix C of Ref. \cite{CP-4loop})
\be
\int {d^2p\over (2\pi)^2} {\sum_\mu \hat{p}^4_\mu\over (\hat{p}^2)^2} \, 
   =\, 1 - {1\over \pi}.
\ee
We get eventually, using Eqs. \reff{I2LT2exp} and \reff{ILTexpansion},
\be 
{\cal I}_{L,T}(m^2) =\, {1\over2\pi} \log L +\, 
          {\cal F}_0(z;\rho) - 
          {z^2\over 16 \pi L^2} (1 - 12\alpha_1 - 16 \alpha_2)\log L + 
         {1\over L^2} {\cal F}_1(z;\rho) ,
\label{sommaasintotica}
\ee
where the neglected terms are of order $O(L^{-4}\log L)$ and 
\begin{eqnarray}
{\cal F}_0(z;\rho) &=& F_0(z;\rho) +\Lambda_0, \\
{\cal F}_1(z;\rho) &=& (1-12 \alpha_1) F_1(z;\rho) + 
     z^2 \left({\alpha_1\over 8\pi} - \Lambda_1\right) 
    + 2 \alpha_2 z^2 F_0(z;\rho) + 
     {\alpha_2\over2} z^3 {\partial F_0\over \partial z} (z;\rho) ;
 \nonumber \\ [-2mm]
{}
\label{F1storto}
\end{eqnarray}
$F_0(z;\rho)$ and $F_1(z;\rho)$ are defined 
in Eqs. \reff{F0explicito} and \reff{F1explicito}. Explicit values of 
$\Lambda_0$ and $\Lambda_1$ for the hamiltonians we have introduced in 
the text are reported in table \ref{tableLambda}.

\begin{table}
\begin{center}
\begin{tabular}{|l|c|c|}
\hline
  &  $\Lambda_0$ & $\Lambda_1$ \\
\hline
$H^{diag}$   & \hphantom{$-$}0.0322658881033520480  & 0.00371978402668476  \\
$H^{Sym}$    & $-$0.0471699346329274140  & 0.00811339924292905    \\
$H^{Sym2}$   & $-$0.0564354728047190420  & 0.00331572798724030    \\
\hline
\end{tabular}
\end{center}
\caption{Values of $\Lambda_0$ and $\Lambda_1$ for various
hamiltonians.
 }
\label{tableLambda}
\end{table}

Using the expansions of $F_0(z;\rho)$ and $F_1(z;\rho)$ 
(see previous section) we can easily 
obtain the asymptotic expansions of ${\cal F}_1(z;\rho)$. For large $z$ we
obtain
\begin{eqnarray}
{\cal F}_1(z;\rho) &=& - (12 \alpha_1 + 16\alpha_2 - 1) 
   {z^2\over 32\pi} \log {z^2\over 32} - 
   {z^2\over 32\pi} (8\alpha_1 + 8 \alpha_2 - 1) - z^2 \Lambda_1 
\nonumber \\
  && - {1\over24} (12 \alpha_1 + 12 \alpha_2 - 1) z^3 \
\left({e^{-z}\over \sqrt{2\pi z}} + {\rho e^{-\rho z}\over \sqrt{2 \pi \rho z}}
   \right) +\, O(z^{3/2} e^{-z}, z^{3/2} e^{-\rho z})  .
\nonumber \\ [-2mm]
{}
\end{eqnarray}
For finite $\rho$ and $z \ll 1$, $z\ll 1/\rho$, neglecting terms of
order $z^4$, we have 
\be
\hskip -2truept
{\cal F}_1(z;\rho)\, =\, 
   (1- 12\alpha_1) F_{10}(\rho) + {\alpha_2\over\rho} + z^2 
   \left[(1-12\alpha_1) F_{11}(\rho) + 2 \alpha_2 F_{00}(\rho) + 
          {\alpha_1\over 8\pi} - \Lambda_1\right]\; ,
\ee
while on the strip, for $z\to 0$, we obtain
\begin{eqnarray}
{\cal F}_1(z;\infty) & = &
   {\pi\over72} (1 - 12\alpha_1) + {z\over 16} (12 \alpha_1 + 12 \alpha_2 - 1)
\nonumber \\ 
    && + z^2 \left[ (1 - 12\alpha_1) \overline{F}_{11} + 
                2 \alpha_2 \overline{F}_{00} + 
                {\alpha_1\over8\pi} - \Lambda_1\right] + O(z^4).
\end{eqnarray}

\subsubsection{Sum for the tensor correlation length}

In this section we describe the computation, in the FSS limit, of 
\be
I_{2,LT}(m^2) = 
   {1\over LT} \sum_{n_x,n_y}
   {1\over [ \widehat{(p_x - p_0)}^2 + \hat{p}^2_y + m^2] 
           [ \hat{p}^2_x + \hat{p}^2_y + m^2]}\; ,
\label{eqB.95}
\ee
where $p_0=2 \pi/L$ and, as before, $p_x=2 \pi n_x/L$ and 
$p_y=2 \pi n_y/T$. 

First of all we rewrite Eq. \reff{eqB.95}  as 
\be
I_{2,LT}(m^2) = {2\over LT} \sum_{n_x,n_y}
   {1\over \widehat{(p_x - p_0)}^2 - \hat{p}^2_x}
   {1\over \hat{p}^2_x + \hat{p}^2_y + m^2} \; ;
\ee
then we sum over $p_y$ to get 
\begin{eqnarray}
&& \hskip-30pt 
  {2\over L} \sum_{n=0}^{L-1} {1\over \widehat{(p - p_0)}^2 - \hat{p}^2}
   \, {1\over \sqrt{\hat{p}^2 + m^2} \sqrt{4 + \hat{p}^2 + m^2} }
    \times \nonumber \\
&& \left\{ 1 + 
    {2\over \exp[2 T \hbox{\rm arcsh} ({1\over2} \sqrt{\hat{p}^2 + m^2})]
        - 1} \right\}\; ,
\end{eqnarray}
where we have simplified the notation using $p$ instead of $p_x$.

The contribution due to the second term in curly brackets is obtained 
by simply expanding in powers of $1/L^2$ (see the 
discussion of Eq. \reff{eqB.40} for $c\not=0$). 
The remaining term requires more 
care. Assuming $L$ even, we rewrite
\begin{eqnarray}
&& \hskip -20pt 
   {2\over L} \sum_{n=0}^{L-1} 
   {1\over \widehat{(p-p_0)}^2 - \hat{p}^2} 
    {1\over \sqrt{\hat{p}^2 + m^2} \sqrt{4 + \hat{p}^2 + m^2} } \nonumber \\
&&  = \,  {2\over L \hat{p}_0^2} 
     {1\over \sqrt{4 + m^2}} \left( {1\over m} - {1\over \sqrt{8 + m^2}}\right)
   \nonumber \\
&& \qquad + {4\over L} \sum_{n=1}^{L/2-1} 
    {1 - \hat{p}^2/2\over (\hat{p}_0^2 - 4 \sin^2 p) } 
    {1\over \sqrt{\hat{p}^2 + m^2} \sqrt{4 + \hat{p}^2 + m^2} }\; .
\label{firsttensor}
\end{eqnarray}
Consider now the last sum and notice that, for $L\to\infty$, $m\to 0$,
beside the singularity at $p=0$, there is an additional singularity 
at $p=\pi$. Using the fact that 
\be
    \sum_{n=1}^{L/2-1} {1\over \hat{p}_0^2 - 4 \sin^2 p} \, =\, 
    - {1\over \hat{p}_0^2} \; ,
\ee
and keeping only those contributions that do not vanish for $L\to\infty$
we rewrite Eq. \reff{firsttensor} as 
\begin{eqnarray}
&& \hskip -40pt 
   {2\over z \hat{p}_0^2} {1\over \sqrt{4 + m^2}}  + 
  {1\over 2\sqrt{2}} {1\over L \hat{p}_0^2} \nonumber \\
&& \quad + {4\over L} \sum_{n=1}^{L/2-1} 
    {1\over \hat{p}_0^2 - 4 \sin^2 p} 
   \left( {1\over 4\sqrt{2}} + 
  { 1 - \hat{p}^2/2\over \sqrt{\hat{p}^2 + m^2} \sqrt{4 + \hat{p}^2 + m^2} }
   \right).
\end{eqnarray}
In this way we have removed the singularity for $p=\pi$. The remaining 
part of the calculation follows the lines we have presented for 
Eq. \reff{eqB.40} when $c=0$. 
We subtract to the sum the first two terms of the asymptotic 
expansion in $1/L^2$ and then replace the sum with the integral. 
Explicitly, if we define
\be
\hskip -2truept
R(p,p_0,m^2) = 
   {1\over 2(p_0^2 - 4 p^2)} \left[ {1\over \sqrt{p^2 + m^2}} 
 \left( 1 + {p_0^2\over 12} - {p^2\over4} - {m^2\over 6} + 
    {1\over24} {m^4\over p^2 + m^2}\right) +\, 
    {1\over 2\sqrt{2}}\right],
\ee
we obtain
\begin{eqnarray}
&& \hskip -20pt 
  {2\over z \hat{p}_0^2} {1\over \sqrt{4 + m^2}}  +
  {1\over 2\sqrt{2}} {1\over L \hat{p}_0^2} \nonumber \\
&& + 4 \int^\pi_0 {dp\over 2\pi} 
   \left[ - {1\over 4\sin^2 p} 
   \left( {1\over 4 \sqrt{2}} + 
    {1 - \hat{p}^2/2 \over \hat{p}\sqrt{4 + \hat{p}^2} }\right) 
   - R(p,0,0)\right]   \nonumber \\
&& + {4\over L} \sum_{n=1}^{L/2-1} R(p,p_0,m^2) +\, O(1/L).
\end{eqnarray}
The last sum can be dealt with following the strategy of section
\ref{secB.1.3}. We get finally 
\begin{eqnarray}
&&\hskip -20pt 
   {2\over L} \sum_{n=0}^{L-1} 
   {1\over \widehat{(p-p_0)}^2 - \hat{p}^2} 
    {1\over \sqrt{\hat{p}^2 + m^2} \sqrt{4 + \hat{p}^2 + m^2} } =\nonumber \\
&&  {1\over z} {L^2\over 4\pi^2} \left( 1 + {\pi^2\over 3L^2}\right) 
   + {1\over 16\pi} \left(\log L + \gamma_E - \log \pi + 
   {1\over2} \log 2 - {2\over3}\right) 
\nonumber \\
&& + {L^2\over 2\pi^2} \sum_{n=1}^\infty 
     {1\over (1-4 n^2) \sqrt{4 \pi^2 n^2 + z^2}} 
\nonumber \\
&& + {1\over 2\pi^2} \sum_{n=1}^\infty \left\{
     {1\over (1-4 n^2) \sqrt{4 \pi^2 n^2 + z^2}}  \left[
     {\pi^2\over3} - \pi^2 n^2 - {z^2\over6} + 
     {1\over 24} {z^4\over 4\pi^2 n^2 + z^2}\right] - 
     {\pi \over 8n}\right\}\; .
\nonumber \\ [-2mm]
&& {} 
\end{eqnarray}
Collecting everything together and introducing the functions 
${\cal H}_k(\alpha)$ and ${\cal M}_{pq}(z;\rho)$ defined in sections
\ref{secA.1} and \ref{secA.2} we have
\be
I_{2,LT}(m^2) =\, 
   L^2 F_3(z;\rho) + {1\over 16\pi}\log L + 
       F_4(z;\rho) + O(\log L/L^2),
\ee
where
\be
F_3(z;\rho) =\, {1\over 4\pi^2 z} + {1\over 4\pi^3} 
    \left[ {\cal H}_0 \left( {z\over 2\pi} \right) + 
           {\cal M}_{1,1}(z;\rho) \right]\; ,
\ee
and 
\begin{eqnarray}
F_4(z;\rho) &=& 
    {1\over 12z} + {1\over 16\pi} 
    \left(\gamma_E - \log \pi + {1\over2} \log 2 - {2\over3}\right)
    - {z\over 32 \pi^2} 
\nonumber \\
&& + {1\over 48 \pi^3} (\pi^2 - 2 z^2) {\cal H}_0\left( {z\over 2\pi} \right)
   + {1\over 16\pi} G_0 \left( {z\over 2\pi} \right) 
   + {1\over 24\pi} \left({z\over 2\pi} \right)^4 
            {\cal H}_1\left( {z\over 2\pi} \right)
\nonumber \\
&& + {1\over 24 \pi} \left( {\cal M}_{1,1}(z;\rho) + 
        M_{1,1}(z;\rho)\right) 
\nonumber \\
&& - {1\over24\pi} \left[ 
   2 {\cal M}_{-1,1}(z;\rho) + 
   2 \left({z\over 2\pi} \right)^2 {\cal M}_{1,1}(z;\rho) -
     \left({z\over 2\pi} \right)^4 {\cal M}_{3,1}(z;\rho) \right] 
\nonumber \\
&& + {\rho\over 12} \left[
   2 {\cal M}_{-2,1}(z;\rho) - 
   2 \left({z\over 2\pi} \right)^2 {\cal M}_{0,1}(z;\rho) +
     \left({z\over 2\pi} \right)^4 {\cal M}_{2,1}(z;\rho) \right.
\nonumber \\
&& \quad  \left.
   + 2 {\cal M}_{-2,2}(z;\rho) - 
   2 \left({z\over 2\pi} \right)^2 {\cal M}_{0,2}(z;\rho) +
     \left({z\over 2\pi} \right)^4 {\cal M}_{2,2}(z;\rho) \right]\; .
\end{eqnarray}
To conclude this section we give the asymptotic expansions of 
$F_3(z;\rho)$ and $F_4(z;\rho)$ for large and small values of $z$. 
The necessary formulae for the derivations are reported in sections 
\ref{secA.1} and \ref{secA.2}. 
For large $z$  we have
\begin{eqnarray}
F_{3}(z;\rho) &=& {1\over 4\pi z^2} \left[
 1 - {2 \pi^2\over 3z^2} + {8 \pi^4\over 15 z^4} - 
  {16 \pi^6\over 35 z^6} + O(z^{-8})\right] ,
\label{F3largez}
\\
F_4(z;\rho) &=& - {1\over32\pi}\left( \log{z^2\over 32} + 2\right) + 
    {\pi\over 48 z^2} - {\pi^3\over 360 z^4} - 
    {\pi^5\over 126 z^6} + O(z^{-8}).
\label{F4largez}
\end{eqnarray}
For finite $\rho$ and $z\ll 1$,  $z\ll 1/\rho$ we have
\begin{eqnarray}
F_{3}(z;\rho) &=& {1\over 2 \pi^2 \rho z^2} + F_{30}(\rho) + O(z^2), \\
F_{4}(z;\rho) &=& {1\over 6 \rho z^2} + F_{40}(\rho) + O(z^2), 
\end{eqnarray}
where
\begin{eqnarray}
\hskip -25pt
F_{30}(\rho) &=& {1\over 4\pi^3} (1 - 2\log 2) + {\rho \over 24 \pi^2}
   + {1\over 2\pi^3} {\cal N}_{1,1} (\rho), \\
\hskip -25pt
F_{40}(\rho) &=& {1\over 16\pi} (\gamma_E - \log \pi) - 
    {1\over 96 \pi} (2 + \log 2) + {\rho\over 72} - 
    {1\over 24 \pi^2 \rho} \nonumber \\
&& \hskip -20pt + {1\over 12\pi} 
   (N_{1,1}(\rho) + {\cal N}_{1,1} (\rho) - 2 {\cal N}_{-1,1} (\rho))
   + {\rho\over 3} ( {\cal N}_{-2,1}(\rho) + {\cal N}_{-2,2}(\rho) ).
\end{eqnarray}
On the strip, for small $z$, we have
\begin{eqnarray}
\hskip -20pt
F_3(z;\infty) &=& {1\over 4\pi^2 z} + {1\over 4\pi^3} (1 - 2\log 2) + 
     O(z^2)  ,
\\
\hskip -20pt
F_4(z;\infty) &=&  {1\over 12 z} + {1\over 16\pi} (\gamma_E - \log \pi) -
    {1\over 96 \pi} (2 + \log 2) - {z\over 32\pi^2} + 
    O(z^2). 
\end{eqnarray}

\section{Asymptotic expansion of lattice integrals}

In this section we want to discuss the asymptotic expansion for 
$m_0^2\to 0$
of the integrals
\begin{eqnarray}
  I_\infty (m_0^2) &=& \int {d^2p\over (2\pi)^2} 
      {1\over \hat{p}^2 + m_0^2}, \\
  {\cal I}_\infty (m_0^2) &=& \int {d^2p\over (2\pi)^2}
      {1\over w(p) + m_0^2}. 
\end{eqnarray}
More general integrals can be discussed following the same 
method and using the results of App. C of Ref. \cite{CP-4loop}.
The expansion of $I_\infty (m_0^2)$ is easily obtained from its expression
in terms of elliptic integrals \cite{GR}
\begin{eqnarray}
   I_\infty (m_0^2) &=& {2\over \pi} {1\over 4+m_0^2} 
     K\left( {4\over 4+m_0^2}\right) \nonumber \\
  &=& - {1\over 4\pi} \log {m_0^2\over 32} +\,
     {m_0^2\over 32\pi} \left( \log {m_0^2\over 32} + 1 \right)
   +\, O(m_0^4 \log m_0^2).
\end{eqnarray}
To obtain the expansion of ${\cal I}_\infty(m_0^2)$ let us proceed
as in section \ref{secB.2.2}. We rewrite
\begin{eqnarray}
{\cal I}_{\infty}(m_0^2) &=& 
    \int {dp\over (2\pi)^2}
      \left[ {1\over w(p) + m^2} - {1\over \hat{p}^2 + m^2}
      + {\alpha_1 \sum_\mu \hat{p}^4_\mu + \alpha_2 (\hat{p}^2)^2
           \over (\hat{p}^2 + m^2)^2 }\right]
   \nonumber \\
&& \qquad + \int {dp\over (2\pi)^2} {1\over \hat{p}^2 + m^2} -
   \, \int {dp\over (2\pi)^2} 
   {\alpha_1 \sum_\mu \hat{p}^4_\mu + \alpha_2 (\hat{p}^2)^2\over 
       (\hat{p}^2 + m^2 )^2}.
\end{eqnarray}
If we want to compute the expansion neglecting terms of order 
$O(m_0^4 \log m_0^2)$ we can expand the first integral in powers of $m_0^2$.
Then, using (see App. C of Ref. \cite{CP-4loop})
\be
\int {dp\over (2\pi)^2}
{\sum_\mu \hat{p}_\mu^4 \over (\hat{p}^2 + m_0^2)^2} =\,
  1 - {1\over \pi} + {m_0^2\over 8\pi} 
   \left( 3 \log {m_0^2\over32} + 2\right) + \,
  O(m_0^4 \log m_0^2),
\ee
we obtain
\begin{eqnarray}
{\cal I}_\infty (m_0^2) &=& 
    - {1\over 4\pi} \log {m_0^2\over32} + \Lambda_0 \nonumber \\
   && \hskip -15pt + {m_0^2\over32 \pi} 
     (1 - 12 \alpha_1 - 16 \alpha_2) \log {m_0^2\over32} \, +\, 
    {m_0^2\over32\pi} (1 - 8\alpha_1 - 8\alpha_2) -
    m_0^2 \Lambda_1,
\label{intasintotico}
\end{eqnarray}
where $\Lambda_0$ and $\Lambda_1$ are defined in 
Eqs. \reff{Lambda0} and \reff{Lambda1}.


\begin{thebibliography}{199}

\bibitem{Fisher_72}
M. E. Fisher, in {\em Critical Phenomena}, Proc. 51st Enrico Fermi Summer
School, Varenna, M. S. Green ed. (Academic Press, New York, 1972).

\bibitem{Fisher-Barber_72}
M. E. Fisher and M. N. Barber, Phys. Rev. Lett. {\bf 28}, 1516 (1972).

\bibitem{Barber_83}
M. N. Barber,
in {\em Phase Transitions and Critical 
Phenomena}, Vol. 8, C. Domb and J. L. Lebowitz eds. (Academic Press, London,
1983).

\bibitem{Cardy_book} 
J. L. Cardy, ed., {\em Finite Size Scaling}, (North Holland, Amsterdam, 1988).

\bibitem{Privman_FSS_book}   V. Privman, ed.,
   {\em Finite Size Scaling and Numerical Simulation of Statistical Systems}\/
   (World Scientific, Singapore, 1990).

\bibitem{Nightingale_76_82}
M. P. Nightingale, Physica {\bf 83A}, 561 (1976); J. Appl. Phys. 
{\bf 53}, 7927 (1982).

\bibitem{Luscher_91}  M. L\"uscher, P. Weisz, and U. Wolff,
   Nucl. Phys. {\bf B359}, 221 (1991).

\bibitem{Kim}  
J.-K. Kim, Phys. Rev. Lett. {\bf 70}, 1735 (1993);
Nucl. Phys. B (Proc. Suppl.) {\bf 34}, 702 (1994);
Phys. Rev. {\bf D50}, 4663 (1994);
Europhys. Lett. {\bf 28}, 211 (1994);
Phys. Lett. {\bf B345}, 469 (1995).

\bibitem{fss_greedy}  S. Caracciolo, R. G. Edwards, S. J. Ferreira,
   A. Pelissetto, and A. D. Sokal, Phys. Rev. Lett. {\bf 74}, 2969 (1995);
   Nucl. Phys. B (Proc. Suppl.) {\bf 42}, 749 (1995).

\bibitem{o3_letter} S. Caracciolo, R. G. Edwards,
   A. Pelissetto, and A. D. Sokal, Phys. Rev. Lett. {\bf 75}, 1891 (1995);
  Nucl. Phys. B (Proc. Suppl.) {\bf 42}, 752 (1995).

\bibitem{MGMC_96} 
   T. Mendes, A.~Pelissetto, and A.~D.~Sokal,
   Nucl. Phys. {\bf B477}, 203 (1996).

\bibitem{SU3_letter}  G. Mana, A. Pelissetto, and A. D. Sokal,
   Phys. Rev. {\bf D54}, R1252 (1996).

\bibitem{SU3_paper} G. Mana, A. Pelissetto, and A. D. Sokal,
   Phys. Rev. {\bf D55}, 3674 (1997).

\bibitem{Sokal-Salas} J. Salas and A. D. Sokal, 
J. Stat. Phys. {\bf 88}, 567 (1997).

\bibitem{Symanzik} K. Symanzik, in {\em ``Mathematical problems in
theoretical physics"}, R. Schrader et al. eds., (Springer, Berlin, 1982);
Nucl. Phys. {\bf B226}, 187 (1983);
{\em ibid.} 205 (1983).

\bibitem{LW} 
M. L\"uscher and P. Weisz, Comm. Math. Phys. {\bf 97}, 59 (1985);
{\em erratum} {\bf 98}, 433 (1985).

\bibitem{perfect} P. Hasenfratz and F. Niedermayer,
Nucl. Phys. {\bf B414}, 785 (1994).

\bibitem{Luscher_LAT97} 
S. Capitani, M. Guagnelli, M. L\"uscher, S. Sint, R. Sommer,
P. Weisz, and H. Wittig,
Nucl. Phys. B (Proc. Suppl.) {\bf 63}, 153 (1998).

\bibitem{Hasenfratz_LAT97} 
P. Hasenfratz, 
Nucl. Phys. B (Proc. Suppl.) {\bf 63}, 53 (1998).

\bibitem{Polyakov_75}  A. M. Polyakov, Phys. Lett. {\bf B59}, 79 (1975).

\bibitem{Brezin_76}  E. Br\'ezin and J. Zinn-Justin,
  Phys. Rev. {\bf B14}, 3110 (1976).

\bibitem{Bardeen_76}
W. A. Bardeen, B. W. Lee, and R. E. Shrock, Phys. Rev. {\bf D14}, 985 (1976).

\bibitem{Stanley}  H.~E. Stanley, Phys. Rev. {\bf 176}, 718 (1968);
{\bf 179}, 570 (1969).

\bibitem{DiVecchia} P. Di Vecchia, R. Musto, F. Nicodemi, R.~Pettorino, 
and P.~Rossi,
  Nucl. Phys. {\bf B235} [FS11], 478 (1984).

\bibitem{Muller}  V. F. M\"uller, T. Raddatz, and W. R\"uhl,
  Nucl. Phys. {\bf B251} [FS13], 212 (1985);
  {\em erratum} Nucl. Phys. {\bf B259}, 745 (1985).

\bibitem{Flyvbjerg}  H. Flyvbjerg, Phys. Lett. {\bf B219}, 323 (1989);
  H. Flyvbjerg and S. Varsted, Nucl. Phys. {\bf B344},
  646 (1990);
  H. Flyvbjerg and F. Larsen, Phys. Lett. {\bf B266}, 92, 99 (1991).

\bibitem{Campostrini_90ab}  P. Biscari, M. Campostrini, and P. Rossi,
  Phys. Lett. {\bf B242}, 225 (1990);
  M. Campostrini and P. Rossi, Phys. Lett. {\bf B242}, 81 (1990).

\bibitem{Zamolodchikov_79}  A. B. Zamolodchikov and A. B. Zamolodchikov,
  Nucl. Phys. {\bf B133}, 525 (1978);
  Ann. Phys. (N.Y.) {\bf 120}, 253 (1979).

\bibitem{Polyakov-Wiegmann_83}  A. Polyakov and P. B. Wiegmann,
  Phys. Lett. {\bf 131B}, 121 (1983).

\bibitem{Hasenfratz-Niedermayer_1}  P. Hasenfratz, M. Maggiore, and
  F. Niedermayer, Phys. Lett. {\bf B245}, 522 (1990).

\bibitem{Hasenfratz-Niedermayer_2} P. Hasenfratz and F. Niedermayer, 
  Phys. Lett. {\bf B245}, 529 (1990); {\bf B268}, 231 (1991).

\bibitem{Wolff_O4_O8}  U. Wolff, Phys. Lett. {\bf B248}, 335 (1990).

\bibitem{MGMC_O4}  R. G. Edwards, S. J. Ferreira, J. Goodman, 
and A. D. Sokal,
  Nucl. Phys. {\bf B380}, 621 (1992).

\bibitem{CEMPS}
S.~Caracciolo, R.~G.~Edwards, T.~Mendes, A.~Pelissetto, and A.~D.~Sokal,
Nucl. Phys. B (Proc. Suppl.) {\bf 47}, 763 (1996).

\bibitem{Alles-Symanzik} 
B. Alles, A. Buonanno, and G. Cella,
Nucl. Phys. {\bf B500}, 513 (1997);
Nucl. Phys. B (Proc. Suppl.) {\bf 53}, 677 (1997).

\bibitem{Falcioni-Treves}  
M. Falcioni and A. Treves, Nucl. Phys. {\bf B265}, 671 (1986).

\bibitem{CP-3loop} S. Caracciolo and A. Pelissetto,
Nucl. Phys. {\bf B420}, 141 (1994).

\bibitem{CP-4loop} S. Caracciolo and A. Pelissetto,
Nucl. Phys. {\bf B455} [FS], 619 (1995).

\bibitem{Barber-Fisher_73}
M. N. Barber and M. E. Fisher, Ann. Phys. (N.Y.) {\bf 77}, 1 (1973).

\bibitem{Brezin_82}
E. Br\'ezin, J. Physique {\bf 43}, 15 (1982).

\bibitem{Luscher_82}
M. L\"uscher, Phys. Lett. {\bf 118B}, 391 (1982).

\bibitem{Privman-Fisher_83} 
V. Privman and M. E. Fisher, J. Phys. {\bf A16}, L295 (1983).

\bibitem{Magnoli-Ravanini}  N. Magnoli and F. Ravanini,
  Z. Phys. {\bf C34}, 43 (1987).   

\bibitem{Pat-Seil_comment}   A. Patrascioiu and E. Seiler, Phys. Rev. Lett.
   {\bf 76}, 1178 (1996).

\bibitem{CEPS_O3_reply_to_Pat-Seil}  S. Caracciolo, R. G. Edwards, 
  A. Pelissetto, and A. D. Sokal, Phys. Rev. Lett. {\bf 76}, 1179 (1996).

\bibitem{Pat-Seil}  A. Patrascioiu and E. Seiler,
    Nucl. Phys. B (Proc. Suppl.) {\bf 30}, 184 (1993);
     Phys. Rev. Lett. {\bf 74}, 1920, 1924 (1995).

\bibitem{Hasenbusch}
M. Hasenbusch, Phys. Rev. {\bf D53}, 3445 (1996).

\bibitem{Catteral-etal}
S. M. Catterall, M. Hasenbusch, R. R. Horgan, and R. L. Renken,
{\em The Nature of the continuum limit in the 2-$D$ $RP^2$ gauge model},
{\tt hep-lat/9801032};
Nucl. Phys. B (Proc. Suppl.) {\bf 63}, 670 (1998).

\bibitem{Nijs-etal}
M.~den Nijs, M. P. Nightingale, and M. Schick,
Phys. Rev. {\bf B26}, 2490 (1982).

\bibitem{Salas-Sokal-antiferromagnet}
J. Salas and A. D. Sokal, 
{\em The 3-state square-lattice Potts antiferromagnet at zero 
temperature}, {\tt cond-mat/9801079}.

\bibitem{BF-Arch-Rat-Mech-Anal}
M. E. Fisher and M. N. Barber, 
Arch. Rat. Mech. Anal. {\bf 47}, 205 (1972).

\bibitem{CP_LAT96} S. Caracciolo and A. Pelissetto,
Nucl. Phys. B (Proc. Suppl.) {\bf 53}, 693 (1997).

\bibitem{Niedermayer_LAT96} F. Niedermayer,
Nucl. Phys. B (Proc. Suppl.) {\bf 53}, 56 (1997).

\bibitem{Bell-Wilson} 
T. L. Bell and K. Wilson, 
Phys. Rev. {\bf B10}, 3935 (1974).

\bibitem{RV} P. Rossi and E. Vicari,
Phys. Lett. {\bf B389}, 571 (1996).

\bibitem{Hasenfratz-Niedermayer_97} 
P. Hasenfratz and F. Niedermayer,
Nucl. Phys. {\bf B507}, 399 (1997).

\bibitem{CEPS_RP}
S. Caracciolo, R. G. Edwards, A. Pelissetto, and A. D. Sokal,
Phys. Rev. Lett. {\bf 71}, 3906 (1993);
Nucl. Phys. B (Proc. Suppl.) {\bf 34}, 129 (1994).

\bibitem{CPS_RP_Dallas}
S. Caracciolo, A. Pelissetto, and A. D. Sokal,
Nucl. Phys. B (Proc. Suppl.) {\bf 34}, 683 (1994).

\bibitem{funzione2pt}
M.~Campostrini, A.~Pelissetto, P.~Rossi and E.~Vicari,
Europhys. Lett. {\bf 38}, 577 (1997);
Phys. Rev. {\bf E57}, 184 (1998);
Nucl. Phys. B (Proc. Suppl.) {\bf 53}, 690 (1997).

\bibitem{CP_lettera} 
S. Caracciolo and A. Pelissetto, 
 Phys. Lett. {\bf B402}, 335 (1997).

\bibitem{Luscher_massa} 
M. L\"uscher, Comm. Math. Phys. 
{\bf 104}, 177 (1986); {\bf 105}, 153 (1986).

\bibitem{Chandra} K. Chandrasekharan,
{\em Elliptic Functions}, (Springer, Berlin, 1985).

\bibitem{Fisher-Burford} 
M. E. Fisher and R. J. Burford, Phys. Rev. {\bf 156}, 583 (1967). 

\bibitem{CPRV_lettera}
M.~Campostrini, A.~Pelissetto, P.~Rossi and E.~Vicari,
Phys. Lett. {\bf B402}, 141 (1997).

\bibitem{Meyer_unp} S.~Meyer, unpublished, quoted in \cite{o3_letter}.

\bibitem{Morse} P. M. Morse and H. Feshbach, 
{\em Methods of theoretical physics}, vol. 1,
(Mc Graw-Hill, New York-Toronto-London, 1953).

\bibitem{GR} I. S. Gradshtein and I. M. Rizhik, 
{\em Tables of Integrals, Series and Products} (Academic Press, Orlando, 1980).

\bibitem{Abramowitz} 
M. Abramowitz and I. A. Stegun, 
{\em Handbook of mathematical functions}, 9th edition, (Dover, New York, 1972).

\bibitem{LW-theorem} M. L\"uscher and P. Weisz, 
Nucl. Phys. {\bf B266}, 309 (1986).







\end{thebibliography}
\end{document}